\documentclass[a4paper,11pt]{article}
\usepackage{pictex}

\DeclareSymbolFont{boldgr}{OML}{cmm}{b}{it}
\DeclareSymbolFontAlphabet{\bi}{boldgr}
\DeclareSymbolFont{boldcl}{OMS}{cmsy}{b}{n}
\DeclareSymbolFontAlphabet{\bcal}{boldcl}
\DeclareMathAlphabet{\Bbb}{U}{msb}{m}{n}
\DeclareMathAlphabet{\euf}{U}{euf}{b}{n}

\DeclareMathSymbol{\bgamma}{0}{boldgr}{"0D}
\DeclareMathSymbol{\bchi}{0}{boldgr}{"1F}
\DeclareMathSymbol{\bphi}{0}{boldgr}{"1E}
\DeclareMathSymbol{\bomega}{0}{boldgr}{"21}
\DeclareMathSymbol{\blambda}{0}{boldgr}{"15}
\DeclareMathSymbol{\bdot}{0}{boldcl}{"01}

\makeatletter
\@addtoreset{equation}{section}
\def\theequation{\arabic{section}.\arabic{equation}}
\def\@begintheorem#1#2{\trivlist\item[\hskip\labelsep{\it #1\ #2.}]}
\def\@opargbegintheorem#1#2#3{\trivlist
                   \item[\hskip\labelsep{\it #1\ #2\ (#3).}]}
\def\eqlabel[#1]{\label{#1}}
\def\beq{\arraycolsep .1em\begin{eqnarray}\@ifnextchar[{\eqlabel}{}}
\def\eeq{\end{eqnarray}}
\def\zl{\nonumber\\}
\def\txt#1{\quad\hbox{#1}\quad}

\newtheorem{exer}{Exercise}[section]
\newtheorem{theo}{Theorem}[section]
\newtheorem{defi}{Definition}[section]
\newenvironment{indented}{\begin{itemize}}
       {\end{itemize}\smallskip\hrule\smallskip}

\def\sref#1{section~\ref{#1}}
\def\fref#1{figure~\ref{#1}}
\def\Fref#1{Figure~\ref{#1}}
\def\eref#1{(\ref{#1})}

\def\Tr{\,\mathrm{Tr}}
\def\i{\mathrm{i}}
\def\d{\mathrm{d}}
\def\ssty{\scriptstyle}

\def\^{{}^}
\def\_{{}\lowind1_}
\def\lowind#1{\@ifnextchar_{\makebox[-#1pt]{ }}{}}
\def\alg#1({\euf{#1}(}
\def\grp#1({\mathsf{#1}(}
\def\cgrp#1({\cov{\mathsf{#1}}(}
\def\follows{\quad\Rightarrow\quad}
\def\equivalent{\qquad\Leftrightarrow\qquad}
\def\ft#1#2{{\textstyle{{#1}\over{#2}}}}
\def\intd#1#2{\int\!\d^{#1}\!#2 \,}
\def\intl#1#2{\int_{#1}^{#2}}
\def\intN#1#2{\intd2\x #1\Tr\big(#2\big)}
\def\intdN#1{\oint \d\phi\,\Tr\big(#1\big)}
\def\intdNl#1#2{\oint_{#1}\d\phi\,\Tr\big(#2\big)}
\def\Pexp{{\cal P}\!\exp}
\def\weak{\approx}
\def\comm#1#2{\big[#1,#2\big]}
\def\pois#1#2{\big\{#1,#2\big\}}
\def\Trr#1{\Tr\big(#1\big)}
\def\Trrr#1{\Tr\Big(#1\Big)}
\def\bei#1#2{\big|_{#1}^{#2}}
\def\opr#1{{}_R#1}
\def\opl#1{{}_L#1}
\def\del{\partial\lowind1}
\def\D{D\lowind1}
\def\nabl{\nabla\lowind1}
\def\deldel#1/#2/{\frac{\del #1}{\del #2}\,}
\def\dd#1/#2/{\frac{\d #1}{\d #2}\,}
\def\deltadelta#1/#2/{\frac{\delta #1}{\delta #2}\,}
\def\ddd{{\,\bdot}}
\def\send{\sum_{\endn}}
\def\e#1#2{e_{#1}\^{#2}}
\def\o{\omega\lowind1}
\def\oo{\bomega\lowind1}
\def\F{F}
\def\FF{{\bi F}}
\def\chris{{\mit \Gamma}}
\def\oophi{\bar\oo_\phi}
\def\gamphi{\bar\gam_\phi}
\def\spin{\bchi}
\def\a{{\bi a}}
\def\b{{\bi b}}
\def\c{{\bi c}}
\def\f{{\bi f}}
\def\g{{\bi g}}
\def\h{{\bi h}}
\def\n{{\bi n}}
\def\u{{\bi u}}
\def\v{{\bi v}}
\def\w{{\bi w}}
\def\tg{\widetilde\g}
\def\th{\widetilde\h}
\def\tu{\widetilde\u}
\def\inv{\!\!\^{-1}}
\def\GRP{G}
\def\eps{\varepsilon}
\def\gam{\bgamma}
\def\eins{{\mathbf{1}}}
\def\teins{{\widetilde \eins}}
\def\tgam{{\widetilde\gam}}
\def\RR{\Bbb{R}}
\def\CC{\Bbb{C}}
\def\MM{\Bbb{M}}
\def\ZZ{\Bbb{Z}}
\def\SS{\Bbb{S}}

\def\LG{{\Upsilon}}

\def\Sspace{{\cal S}}
\def\Slight{{\cal L}}
\def\Stime{{\cal T}}
\def\prodf{{\cal G}}
\def\M{{\cal M}}
\def\N{{\cal N}}
\def\X{{\bi x}}
\def\Y{{\bi y}}

\def\x{{x}}
\def\y{{y}}
\def\cov#1{\widetilde{#1} }

\def\xo{{\bar\x}}
\def\proj{\pi}
\def\kette{\bullet}
\def\shift{b}
\def\defangle{\zeta}
\def\scher{l}
\def\curva{\alpha}
\def\curvb{\beta}
\def\curvc{\gamma}

\def\loopb{\kappa}
\def\loopa{\lambda}
\def\loopc{\varrho}
\def\loopd{\varsigma}
\def\loopn{\eta}
\def\U{U}
\def\tU{\widetilde \UU}
\def\UU{{\bi U}}
\def\V{V}
\def\VV{{\bi V}}
\def\UUU{{\cal U}}
\def\VVV{{\cal V}}
\def\UUUU{{\bcal U}}
\def\VVVV{{\bcal V}}
\def\endn{\ell}
\def\endm{{\ell'}}
\def\endnull{\mathrm{o}}
\def\endeins{\mathrm{i}}
\def\Klasse{{\cal C}}
\def\winkel{\theta}
\def\Winkel{{\mit \Theta}}
\def\lange{v^2}
\def\Delt{{\mit \Delta}}
\def\pdelta{\widehat\delta}
\def\stepf{{\mit \Delta}}
\def\gpar{\xi}
\def\diag{\mathrm{diag}}
\def\sign{\mathrm{sign}}
\def\Wir{I}

\def\Lag{L}
\def\Lagb{L'}
\def\Lagkin{L_{\mathrm{kin}}}
\def\Lagint{L_{\mathrm{int}}}
\def\Lagbnd{L_{\mathrm{bnd}}}

\def\Ham{H}
\def\L{{\bi L}}
\def\H{{\bi H}}
\def\const{{\bi C}}
\def\lp{\blambda}   
\def\FS{{\cal F}\lowind2}   
\def\Obs{{\cal O}}
\def\where{\txt{where}}
\def\bra#1{\big\langle#1\,\big|}

\def\ket#1{\big|\,#1\big\rangle}
\def\Ket#1{\Big|\,#1\Big\rangle}
\def\kt#1{|#1\rangle}
\def\braket#1#2{\big\langle #1 \,\big|\, #2 \big\rangle}
\def\brkt#1#2{\langle #1|#2\rangle}
\def\hyperbel{\beginpicture
\plot -2.0 2.236
     -1.8 2.059
     -1.6 1.887
     -1.4 1.72
     -1.2 1.562
     -1.0 1.414
     -0.8 1.281
     -0.6 1.166
     -0.4 1.077
     -0.2 1.02
     0 1.0
     0.2 1.02
     0.4 1.077
     0.6 1.166
     0.8 1.281
     1.0 1.414
     1.2 1.562
     1.4 1.72
     1.6 1.887
     1.8 2.059
     2.0 2.236 /
\plot
     -2.0 -2.236
     -1.8 -2.059
     -1.6 -1.887
     -1.4 -1.72
     -1.2 -1.562
     -1.0 -1.414
     -0.8 -1.281
     -0.6 -1.166
     -0.4 -1.077
     -0.2 -1.02
     0 -1.0
     0.2 -1.02
     0.4 -1.077
     0.6 -1.166
     0.8 -1.281
     1.0 -1.414
     1.2 -1.562
     1.4 -1.72
     1.6 -1.887
     1.8 -2.059
     2.0 -2.236  /
\endpicture}
\def\coskurve{\beginpicture
\plot -1.000 .00
      -1.003 .04
      -1.015 .08
      -1.036 .12
      -1.065 .16
      -1.105 .20
      -1.156 .24
      -1.221 .28
      -1.302 .32
      -1.402 .36
      -1.527 .40
      -1.684 .44
      -1.881 .48
      -2.000 .50
      -1.881 .52
      -1.684 .56
      -1.527 .60
      -1.402 .64
      -1.302 .68
      -1.221 .72
      -1.156 .76
      -1.105 .80
      -1.065 .84
      -1.036 .88
      -1.015 .92
      -1.003 .96
      -1.00 1.00 /
\endpicture}
\def\Pfeil#1 #2 #3 #4 {\arrow <2ex> [.2,.5] from #1 #2 to #3 #4 }
\def\pfeil#1 #2 #3 #4 {\arrow <1.4ex> [.2,.5] from #1 #2 to #3 #4 }
\makeatother

\begin{document}
\begin{titlepage}
\begin{flushright}
gr-qc/9506069\\
KCL-TH-95-5
\end{flushright}
\begin{center}
{\LARGE  Three Dimensional Canonical Quantum Gravity}\\[3ex]
{\large Hans-J\"urgen Matschull\\[1ex]
        Department of Mathematics,
	King's College London,\\
	Strand,
	London WC2R 2LS,
	England\\[2ex]
        June 28, 1995}\\[2ex]
\end{center}
\begin{abstract}
General aspects of vielbein representation, ADM formulation and canonical
quantization of gravity are reviewed using pure gravity in three dimensions as
a toy model. The classical part focusses on the role of observers in general
relativity, which will later be identified with quantum observers. A precise
definition of gauge symmetries and a classification of inequivalent solutions
of Einstein's equations in dreibein formalism is given as well. In the quantum
part the construction of the physical Hilbert space is carried out explicitly
for a torus and cylinder type space manifold, which has not been done so far.
Some conceptual problems of quantum gravity are discussed from the point of
view of an observer sitting inside the universe.
\end{abstract}
\tableofcontents
\end{titlepage}

\section*{Why three dimensional canonical gravity?}
When you started to learn quantum mechanics as a student, you have certainly
studied some of the ``toy models'' appearing in every introductory lecture and
every book on quantum physics. These standard examples not only provide very
useful models to explain the ``canonical construction'' for the quantum state
space of a system which is classically defined by an action principle. They
also played an important role in the development of quantum mechanics itself.
Nowadays we are looking for a quantum theory of gravity, and maybe we can learn
something from quantum mechanics.
What makes the development of quantum gravity a bit different, is that up to
now, and possibly also in the near future, there are no ``key observations''
waiting to be explained. Instead, it is merely the lack of a mathematically
consistent theory of physics at the Planck scale that forces us to look for a
quantum theory of gravity. It is Einstein's classical theory that is in
contradiction with quantum physics at small length or high energy scales.
Nevertheless there are a few standard toy models which are not so different
from those considered in quantum mechanics, and studying these can help us to
get some ideas of how quantum gravity might work and what its main principles
are.

The hydrogen atom of quantum gravity is certainly the radiating black hole.
Explaining the Hawking radiation, which might in some sense be considered as a
key {\em observation}, within the framework of a mathematically consistent
model would probably be one of the main steps towards a quantum theory of
gravity, just like the explanation of the stability and discrete energy states
of the hydrogen atom was a milestone for quantum mechanics. What we have so far
are various semi-classical or perturbative models describing Hawking radiation
as quantum effects occuring at horizons in classical spacetimes, or lower
dimensional toy models with very special matter fields that allow a consistent
quantization and behave similar to black holes in four dimensions. Coming back
to quantum mechanics, these models should be compared with the Bohr model of
the hydrogen atom. It already gave an intuitive understanding of some
properties of atoms, but it was not based on a fully consistent theory.
However, this article is not about the hydrogen atom. It is about the harmonic
oscillator.

Like the harmonic oscillator in quantum mechanics, three dimensional pure
gravity is a very nice model to study some of the main features of quantum
gravity. It has all the typical features of gravity which are not present in
other field theories like, e.g., quantum electrodynamics. On the other hand,
all the technical difficulties normally appearing in quantum field theories are
absent. For example, we do not need to renormalize the fields, all relevant
operators will be well defined without regularization. In addition, complete
solutions at both the classical and the quantum level can be given explicitly,
so we can study the principal questions having explicit classical solutions and
a well defined Hilbert space of quantum states (for a review on principal
problems of quantum gravity see, e.g., \cite{isham:91,kuchar:92,isham:93}). In
fact, some of these conceptual problems of quantum gravity can be solved for
this toy model, and one can expect that similar solutions also exist in the
four dimensional theory.

Gravity is different from other field theories mainly because of its background
structure. Typical field theories, including the standard model, are based on a
flat Minkowski space, and the resulting global Poincar\'e symmetry also plays a
crucial role in any explicit calculation in these theories. It is especially
the existence of particles and the perturbative approach based thereon that
made the standard model so successful. There are several reasons why many
attempts to quantize gravity in a similar way failed. First of all, there is no
such background structure like Minkowski space in general relativity. There are
no particles and no special field configuration which could serve as a starting
point for perturbation theory. Moreover, even if one artificially introduces a
background and describes gravity as a perturbation around a fixed metric, the
resulting quantum theory turns out not to be renormalizable. Whereas this could
be merely a technical difficulty, there is another and maybe deeper reason for
the failure of these ``covariant'' quantization methods. When applied to a
gauge theory, most of them require some kind of gauge fixing, but one always
tries to keep the manifest Poincar\'e symmetry. It has even become a criterion
for a good quantum field theory to be manifestly Poincar\'e covariant. It is a
priori not clear how this could work in general relativity, where the
Poincar\'e group {\em is\/} the gauge group and a global Poincar\'e symmetry
does not exist in general.

So we should either look for something completely new, like string theory, or,
and this is what we are going to do here, go back to the early days of quantum
physics, forgetting everything we know about quantum field theory in Minkowski
space, and try to construct a quantum theory of Einstein gravity. Thereby we
will use the main ideas of quantum physics but not the sophisticated methods of
modern quantum field theory. Our ``old fashioned'' way of doing quantum physics
is known from non-relativistic quantum mechanics. Starting from a classical
action principle, one derives the Hamilton Jacobi formulation, translates phase
space coordinates into quantum operators, defines quantum states as wave
functions on which these operators act, and finally physics is described in
terms of expectation values of observables, defined with respect to a scalar
product on the state space. This procedure is know as the ``Dirac
programme''~\cite{dirac:65} or ``canonical quantization.'' There have been some
early attempts to use this method to quantize gravity, based on the Hamilton
Jacobi or ``ADM'' formalism derived by Arnowitt, Deser and
Misner~\cite{arnowitt.deser.misner:62,misner.thorne.wheeler:73}. The main
result was the famous ``Wheeler~DeWitt
equation''~\cite{wheeler:64,dewitt:67a,dewitt:67b,%
christensen:84}, but for a long time these approaches had no success either,
and they seemed to be technically even more difficult than the perturbative
methods.

With two important discoveries the situation changed drastically and the
``canonical quantization'' of gravity became a promising field of research
again. First it was shown by Ashtekar~\cite{ashtekar:86,ashtekar:87} that there
is a new set of variables for Einstein gravity which simplifies the canonical
treatment and hence the Dirac programme considerably. Closely related to this
Witten~\cite{witten:88} showed that using the dreibein and spin connection
instead of the metric as the basic fields variables, three dimensional gravity
can be quantized by straightforward application of Dirac's method. After that,
other methods were shown to produce similar results~\cite{carlip.nelson:94a,%
carlip.nelson:94b}.

The aim of this article is to provide a self-contained presentation of the
Dirac programme applied to three dimensional gravity, emphasizing mostly
physical aspects which are interesting for quantum gravity in general. All you
have to know to read this article is how to quantize systems with gauge degrees
of freedom in Dirac's formalism~\cite{dirac:65,hanson.regge.teitelboim:76}, and
some basic knowledge about general relativity, vielbein formalism, and the ADM
formulations of Einstein gravity will be useful,
too~\cite{misner.thorne.wheeler:73}. The new results presented here are mainly
based on earlier work~\cite{dewit.matschull.nicolai:93,%
matschull:doc}, and also related to similar approaches in four
dimensions~\cite{matschull:95a}.
It turns out that the combination of dreibein gravity, ADM formalism, and Dirac
quantization is an ideal tool helping us to understand the physics of quantum
gravity. Our approach will be a little bit unconventional, which hopefully
makes this article interesting for both beginners looking for an introductory
review and for experts, who might get some new insight into old questions
considered from a different point of view. Our point of view will be that of an
{\em observer\/} sitting inside the universe rather than looking at it from the
outside, which is not restricted to three dimensions. One of the central
questions will be whether we can make this observer a relativistic and a
quantum observer at the same time.

Moreover, instead of the ``standard'' example of a torus or higher genus
manifold, we will consider non-compact space manifolds. This makes the
mathematical treatment more complicated at first sight, because we have to
introduce boundary conditions, but on the other hand it will be much easier to
give a complete classification of solutions and, which is even more important,
to extract the physical properties of these solutions. This is because the
``boundaries'' of spacetime will behave like ``asymptotically flat ends,'' and
associated with them there will be something like a ``fixed stars background,''
which can be used to synchronize clocks, measure velocities using a generalized
``redshift'' and for many other ``astronomical'' measurements. So we really
want to do physics in our three dimensional universe, and notions like
``observer'' and ``measurement'' already appearing on the classical level will
help us to get some insight into the quantum theory.

Nevertheless we will also focus on some mathematical subtleties, which so far
have not been considered in detail. They turn out to be important when dealing
with non-compact manifolds. The first section will entirely be reserved to some
helpful mathematical tools. Having exact mathematical definitions of, e.g., a
covering manifold and the loop group, will enable us to give more precise
theorems to classify solutions of the equations of motions and to construct the
physical quantum state space. Sometimes it is also overlooked that the simply
connected group associated with the Lorentz algebra in three dimensions is
neither $\grp SO(1,2)$ nor $\grp SL(2)$, so we will also discuss this point in
more detail before going over to physics.

In \sref{kinematics} we will present our ``unconventional'' approach to general
relativity, which makes use of both the ADM formulation and the dreibein and
connection field as the basic variables. The central object in this approach is
the ``local observer'' sitting inside the universe and making experiments in
his ``local laboratory.'' Gravity will be considered as the theory that
describes how these observers communicate with each other. As an additional
feature of this interpretation of gravity, it will not break down if we allow
exactly those kinds of singular metrics which are needed to quantize the theory
based on Witten's idea. An explicit example will be given to show what happens
at spacetime singularities.

We will also introduce some non-local objects like parallel transport operators
and geodesic distances, which are closely related to the astronomical
measurements mentioned above.
These non-local objects are needed because three dimensional gravity is not
really a field theory. Einstein's equations in three dimensions imply that
spacetime is locally flat (see, e.g.,~\cite{brown:88} for an introduction to
lower dimensional gravity and many references). Hence, there are no
gravitational waves or forces.
Spacetime locally looks like Minkowski space (unless we are in the
neighbourhood of a singularity), and no local observables like curvature exist.
Non-trivial solutions to the field equations only occur if the spacetime
manifold is not simply connected. Then, e.g., transporting a vector once around
a non-contractible loop will in general result in a rotated vector, and the
``angle'' of this rotation is one of the finitely many degrees of freedom of
gravity in three dimensions.

To identify all such parameters classifying a solution of Einstein's equations
modulo gauge transformations will be the aim of \sref{class}. Here we will need
all the mathematical structures introduced in \sref{mathematics}. We will find
that this classification is much simpler for space manifolds with ``asymptotic
ends'' than for those without like the torus. As an example, we will choose an
infinitely long cylinder as space and construct some ``funny spacetimes,''
showing that there is a rich set of inequivalent solutions even for the
simplest non-trivial manifold. This will also give us some understanding of the
physical content of the gauge invariant parameters found to classify the
solutions.

In \sref{quantization} we are going to quantize the model using Dirac's method.
As at the classical level, the reduction from a field theory to a state space
with finitely many degrees of freedom will come out automatically. It will be
even simpler to identify the quantum state space than to find the classical
solutions. The result will be a physical state space consisting of ``wave
functions'' depending on finitely many variables like in ordinary quantum
mechanics, and a set of ``observables'' acting thereon as linear operators. We
will also discuss some problems arising with products of non-commuting
operators. As a consequence, for some of the observables the quantum operator
will be defined up to ordering ambiguities only, or equivalently up to
corrections of the order $\hbar$. However, we will find that the ordering is
unique if we want the observables to become hermitian.

In the last section we will briefly discuss some conceptual problems of quantum
gravity, and show how to solve them for our toy model. For example, we will
describe what the physical content of the observables is, and how the ``problem
of time'' can be solved by using special phase space functions that behave like
clocks. For a cylinder manifold we will find a unique Hilbert space such that
all observables become hermitian operators, and we will also see that there is
a unique physical Hilbert space for the torus as well. Though the torus has
always been the standard example, an explicit construction of the scalar
product on the physical state space has not been carried out so far. We will
find that up to rescaling there is only one reasonable product such that all
standard observables become hermitian.

As we want to be mathematically as precise as possible, you will find some
typically mathematical structures in this article. The main objects are
introduced in ``definitions,'' and a few important results are summarized in
``theorems.'' This makes it somehow easier to refer to them later on, and it
also helps to organize the article. In addition, there are lots of
``exercises.'' They mainly split into two categories. The first type just
summarizes some simple properties of new symbols or notions, which are either
easy to prove or interesting but actually not required for our special purpose.
 Another kind of exercise typically consists of a ``physical'' question. Here
it is sometimes better to ask a precise question than to describe things
lengthly. Some exercises also suggest generalizations or extensions, which are
mostly straightforward but beyond the scope of this article.

 \section{Mathematics}
\label{mathematics}
In the main part of this article we want to focus on the physical aspects of
three dimensional gravity, so maybe it's best to introduce all the mathematical
notions necessary for this purpose in a separate section. We then have all the
structures at hand when we need them. There are mainly two mathematical objects
playing a crucial role: the loop group (or first fundamental group) of some
manifold and the three dimensional Poincar\'e group, which is the simply
connected Lie group associated with the Lie algebra $\alg iso(1,2)$.
Introducing these two objects will give us all the relevant mathematical
structure.

In principle we should also introduce a number of fibre bundles to allow exact
differential geometrical definitions for the various ``fields''. However, we
will restrict our considerations to models based on manifolds with trivial
tangent bundles. This class of models is already interesting enough for our
purpose. All the bundles will then become trivial and sections therein may be
treated as fields, coordinates are always global etc. The notation (and also
the mathematical language) simplifies considerably. All the constructions made
below can nevertheless be extended to ``higher genus'' manifolds with
non-trivial tangent bundles, and for those familiar with fibre bundles,
sections and connections on them, here is a good exercise:
\begin{exer}[Extension to general manifolds]\label{difgeo}
For all the ``fields'' introduced in this article, give the correct definition
in differential geometrical terms (some of them are section, others are
connections or maps from one fibre into another), and show that the general
results are correct for manifolds with non-trivial tangent bundles as well.
Sometimes one has to make other restrictions, e.g.\ the manifold has to be
connected, orientable, etc.
\end{exer}

\subsection{Covering manifold and loop group}\label{covering}
Given a manifold $\N$, we introduce its covering manifold $\cov\N$. Physicists
often define it by somehow ``unwrapping'' $\N$, so that the result is a simply
connected manifold which is locally isomorphic to $\N$. This is rather
illustrative, but there is a better ``mathematical'' definition, which turns
out to be exactly the right thing for us here and, in addition, which will give
us a second structure, the loop group, for free.
\begin{defi}[Covering Manifold]
Let $\N$ be a connected manifold (in all our applications two dimensional,
orientable). A path in $\N$ is a (smooth) map $\curva:[0,1]\mapsto\N$. Two
paths are equivalent (or homotopic) if they can be (smoothly) deformed into
each other without varying the end points, i.e.
\beq
  \curva \sim \curvb  \quad\Leftrightarrow\quad
     \exists f:[0,1]^2\to\N, \quad & &
     f(0,s)=\curva(s), \quad  f(t,0)=\curva(0)=\curvb(0), \zl &&
     f(1,s)=\curvb(s), \quad  f(t,1)=\curva(1)=\curvb(1).
\eeq
If two paths with the same end points are always homotopic, the manifold is
called simply connected. Now fix a point $\xo\in\N$, then the covering manifold
is the set of all homotopy classes of paths beginning at $\xo$:
\beq
   \cov\N = \big\{ \curva \,\big| \, \curva(0)=\xo  \big\} \big/ \mbox{$\sim$}.
\eeq
\end{defi}
By ``smooth'' we always mean sufficiently often differentiable.
The covering manifold is in fact a (discrete) bundle over $\N$ and thus you can
think of $\cov\N$ as somehow wrapped around $\N$ such that when you go once
around a non-contractible loop you are on a different sheet of $\cov\N$. But
the definition as homotopy classes of paths is more suitable as paths will
naturally appear whenever we deal with $\cov\N$.
There are some natural maps on $\cov\N$, the first is the projection back onto
$\N$:
\begin{defi}[Projection]
\beq[projection]
  \proj : \cov\N \to \N , \qquad
  \proj(\curva) = \curva(1).
\eeq
\end{defi}
We use the same symbols $\curva,\curvb,\dots$ for paths and for their homotopy
classes, i.e.\ points in $\cov\N$; this should not cause trouble. We only have
to check that expressions are invariant under deformations whenever $\curva$
denotes a special path. In \eref{projection} $\curva$ denotes a homotopy class
(an element of $\cov\N$) on the left hand side, but a special path on the right
hand side. However, $\curva(1)$ is the same for all path belonging to a given
homotopy class.

An important feature of paths is that they may be linked together, simply by
going along one first and then along the other:
\begin{defi}[Link]\label{link-def}
For two paths $\curva:s\mapsto\curva(s)$ and $\curvb:s\mapsto\curvb(s)$ such
that $\curva(1)=\curvb(0)$, the link $\curva\kette\curvb$ is given by
\beq[link]
    \curva\kette\curvb : s \mapsto \cases{ \curva(2s) & for $s\le\ft12$ , \cr
                                        \curvb(2s-1) & for $s\ge\ft12$.}
\eeq
\end{defi}
With the help of this operation one can show that the structure of the covering
manifold does not depend on the choice of the arbitrary point $\xo$, if $\N$
itself is connected, i.e.\ for any two points there must be a path connecting
them.
\begin{exer}
The projection $\proj$ is locally one-to-one, and by using the same coordinates
$\cov\N$ becomes a differentiable manifold. If another origin $\xo'$ defines
another covering manifold $\cov\N'$, then there is a one-to-one diffeomorphism
$f:\cov\N\to\cov\N'$ such that $\proj'\circ f=\proj$.
\end{exer}
The link is also well defined for homotopy classes of paths, and thus it also
acts on $\cov\N$. However, elements of $\cov\N$ can only be linked if the first
element is a loop, i.e.\ a path beginning and ending at $\xo$. This subset of
$\cov\N$ already provides the loop group of $\N$:
\begin{defi}[Loop Group]
The set of loops $\LG\subset\cov\N$ consists of all (homotopy classes of) paths
$\loopa$ ending at $\xo$, and the group multiplication is $\kette$. The unity
$\loopn$ and the inverse $\loopa^{-1}$ are given by (the homotopy classes of)
the constant loop and the reversed loop, respectively,
\beq
  \loopn(s)= \xo , \qquad
  \loopa^{-1}(s) = \loopa(1-s).
\eeq
\end{defi}
The last equation can also be used to define an inverse of any path $\curva$,
or of any homotopy class of paths, but for $\curva\in\cov\N$ the inverse
$\curva^{-1}$ is in general no longer an element of $\cov\N$.
As a set we have $\LG=\proj^{-1}(\xo)$, i.e.\ $\LG$ consists of all points of
$\N$ lying ``over $\xo$''. Our definition of the loop group does not require
any additional structure: it is just a subset of $\cov\N$ and the
multiplication is \eref{link}.

We can not only link a loop with another loop but also with another path
starting at $\xo$. This gives a second natural map on $\cov\N$:
\begin{defi}[Shift]
For every loop $\loopa\in\LG$ there is a diffeomorphism
\beq[shift]
  \shift_\loopa : \cov\N \to \cov\N , \qquad
     \curva \mapsto \loopa \kette \curva.
\eeq
\end{defi}
If the manifold $\N$ is not too complicated, i.e.\ it does not have infinitely
many ``holes'' or something like that, than the loop group has a simple
structure. In all our examples it will be a discrete group generated by a
finite number of elements.
\begin{defi}[Primitive Loops]\label{primloop}
A finite set $\LG_0\subset\LG$ is called a set of primitive loops, if every
$\loopa\in\LG$ can be written as a finite product of elements of $\LG_0$ and
their inverses. The loop group itself is then given by the free group generated
by $\LG_0$, i.e.\ the set of all possible products of elements, modulo some
relations, i.e.\ there may be products of primitive loops which are
contractible.
\end{defi}
A well know example is the torus, where we have two primitive loops
$\LG_0=\{\loopa,\loopb\}$, and one relation $\loopa \kette \loopb \kette
\loopa^{-1} \kette \loopb^{-1} = \loopn$.
An example for a manifold with boundaries will be given below.

There are some other useful notions we introduce here without motivation. They
will become important later on, but here is a better place to give their
definitions. The first is rather simple.
\begin{defi}[Periodic Function]
A function $f$ on $\cov\N$ is called periodic if $f\circ\shift_\loopa=f$ for
all $\loopa\in\LG$.
\end{defi}
Obviously, the projection $\proj$ is periodic, and every periodic function can
be written as $f\circ\proj$ with some function $f$ on $\N$. To simplify the
notation, we always use the same symbol for a periodic function on $\cov\N$ and
the corresponding function on $\N$. More interesting than periodic functions
are quasiperiodic functions.
\begin{defi}[Quasiperiodic Function]\label{quasi-def}
Let $\GRP$ be a group. A map $\g:\cov\N\to\GRP$ is called quasiperiodic if
there are elements $\g_\loopa\in\GRP$ such that
$\g\circ\shift_\loopa=\g_\loopa\g$ for all $\loopa\in\LG$.
\end{defi}
Hence, the shifted function $\g\circ\shift_\loopa$ differs from $\g$ itself by
multiplication with a constant from the left only. It follows immediately that
$\loopa\mapsto\g_\loopa$ is a group homomorphism $\LG\to\GRP$.
\begin{defi}[Normalized Quasiperiodic Function]\label{norm-def}
A quasiperiodic function is called normalized if $\g_\loopa=\g(\loopa)$, or
equivalently if $\g(\loopn)=\eins$, the unity element of $\GRP$.
\end{defi}
Such functions will appear naturally as parameters for the solutions of the
equations of motion, and we will also need them as auxiliary fields necessary
to deal with boundaries.

The last construction in this section also has to do with boundaries. Assume
that $\N$ is a compact 2-manifold with some boundaries. Each boundary labeled
by an index $\endn$ is a circle (it is a 1-manifold, but it cannot be a real
line because $\N$ is compact). How does the boundary of $\cov\N$ look like? To
see this, we introduce a periodic coordinate $\phi$ on the boundary of $\N$
(with period $2\pi$). Then we can construct a family of paths
$\curva_{\endn,\phi}\in \cov\N$, all starting at $\xo$ and ending at the
boundary point with coordinate $\phi$. Once we fixed the path
$\curva_\endn=\curva_{\endn,0}$, all other paths are unique up to smooth
deformations. \Fref{boundpath} shows how the family looks like.
\begin{figure}[t]
\caption{The family of paths $\curva_{\endn,\phi}$ associated with a boundary
of $\N$ (the bold circle). The family is fixed by choosing the homotopy class
of $\curva_\endn$. There is also a loop $\loopa_\endn\in\LG$ associated with
the boundary, which is given by
$\loopa_\endn=\curva_{\endn,\phi+2\pi}\kette\curva_{\endn,\phi}\inv$ for any
$\phi$.}
\label{boundpath}
\begin{indented}
\item[] \quad
\beginpicture
\setcoordinatesystem units <3em,3em>
\put {$\bullet$} at 0 0
\put {$\xo$} [br] at -.1 0
\pfeil 4.6 -.2 4.25 0
\put {$\N$} at .5 1.5
\put {boundary}  at 5.5 .7
\put {``$\endn$''} at 5.5 .3
\put {$\phi=0,2\pi,\dots$} [lt] at 4.6 -.2
\pfeil 0 0 2 0
\plot 2 0 4.2 0 /
\put {$\curva_\endn$} [tl] at 2 -.1
\plot 2 .81 3.1 1.25 /
\pfeil 0 0 2 .81
\ellipticalarc axes ratio 2:1 -100 degrees from 3.1 1.25 center at 3.7 .8
\put {$\curva_{\endn,\phi}$} [tl] at 2 .71
\pfeil 0 0 2 1.6
\put {$\curva_{\endn,\phi+2\pi}$} [tl] at 2 1.5
\ellipticalarc axes ratio 4:1 -130 degrees from 2 1.6 center at 4.3 1.4
\ellipticalarc axes ratio 4:3 -180 degrees from 6.4 1.71 center at 5.3 -.1
\ellipticalarc axes ratio 3:5 -180 degrees from 4.2 -1.91 center at 4.44 -.45
\pfeil 0 0 1.6 1.8
\put {$\loopa_\endn$} [br] at 1.6 1.9
\plot 0 0 -.245 -.86 /
\ellipticalarc axes ratio 3:1 -120 degrees from 1.6 1.8 center at 4.9 1.4
\ellipticalarc axes ratio 3:2 -140 degrees from 7.6 2.15 center at 6 0
\ellipticalarc axes ratio 3:1 -126 degrees from 6.82 -2.33 center at 4.4 -1
\put { } at -.2 2.5
\put { } at -.2 -2.5
\put { } at 8 2.9
\put { } at 8 -2.9
\setplotsymbol ({\bf .})
\circulararc 360 degrees from 4.2 0 center at 5.5 0
\endpicture
\end{indented}
\end{figure}
Starting from $\curva_{\endn}$ the paths $\curva_{\endn,\phi}$ are obtained by
moving the end point along the boundary. All these paths represent points on
the boundary of $\cov\N$. The family of paths is not unique: there are in
general several boundaries of $\cov\N$ projected onto the same boundary of
$\N$, as choosing another $\curva_\endn$, say $\loopa\kette\curva_\endn$ for
some $\loopa\in\LG$, we end up with a different family of paths in
\fref{boundpath}, corresponding to a different boundary of $\cov\N$. However,
there is a special element of the loop group for which we do not get a
different family but only a reparameterization of the same family. Calling this
$\loopa_\endn$, we have
$\loopa_\endn\kette\curva_{\endn,\phi}=\curva_{\endn,\phi+2\pi}$.

If we now assume that this $\loopa_\endn$ is not contractible, and that this is
also true for any power of it, i.e.\ $(\loopa_\endn)^z\ne\loopn$ for $z\ne0$,
then two paths from the same family are never homotopic.
Hence, the boundary of $\cov\N$ is a real line parameterized by $\phi\in\RR$.
For each $\loopa\in\LG$, which is not a power of $\loopa_\endn$, we get a
different line represented by $\loopa\kette\curva_{\endn,\phi}$, $\phi\in\RR$.
In general there are infinitely many such lines, so the boundary of $\cov\N$ is
rather complicated. However, for our purposes it is sufficient to know that for
each boundary of $\N$ we can construct a {\em special\/} family of paths
$\curva_{\endn,\phi}$, which is not unique, but once fixed there is a unique
loop $\loopa_\endn$ that goes exactly once around that boundary.

As an example, consider a sphere with $N\ge2$ holes cut out, so that we have
$N$ boundaries like in \fref{boundpath}. We can arrange the paths
$\curva_\endn$ such that for $\endn=1,\dots,N$ they leave $\xo$ ``clockwise''
and do not intersect. Then we can take the $\loopa_\endn$ to be the primitive
loops and the loop group of $\N$ is given as described in
definition~\ref{primloop} with the relation
$\loopa_{1}\kette\cdots\kette\loopa_{ N}=\loopn$, as linking all the loops
together we get a single loop going once around all the holes, which can be
contracted. For $N=2$ the two families form the complete boundary of $\cov\N$
($\N$ is a cylinder and $\cov\N$ is a infinitely long strip). But $\cov\N$ has
infinitely many boundaries for $N\ge3$.
\begin{exer}\label{loop-free}
Of course, we can drop the loop $\loopa_N$ from $\LG_0$. Then the loop group
becomes the free group generated by $\LG_0=\{\loopa_1,\dots,\loopa_{N-1}\}$.
More general: show that the loop group of a manifold consisting of a genus $g$
surface, with $N$ holes like in \fref{boundpath} cut out, is a free group
generated by $2g+N-1$ primitive loops.
\end{exer}
Finally, here are some useful features of paths and loops, some of which have
already been used:
\begin{exer}
Show that
\begin{enumerate}
\item $\cov\N$ is simply connected, or equivalently
$\,\cov{\!\cov\N\,}\!=\cov\N$;
\item definition~\ref{link-def} can be changed such that $\curva\kette\curvb$
is smooth at $s=1/2$;
\item the link is well defined on homotopy classes of paths;
\item $\LG$ is a group;
\item the shift maps provide a representation of $\LG$:
     $\shift_\loopn={\bf id}, \quad
       \shift_\loopa \circ \shift_\loopb = \shift_{\loopa\kette\loopb}$.
\end{enumerate}
\end{exer}

\subsection{Symmetry algebras and groups}
\label{lorentz-group}
Three dimensional gravity can be formulated as a gauge theory, the gauge group
being the Poincar\'e group. The basic fields can be combined into a single
connection and the equations of motion then require the field strength of this
connection to vanish. Thus, to understand three dimensional gravity it is
essential to know how the Poincar\'e algebra and especially the corresponding
group looks like. The basic observables, i.e.\ the gauge invariant parameters
of the solutions of the equations of motion, will turn out to be elements of
this group.
\begin{defi}[Poincar\'e Algebra]
The Poincare algebra in three dimensions is six dimensional, and a suitable
basis is given by the three Lorentz rotations $\L^a$ together with three
translations $\H^a$, $a=0,1,2$ labeling the coordinates of three dimensional
Minkowski space $\MM^3$. Their commutation relations read
\beq
  \comm{\L^a}{\L^b}=-\eps^{ab}\_c \L^c , \quad
  \comm{\L^a}{\H^b}=-\eps^{ab}\_c \H^c , \quad
  \comm{\H^a}{\H^b}=0,
\eeq
where the structure constants are given by the Levi Civita symbol obeying
$\eps^{012}=-\eps_{012}=+1$, and indices are raised and lowered using the
Minkowski metric $\eta_{ab}=\diag(-1,1,1)$. Small Latin indices $a,b,c,\dots$
are also called flat indices, referring to the flat metric $\eta_{ab}$.
\end{defi}
Let us first have a look at the Lorentz subalgebra spanned by $\L^a$. The
identity $2\eta^{ab}=\eps^{ac}\_d\eps^{bd}\_c$ tells us that $\eta_{ab}$ can be
chosen to be the metric on the Lorentz algebra, so we can identify it with
three dimensional Minkowski space $\MM^3$ (as a vector space). We need two
matrix representations of this algebra:
\begin{defi}[Representations of the Lorentz Algebra]
\label{lorentz-rep}
The vector (or spin 1) representation of the Lorentz algebra is $\alg so(1,2)$
and given by
\beq[vec-rep]
   \L^a \mapsto \eps^{a\ddd}\_\ddd, \qquad
    \comm{\eps^{a\ddd}\_\ddd}{\eps^{b\ddd}\_\ddd} =
        - \eps^{ab}\_c \eps^{c\ddd}\_\ddd,
\eeq
where the dots denote flat indices to be treated as matrix indices. The spinor
(or spin 1/2) representation is $\alg sl(2)$ and the basis is given by
\beq[spin-rep]
   \L^a \mapsto \ft12 \gam^a, \qquad
    \comm{\ft12\gam^a}{\ft12\gam^b} =
        - \eps^{ab}\_c\, \ft12\gam^c ,
\eeq
where the gamma matrices form a Clifford algebra
\beq[cliff]
   \gam_a \gam_b = \eta_{ab} \, \eins - \eps_{abc} \, \gam^c,
\eeq
and they are explicitly given by
\beq[gamma-def]
   \gam_0 = \pmatrix{0&1\cr -1&0}, \quad
   \gam_1 = \pmatrix{0&1\cr 1&0}, \quad
   \gam_2 = \pmatrix{1&0 \cr0&-1}.
\eeq
\end{defi}
The representations act on vectors $v^a\in\MM^3$ and (Majorana) spinors
$\spin\in\SS^2$, where $\SS^2$ is some real two dimensional vector space. We
will avoid products of spinors so it does not matter whether spinors are
commuting or anticommuting, and no additional structure is needed on $\SS^2$.
To keep the notation consistent, we don't write out spinor indices explicitly.
Vector indices are sometimes ``abbreviated'' by a dot to indicate that they are
used like matrix indices. We also use the gamma matrices to map $\MM^3$ onto
$\alg sl(2)$ and vice versa:
\beq
   \v = v^a \, \gam_a, \qquad
   v^a = \ft12 \Trr{\v\gam^a} , \qquad
   v^a w_a = \ft12 \Trr{ \v \w } ,
\eeq
showing explicitly that $\MM^3\simeq\alg sl(2)$. We use this isomorphism
frequently and also speak of a ``vector'' $\v$ which is actually a matrix. We
call it spacelike, timelike or lightlike, or orthogonal to some other matrix,
depending on the corresponding property of $v^a$.

Now it is in principle straightforward to define the Lorentz group as the
unique simply connected Lie group associated with the Lorentz algebra.
Unfortunately this is neither $\grp SO(1,2)$ nor $\grp SL(2)$, which would be
similar to  $\grp SU(2)$ as the simply connected Lie group associated with
$\alg so(3)\simeq\alg su(2)$. In fact, $\grp SL(2)$ is a two-fold covering of
$\grp SO(1,2)$, but it is not yet simply connected. To see this, let's have a
closer look at $\grp SL(2)$.

\begin{figure}[t]
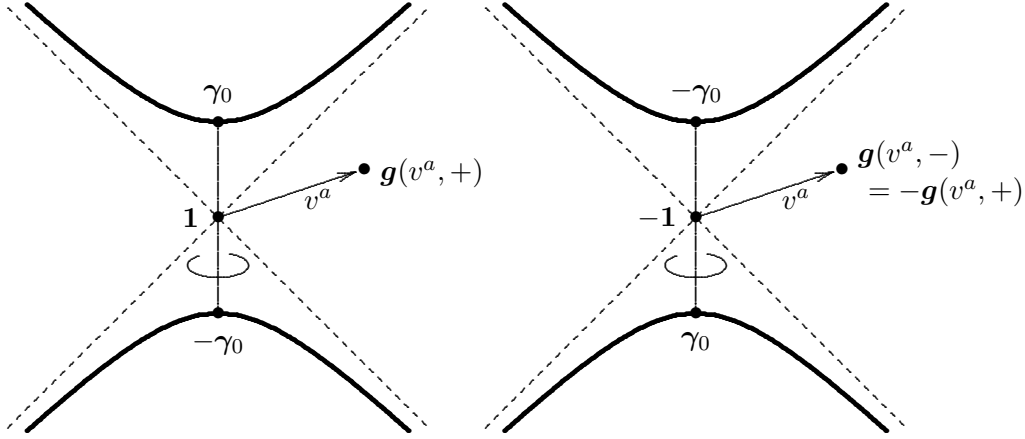

\caption{The group $\grp SL(2)$ consists of two copies of the part of $\MM^3$
between the hyperboloids of proper length squared $-1$. The upper hyperboloid
of the right part is glued to the lower one on the left and vice versa. The
three dimensional manifold is obtained by rotating the picture around the
vertical axis. The covering group $\cgrp SL(2)$ consists of a series of such
parts, each one's upper boundary glued to the next's lower boundary.}
\label{sl2}
\begin{indented}
\item[] \quad
\beginpicture
\setcoordinatesystem units <3.3em,3.3em>
\plot 0 -1 0 1 /
\plot 5 -1 5 1 /
\ellipticalarc axes ratio 5:2 300 degrees from -.2 -.4 center at 0 -.5
\ellipticalarc axes ratio 5:2 300 degrees from 4.8 -.4 center at 5 -.5
\setdashes <.5ex>
\plot -2.2 -2.2 2.2 2.2 /
\plot -2.2 2.2 2.2 -2.2 /
\plot 2.8 -2.2 7.2 2.2 /
\plot 2.8 2.2 7.2 -2.2 /
\put { } at -2.2 -2.4
\put { } at -2.2 2.4
\put { } at 7.2 2.4
\put { } at 7.2 -2.4
\setsolid
\put {$\bullet$} at 0 0 \put {$\eins$} [r] at -.2 0
\put {$\bullet$} at 5 0 \put {$-\eins$} [r] at 4.8 0
\put {$\bullet$} at 0 1 \put {$\gam_0$} [b] at 0 1.2
\put {$\bullet$} at 5 1 \put {$-\gam_0$} [b] at 5 1.2
\put {$\bullet$} at 0 -1 \put {$-\gam_0$} [t] at 0 -1.2
\put {$\bullet$} at 5 -1 \put {$\gam_0$} [t] at 5 -1.2
\pfeil 0 0 1.44 .48
\put {$v^a$} [tl] at .9 .3
\put {$\bullet$} at 1.53 .51
\put {$\g(v^a,+)$} [l] at 1.7 .47
\pfeil 5 0 6.44 .48
\put {$v^a$} [tl] at 5.9 .3
\put {$\bullet$} at 6.53 .51
\put {$\g(v^a,-)$} [lb] at 6.7 .51
\put {$\ =-\g(v^a,+)$} [lt] at 6.7 .42
\setplotsymbol ({\bf .})
\put {\hyperbel} at 0 0
\put {\hyperbel} at 5 0
\endpicture
\end{indented}
\end{figure}

It consists of all $2\times2$ matrices with determinant one. The gamma matrices
together with the unity matrix $\eins$ provide a basis of all $2\times2$
matrices, so we can write an element of $\grp SL(2)$ as
\beq[sl2-mat-gamma]
  \g = \hat v \,  \eins + v^a \gam_a , \quad \hat v \in \RR, \quad v^a \in
\MM^3.
\eeq
Calculating the determinant yields the restriction
\beq[sl2-restr]
  \hat v^2 - v^a v_a = 1 \follows \hat v = \pm \sqrt{\lange +1},
\eeq
where $\lange=v^av_a$.
Hence, we must have $\lange\ge-1$: the vector $v^a\in\MM^3$ must lie in the
region between the two timelike hyperboloids of length squared $-1$. For each
vector there are two elements of $\grp SL(2)$, corresponding to the sign in
\eref{sl2-restr}, except for $\lange=-1$, where the two signs actually yield
the same group element. Changing \eref{sl2-mat-gamma} slightly we introduce a
useful set of coordinates on the group:
\beq[sl2-par]
   \g(v^a,\pm)  &=&  \pm \big( \hat v \,   \eins +  v^a \gam_a  \big), \qquad
   \lange\ge -1, \quad \hat v = + \sqrt{\lange+1}, \zl
   &&\follows \g(v^a,+) = \g(-v^a,-) \txt{for} \lange=-1.
\eeq
A picture of this coordinate system is drawn in \fref{sl2}.
We call the matrices $\g(v^a,\pm)$ spacelike, lightlike, timelike, or null
depending on the length squared of the vector $v^a$. The null matrices are
obviously $\eins$ and $-\eins$, and for all others we have $|\ft12\Tr(\g)|$
greater, equal, or less than $1$ for spacelike, lightlike, or timelike
matrices, respectively.

Two show that $\grp SL(2)$ is a two-fold covering of $\grp SO(1,2)$, we have to
look for a group homomorphism that maps $\g$ onto a matrix $g^a\_b$ satisfying
$g^a\_bg^c\_d\eta_{ac}=\eta_{bd}$. The solution is given here:
\begin{exer}
\beq[homo]
\grp SL(2) \to \grp SO(1,2)_+ : \quad
 \g \mapsto g^a\_b = \ft12 \Tr (\g^{-1} \gam^a \g \gam_b )
\eeq
is a two-to-one group homomorphism mapping $\g$ and $-\g$ onto the same
$g^a\_b$.
\end{exer}
If an $\grp SL(2)$ matrix is denoted by a bold letter, we denote the projected
$\grp SO(1,2)$ matrix by adding indices to the ``un-bolded'' symbol. We also
drop the $\ssty+$ index from $\grp SO(1,2)$, which always means the set of
proper Lorentz transformations with $g^0\_0>0$. The elements $\g$ and $-\g$ are
located at the same point in \fref{sl2} but in different parts, so $\grp
SO(1,2)$ just consists of one such part only, the upper and lower hyperboloid
identified.

In \fref{sl2} we see that $\grp SL(2)$ is not simply connected, i.e.\ it is not
the unique simply connected Lie group associated with $\alg sl(2)\simeq\alg
so(1,2)$. There is a non-contractible loop going along the central axis from
$-\gam_0$ to $\gam_0$ in one part of the picture and then back on the other
side. This is the $\grp SO(2)$ subgroup of $\grp SL(2)$. With the tools
introduced in the previous section, we know how to obtain the simply connected
group we are looking for. It is the covering manifold of $\grp SL(2)$. It is
obvious how this looks like if we draw a picture like \fref{sl2}. We get a
whole series of parts instead of just two and they are glued together to a
chain where each upper hyperboloid is glued to the lower boundary of the next
part.
\begin{defi}[Lorentz Group]
The Lorentz group $\cgrp{SL}(2)$ is the covering manifold of $\grp SL(2)$ (or
$\grp SO(1,2)$) with basepoint $\bar\x=\eins$, i.e.\ its elements are homotopy
classes of paths in $\grp SL(2)$, denoted by $\tg:s\mapsto \g(s)$, with
$\g(0)=\eins$. The product of two paths is defined ``pointwise'':
\beq
   \tg : s\mapsto \g(s), \quad
   \th : s\mapsto \h(s) \follows
   \tg \th: s \mapsto  \g(s) \h(s).
\eeq
\end{defi}
The natural projection $\cgrp SL(2)\to\grp SL(2)$ is denoted by dropping the
tilde from the symbol. Note that in addition to the group multiplication there
is also the link defined on $\cgrp SL(2)$, and there is in fact a simply
relation between them.
\begin{exer}
Show that the group multiplication coincides with the link $\kette$, i.e.\
$\tg\th=\tg\kette\th$, if $\tg$ is a loop. The loop group of $\grp SL(2)$ is a
discrete subgroup of $\cgrp SL(2)$.
\end{exer}
Unfortunately there is no faithful matrix representation for the covering
group, so the definition as homotopy classes of paths in $\grp SL(2)$ is all we
have. However, we can find some useful coordinates on the covering group. We
can extend \eref{sl2-par} using an integer instead of a sign, leading to the
coordinates
\beq[csl2-par]
   \tg(v^a,z), &&\quad
   \lange\ge -1, \quad
   z\in\ZZ,\zl
   \tg(v^a,z)&=&\tg(-v^a,z+1) \txt{for} \lange=-1, \ v^0 > 0 .
\eeq
The projection onto $\grp SL(2)$ reads $\tg(v^a,z)\mapsto\g(v^a,(-)^z)$.
Another set of coordinates is provided by exponentiating the algebra.
\begin{exer}
Show that every element of $\grp SL(2)$ can be written in the form
\beq[alg-sl2]
  \g = \pm \exp ( n^a \gam_a ),
\eeq
but the spacelike sector of the right part in \fref{sl2} cannot be ``reached''
if the sign is positive, i.e.\ there are group elements which are not
exponentials of elements of the algebra.
\end{exer}
For the covering group we define
\beq[exp-csl2]
  \exp ( n^a \tgam_a ) :s \mapsto \exp (s \, n^a \gam_a ),
\eeq
which provides the natural exponential map from the algebra into $\cgrp SL(2)$.
Of course, the notation is only formal, as there are no matrices $\tgam_a$ such
that this map is given by exponentiating them. We can only define it as a
homotopy class of paths in $\grp SL(2)$.
Observe that $\exp(2\pi\gam_0)$ is the unity matrix but $\exp(2\pi\tgam_0)$ is
not the unity of $\cgrp SL(2)$.
Similar to \eref{alg-sl2} every element of $\cgrp SL(2)$ can be written as
\beq[alg-csl2]
  \tg = \teins_z \exp ( n^a \tgam_a ), \txt{where}
  \teins_z  = \exp ( z \pi \tgam_0 ) =  \tg(0,z).
\eeq
The group element $\teins_z$ is the origin of part no.~$z$ in \fref{sl2}. It is
a central element of $\cgrp SL(2)$, i.e.\ it commutes with all other group
elements, and $\teins_0=\teins$ is the unity. The $\teins_z$ form the loop
group of $\grp SO(1,2)$, and the even elements $\teins_{2z}$ build up the loop
group of $\grp SL(2)$.
\begin{exer}\label{smooth-deform}
There is also a {\em global\/} coordinate system on $\cgrp SL(2)$, which can be
chosen to be
\beq
   \tg(w,x,y) =  \exp ( x \tgam_1 + y \tgam_2 ) \exp ( w \tgam_0 ), \quad
     w,x,y \in \RR.
\eeq
As a consequence, show that for any smooth field $\th:\N\to\cgrp SL(2)$ on some
manifold, and its projection $\h:\N\to\grp SL(2)$, there exists a smooth
deformation $\h_s:\N\to\grp SL(2)$ such that $\h_1=\h$ and $\h_0=\eins$.
\end{exer}
This property of $\cgrp SL(2)$ valued fields will be important when defining
gauge transformations: it will tell us whether two fields configurations are
gauge-equivalent or not.

Our coordinates can also be used to read off whether two group elements commute
or not. Using \eref{sl2-par} and the Clifford algebra of the gamma matrices, we
find that $\g(v^a,\pm)$ and $\g(w^a,\pm)$ commute if and only if the two
vectors $v^a$ and $w^a$ are proportional to each other, and the same holds for
$\cgrp SL(2)$:
\begin{exer}\label{commute}
Two elements $ \tg(v^a,z),\tg(w^a,z')\in\cgrp SL(2)$ commute if and only if
\beq[commute-eq]
  \eps_{abc} v^a w^b = 0
  \equivalent
  \exists r,s\in\RR :rv^a=sw^b.
\eeq
\end{exer}
Given an element $\v=v^a\gam_a\in\alg sl(2)$, then there is the adjoint action
of the group element $\u\in\grp SL(2)$ on $\v$:
\beq[gfg]
   \v=v^a\gam_a \mapsto \u^{-1} \v \u = v^a u_a\^b \gam_b,
\eeq
which is an $\grp SO(1,2)$ Lorentz rotation $u_a\^b$ acting on the vector
$v^a\in\MM^3$. Similarly, an inner automorphism of the group $\grp SL(2)$ acts
as a Lorentz rotation on the vector $v^a$ in \fref{sl2}:
\beq
   \u^{-1} \g(v^a,\pm)\, \u
   &=& {} \pm \u^{-1} \big( v^a \gam_a+\hat v \, \eins\big) \u \zl
   &=& {} \pm \big( v^b u_b\^a \gam_a +  \hat v \,  \eins \big)
   = \g(v^b u_b\^a,\pm) .
\eeq
This transformation is obviously well defined for the covering group $\cgrp
SL(2)$, too, as the length of $v^a$ is unaffected and we always stay in the
same part of \fref{sl2}. Hence, there is a well defined map
\beq[adjoint-group]
  \cgrp SL(2) \to \cgrp SL(2) :
  \tg \mapsto \u^{-1} \tg \u  , \where \u \in\grp SL(2).
\eeq
Note that you cannot use group multiplication to define this map, because you
cannot multiply elements of $\grp SL(2)$ and $\cgrp SL(2)$ with each other.
There are three alternative ways to define this map properly.
\begin{defi}[Automorphisms of\/ $\cgrp SL(2)$]\label{auto}
For each $\u\in\grp SL(2)$ there is an inner automorphism of $\cgrp SL(2)$,
which is either defined by
\beq
   \u^{-1} \tg \u = \tu^{-1} \tg \tu
\eeq
for some $\tu\in\cgrp SL(2)$ whose projection is $\u$, or by
\beq
   \u^{-1} \tg \u :s \mapsto \u^{-1} \g(s) \u ,
\eeq
where $s\mapsto\g(s)$ is a path representing $\tg$, or using the coordinates
from above we can write
\beq
 \u^{-1} \tg(v^a,z) \, \u =  \tg(v^b u_b\^a,z)
\eeq
\end{defi}
\begin{exer}
Show that the right hand sides of the first two definitions are independent of
the special $\tu$, and of the special path $\g(s)$, respectively, that all
three definitions are equivalent, and that these are {\em all\/} inner
automorphisms of $\cgrp SL(2)$.
\end{exer}
Now that we have the Lorentz group it is not difficult to get the Poincar\'e
group, which is obtained by attaching the algebra to the group and define an
appropriate multiplication.
\begin{defi}[Poincar\'e Group]\label{poincare}
The Poincar\'e group is given by $\cgrp ISO(1,2)=\cgrp SO(1,2) \times \alg
so(1,2)\simeq\cgrp SL(2)\times\alg sl(2)$, i.e.\ its elements are pairs
$(\tg,\f)$, and the multiplication reads
\beq[poin-mult]
  (\tg,\f)\, (\tg',\f') &=&
 (\tg \tg' , \, \f + \g \f' \g^{-1} ), \zl
  (\tg,\f)^{-1} &=& (\tg^{-1} , - \g^{-1} \f \g ).
\eeq
\end{defi}
\begin{exer}
Show that this is in fact the group $\cgrp ISO(1,2)$, i.e.\ the covering group
of $\grp ISO(1,2)$ consisting of Lorentz rotations and translations of three
vectors, and whose elements are $4\times4$ matrices
\beq
   \pmatrix{ g^a\_b & f^a \cr
             0       &  1 \cr },
\eeq
which is the projection of $(\tg,\f)\in\cgrp ISO(1,2)$ onto $\grp ISO(1,2)$
(remember that $g^a\_b\in\grp SO(1,2)$ is obtained from $\tg$ by first
projecting onto $\g\in\grp SL(2)$ and then using \eref{homo}; $f^a$ is obtained
by expanding $\f=f^a\gam_a$).
\end{exer}
This group is the full gauge group of dreibein gravity. However, we will not
make much use of the fact that we can write it as a single group or single
algebra. This is quite useful when considering only classical solutions of
Einstein gravity in three dimensions. But when going over to Hamilton Jacobi
formalism and then to quantum theory, we have to split the group into its
components consisting of rotations and translations anyway. That's because the
gauge fields corresponding to the different parts will become canonically
conjugate quantities and non-commuting operators, respectively, so that it is
not easy to keep the ``one group'' notation. In addition, the physical
interpretation of the two sets of gauge fields is quite different, so to
extract the physical content of various quantities becomes easier if we keep
the ``component notation'' $(\tg,\f)$ for $\cgrp ISO(1,2)$.

\section{Kinematics}
\label{kinematics}
The aim of this section is to present an unusual approach to general
relativity. It will turn out to have two very different advantages. The first
is rather technical: it will be possible to extend spacetime beyond the conical
singularities arising in three dimensional gravity. As a consequence, the
solutions are always globally defined and we do not have to exclude points from
the spacetime manifold. Nevertheless spacetime will look as if it had
singularity. All this will become possible after allowing certain kinds of
singular metrics.

Another feature of this approach is that in some sense we describe physics
``from inside'', i.e.\ from the point of view of an observer sitting somewhere
on the manifold. Associated with a local observer there will be a local three
dimensional Minkowski space, which may be considered as his ``laboratory'',
where he can make experiments with test particles. All measurements, i.e.\ all
physical entities, will be expressed as objects in this ``local Lorentz
frame''. With such an observer at each point in space, gravity (or geometry)
becomes the theory of their interaction or rather their communication. We will
even go further and forget that the points on the ``spacetime manifold'' are
events. Each point just labels an observer at a specific time. This may sound a
little bit strange but we will see that this interpretation is able to deal
with exactly those singular metrics mentioned above, and ``physics'' does not
change at the singularities.

This approach to gravity is not restricted to three dimensions, except that
only there it is possible to eliminate {\em all\/} singularities. Technically,
what one has to do is to replace the metric by a vielbein and the spin
connection as basic gravitational fields, then the Einstein Hilbert action is
polynomial in both and well defined for singular vielbeins. Below we will do
this for $d=3$ and we will find a rather simple interpretation for the singular
metrics: if there is a line of ``length'' zero (not to be confused with a
lightlike line), or an area or volume of vanishing measure, this just means
that all the observers associated with this line are actually at the same
``physical event''. The underlying manifold is no longer identical to the
physical spacetime. Instead, spacetime is given by identifying those points of
the manifold which have vanishing distance.

\subsection{Dreibein gravity as a theory of local observers}
\label{observers}
One usually derives the dreibein formalism of gravity from the metric formalism
by writing the metric as a function of the dreibein, introducing a spin
connection via the dreibein postulate etc. That's not the way we will follow
here. Our starting point is the ``set of local observers'', which is a
2-manifold $\N$. As already mentioned, the framework presented here is rather
general, but we restrict $\N$ to have a trivial tangent bundle. Typical
examples for such manifolds are two-spheres or tori with arbitrarily many holes
(but at least one for a sphere) cut out.

Now each point $\x\in\N$ shall corresponds to, or rather {\em is}, an
individual local observer, who has a ``laboratory'' represented by a three
dimensional Minkowski space $\MM^3$. It is called the ``local Lorentz frame''
of $\x$. Vectors in $\MM^3$ are denoted by $v^a$, $a=0,1,2$, and the Lorentz
metric $\eta_{ab}$ is used to raise and lower indices, which has been used
already in \sref{lorentz-group}. Every measurement an observer makes has to be
related to a quantity in his local frame. In general, all these local observers
together form an $\MM^3$ bundle over $\N$, but with our restriction above it is
reasonable to assume that this is trivial too, so that the ``flat'' indices
$a,b,\dots$ are also global. As part of exercise~\ref{difgeo} the reader may
think about the strange geometry arising if this bundle is not trivial (or not
isomorphic to the tangent bundle of spacetime, see below).

So far we considered a two dimensional manifold which has something to do with
``space''. Spacetime is now obtained by introducing a ``time evolution'' of
space. Hence, our interpretation of gravity as theory of local observers is in
principle the same as the Hamiltonian or Arnowitt Deser Misner (ADM) formalism
of gravity~\cite{arnowitt.deser.misner:62,misner.thorne.wheeler:73}, except
that not spacetime but the ``space'' $\N$ is our basic background structure
with its interpretation as a set of observers.

Using the language from above, what we do is the following. For each observer,
we introduce a variable $t$ called ``time'', but which does not have to do
anything with the physical time. You can think of it as some ``bad'' clock.
i.e.\ it shows an increasing real number $t$ which can be used to label events
happening in the laboratory of the observer, but it is not suitable to measure
the physical time. ``Increasing'' of course means that there is a relation to
the physical time $\tau$, namely that $\tau$ is a monotonic function of $t$. As
we will see later on, this can always be established by ``gauge fixing''. An
event is now fixed by specifying where it happens, i.e.\ at which observer
$\x\in\N$, and at which time $t\in\RR$. Hence, we define spacetime to be
$\M=\N\times\RR$. Our interpretation of space as a given set of observers
requires the same restriction on spacetime as the ADM formulation: spacetime
must be globally hyperbolic, in order to get a well defined initial value
problem. In the spacetime picture, the individual observer $\x\in\N$ becomes a
worldline $\X(t)=(\x,t)\in\M$. By a gauge choice we will be able to make this a
timelike path, so that the interpretation as real physical observers is always
possible.

The parameter time $t$ can also be considered as a {\em  global\/} time
coordinate: we can look at the universe from the outside and what we see is a
space $\N$, which is given by the set of all observers at a specific time $t$,
and this space, i.e.\ the fields defined on $\N$, becomes time-dependent. The
global coordinate $t$ will enter into the Hamilton Jacobi formalism as time
variable: the action will become a ``time'' integral of some Lagrangian, which
is a functional of the fields and their ``velocities''. The Hamiltonian and
Poisson brackets are defined such that they describe the evolution of the
universe with respect to this parameter time. For the moment, however, we are
only considering the kinematics of gravity, so yet we don't have actions,
Hamiltonians and things like that. All we have so far is a topological
structure, i.e.\ we know which observers are in a neighbourhood of a special
observer. We have to introduce some fields to describe the ``interaction'' of
these observers. In particular, an observer has to know how far away another
observer is and how other local Lorentz frames are related to his own. It is
exactly this what is described by the dreibein and the spin connection.

In the following, we will sometimes switch between the ``canonical'' or ADM
formalism, where the universe is described by fields on a space $\N$ evolving
in time, and the ``spacetime'' point of view, where it is just given by the
same fields on a manifold $\M=\N\times\RR$. Tangent indices on $\N$ are denoted
by $i,j,k,\dots$, and they can be considered as global indices, as we are
assuming that $\N$ has a trivial tangent bundle. For tangent indices on $\M$ we
use small Greek letters $\mu,\nu,\rho,\dots$, and they ``take the values'' $t$
and $i,j,k,\dots$. A covector or one-form $v_\mu$ on $\M$ splits into a scalar
$v_t$ and another one-form $v_i$ on $\N$, and similar relations hold for
arbitrary tensors. As the first gravitational field the dreibein is now
introduced as a field in $\M$:
\begin{defi}[Dreibein]\label{def-dreibein}
The dreibein is a linear map from the tangent bundle of $\M$ into the observer
bundle, or simply a matrix that maps a tangent vector $v^\mu(\X)$ of $\M$ onto
a vector $v^a(\X)=v^\mu(\X) \e \mu a(\X)$ in the local Lorentz frame of the
observer at $\X$. The map is represented by a field $\e \mu a(\X)$ which we
only require to be a smooth function of $\X$.
\end{defi}
No other restrictions are made, in particular, the matrix need not be
invertible. In case of a more general manifold, $\e \mu a$ exists locally only
and it is possible to define an invertible dreibein only if the observer bundle
and the tangent bundle are isomorphic, simply because then the dreibein itself
is an isomorphism. Thus assuming the observer bundle to be trivial is natural
if the tangent bundle is trivial, otherwise the geometry would become rather
strange (see above).

{}From the canonical point of view, the dreibein splits into a set of one-forms
$\e ia(\x,t)$ on $\N$, a $2\times3$ matrix called the ``spatial dreibein'', and
a 3-vector $\e ta(\x,t)$. The physical interpretation of the dreibein is as
follows. Given an observer at a specific time $\X$ with local coordinates
$\X^\mu=(\x^i,t)$ and another one with coordinates $\X^\mu+\d\X^\mu$. Then
$\d\X^a=\e \mu a \d\X^\mu$ is a vector in the local frame of $\x$ that tells
him in which direction and distance $\X+\d\X$ is. We call this vector the
``distance'' from $\X$ to $\X+\d\X$. In particular, the length of the path from
$\X$ to $\X+\d\X$ is given by the length of this vector:
\beq[g=ee]
\d s^2=\d\X^a\d\X_a=g_{\mu\nu}\d\X^\mu\d\X^\nu, \txt{where}
    g_{\mu\nu} = \e \mu a \e \nu b \eta_{ab}.
\eeq
Restricting this consideration to a specific moment $t=t_0$, we also get a
metric on $\N$. The distance of the two observers at the same time is given by
the spatial dreibein and reads
\beq
  \d s^2 = \d\x^i \d\x^j g_{ij}, \txt{where}
      g_{ij} = \e ia \e jb \eta_{ab}
\eeq
is the induced metric on $\N$ as a submanifold of $\M$ defined by $t=t_0$. Note
that this is not necessarily positive definite, and $g_{\mu\nu}$ is in general
not invertible, as there is no such restriction on the dreibein. The time
component $\e ta(\x,t)$ describes the relation between the parameter time $t$
used by $\x$ and his physical time $\tau$. The proper time elapsing when $t$
increases by $\d t$ is
\beq
  \d \tau^2 = {} - \d t^2 \,  \e ta \e tb \eta_{ab} = {} - \d t^2 g_{tt}.
\eeq
For $\tau$ to be well defined, $\e ta$ must be a timelike vector, which is
equivalent to requiring that the worldline on the observer is timelike. As $\e
ta$ will turn out to be a pure gauge degree of freedom, we could restrict its
values to any {\em open\/} subset of $\MM^3$ (otherwise we might loose
equations of motion). However, we can equally well keep it arbitrary, and
whenever we have a special configuration we can find a gauge transformation
such that $\e ta$ becomes timelike and we can read off the physical properties
of that configuration, which are by definition invariant under gauge
transformation.

It is now straightforward to interpret singular dreibein fields. If $\e \mu a$
has a zero eigenvalue one can find $\d\X^\mu=(\d t,\d\x^i)$ such that
$\d\X^a=0$. Using our interpretation from above this means that the distance
between the two observers (possibly at different parameter time, if $\d t\ne0$)
is zero, they actually sit at the same spacetime event. The point $\X\in\M$
labels an observer at a given moment, not a point in spacetime. More precise,
if $\d\X^a=0$, then the same physical event is seen by observer $\x$ at time
$t$ and observer $\x+\d\x$ at {\em his\/} time parameter $t+\d t$: the two (or
more) observers meet each other at this event. Physical spacetime is defined as
the set of equivalence classes of points in $\M$, where two points are
equivalent (i.e.\ two observers see the same event), if there exists a path
connecting them such that $\d\X^a=0$ everywhere on the path. The resulting
spacetime metric is non-singular in the sense that two different events always
have finite distance. Note that $\d\X^a=0$ is different from $\d s^2=0$, which
states that a path is lightlike. Observe also, that now the topology of
spacetime is no longer fixed: different dreibein fields and thus different
equivalence classes may lead to different topologies. An illustrating example
will be given below.
\begin{defi}[Spin Connection]
As a second gravitational field we introduce the spin connection by defining a
covariant derivative of vectors in the local Lorentz frames. It is represented
by an $\alg so(1,2)$ gauge field $\o_{\mu ab}=-\o_{\mu ba}$. Given a vector
field $v^a$, then the covariant derivative reads
\beq[D=d+o]
\D_\mu v^a = \del_\mu v^a + \o_\mu\^a\_b v^b.
\eeq
\end{defi}
We can expand $\o_{\mu ab}$ in terms of the generators of the vector
representation of the Lorentz algebra (see definition~\ref{lorentz-rep}) to get
\beq
\o_{\mu ab} = \o_{\mu c}\,\eps^c\_{ab}  \equivalent
\o_{\mu c} = -\ft12 \eps_c\^{ab}\, \o_{\mu ab}.
\eeq
The covariant derivative then becomes
\beq[D=d+eps o]
\D_\mu v^a = \del_\mu v^a + \o_{\mu c} \, \eps^{ca}\_b v^b.
\eeq
Thus $\o_{\mu a}$ is the abstract Lie algebra notation for the spin connection
($a=0,1,2$ labeling the generators), whereas $\o_{\mu ab}$ is its vector
representation. The field strength or curvature of the spin connection is
defined as the commutator of two covariant derivatives:
\beq[F=DD]
  \D_{[\mu} \D_{\nu]} v^a = \ft12 \F_{\mu\nu}\^a\_b v^b .
\eeq
It reads explicitly
\beq[F-undual]
  \F_{\mu\nu ab} = \del_\mu \o_{\nu ab}
                 - \del_\nu \o_{\mu ab}
                 + \o_{\mu a}\^c \o_{\nu cb}
                 - \o_{\nu a}\^c \o_{\mu cb},
\eeq
in the vector representation, and the abstract algebra representation is
\beq[F=do+oo]
 \F_{\mu\nu a} = \del_\mu \o_{\nu a} - \del_\nu \o_{\mu a}
                 - \eps_{abc} \o_\mu\^b \o_\nu\^c ,
\eeq
where the structure constants $-\eps_{abc}$ appear. A similar relation as for
the spin connection itself holds:
\beq[F-dualize]
\F_{\mu\nu ab} = \eps_{ab}\^c \F_{\mu\nu c} \equivalent
\F_{\mu\nu a} = -\ft12 \eps_a\^{bc} \F_{\mu\nu bc}.
\eeq
Like the dreibein, the spin connection and its field strength split naturally
into its space and time components, simply by inserting $i,j$ or $t$ for the
indices $\mu,\nu$. The spatial component describes how vectors in the local
Lorentz frames of two observers can be compared. Consider vectors $v^a(\x)$ and
$v^a(\x+\d\x)$ measured by different observers (at the same time $t$). The
difference of these vectors, measured in the local frame of $\x$ is given by
$\d \x^i \D_i v^a$. Similarly, the difference of two vectors measured by the
same observer $\x$ at different times, $v^a(t)$ and $v^a(t+\d t)$ reads $\d
t\,\D_t v^a$. We will give more precise definitions for how to compare
measurements made by different observers or at different times below by
introducing transport operators. However, as we already introduced all the
fields, we can see what dreibein gravity is: the theory that describes how
measurements made by different observers can be compared and what the physical
distance between them is.

\subsection{Transport operators and geodesics}
In this section we will construct non-local objects as functions of the fields
$\e \mu a$ and $\o_{\mu ab}$, describing concrete finite ``measurements'' made
by our observers. In metric general relativity one defines a parallel transport
operator as follows. Given a path $\curva:[a,b]\rightarrow\M$. The transport
operator maps a tangent vector $v^\mu(\curva(a))$ at the beginning of the path
onto another tangent vector at the end:
\beq[trans-coor]
v^\mu(\curva(b)) = v^\nu(\curva(a)) \, \U_\curva(a,b)_\nu\^\mu.
\eeq
The matrix $\U_\curva(a,b)_\nu\^\mu\in\grp GL(3)$ is defined such that the
vector is covariantly constant along the path, i.e.\ its covariant derivative
with respect to $b$ vanishes. The full definition is
\beq[trans-coor-def]
\U_\curva(a,a)_\nu\^\mu &=& \delta_\nu\^\mu ,  \zl
\nabl_s \U_\curva(a,s)_\nu\^\mu &=&
\deldel/s/ \U_\curva(a,s)_\nu\^\mu  +
  \U_\curva(a,s)_\nu\^\rho \chris^\mu\_{\sigma\rho} \curva'^\sigma = 0.
\eeq
Note that the two indices of $\U$ are tangent indices sitting at different
points and transform differently under coordinate transformations, so that the
covariant derivative acts on the index located at $\curva(s)$ only. The
notations here is as follows: if the path's tangent vector $\curva'^\mu$ or a
field appears in a product with such a transport operator, the ordering of the
product indicates where the fields are to be taken; i.e.\ in
\eref{trans-coor-def} the tangent vector has to be taken at $s$ and the
Christoffel connection at $\curva(s)$.

A geodesic can now be defined as a path whose tangent vector is covariantly
constant. We can write this as a local differential equation or as a non-local
equation using the transport operator
\beq[geo-metr]
\nabl_s \curva'^\mu(s) = 0 \equivalent
\curva'^\mu(b) = \curva'^\nu(a) \, \U_\curva(a,b)_\nu\^\mu.
\eeq
More precise, it is sufficient that there is a reparameterization of  the path
such that this equation holds.
This definition is equivalent to the ``minimum length'' or ``maximum proper
time'' requirement (if no torsion is present), but it is more general as it
applies to lightlike paths as well. Unfortunately, it is not suitable for our
formalism, because we are dealing with local Lorentz vectors rather than
tangent vectors of the manifold and also because the definition does not work
for non-invertible metrics. However, as we introduced a connection as a basic
field, we can define a transport operator with respect to this connection:
\begin{defi}[Transport operator]
Given a path $\curva:[a,b]\rightarrow\M$ (always assumed to be smooth) we
define a matrix $\U_\curva(a,b)_a\^b$ mapping a local Lorentz vector at
$\curva(a)$ onto one at $\curva(b)$. It is uniquely given by
\beq[o-trans]
\U_\curva(a,a)_a\^b &=& \delta_a\^b ,  \zl
\deldel/s/ \U_\curva(a,s)_a\^b  &=&
  \U_\curva(a,s)_a\^c \o_{\mu c}\^b  \curva'^\mu .
\eeq
\end{defi}
As $\o_{\mu a}\^b$ is an $\alg so(1,2)$ connection, this gives an $\grp
SO(1,2)$ matrix, i.e.\ a proper Lorentz rotation. One can also write $\U$ as a
path ordered integral
\beq[U-int]
  \U_\curva(a,b)_\ddd\^\ddd = \Pexp \Big( \intl ab \d s \, \curva'^\mu
                           \,    \o_{\mu\ddd}\^\ddd \Big),
\eeq
where path ordering is defined such that factors nearer to the lower boundary
of the integral appear to the left. We do not require any ordering like $a<b$
for the parameters. Transport operators may be linked together and they are
``symmetric'':
\beq[kette-U]
  \U_\curva(a,b)_a\^b \, \U_\curva(b,c)_b\^c =
  \U_\curva(a,c)_a\^c, \qquad
  \U_\curva(a,b)_{ab} = \U_\curva(b,a)_{ba}.
\eeq
The second equation follows from the first if we set $a=c$ and use the fact
that the inverse of an $\grp SO(1,2)$ matrix is given by its transpose,
multiplied by the flat metric from both sides. In \sref{covering} we always
used the interval $[0,1]$ as the range for $s$, so we introduce
$\U_{\curva,a}\^b$ as an abbreviation for $\U_\curva(0,1)_a\^b$, and instead of
\eref{kette-U} we can write
\beq[U-kette]
  \U_{\curva,a}\^b \, \U_{\curvb,b}\^c =
  \U_{\curva\kette\curvb,a}\^c , \qquad
  \U_{\curva,ab} = \U_{\curva^{-1},ba}.
\eeq
How is this transport operator related to the metric operator
$\U_\curva(a,b)_\mu\^\nu$ defined above? A priori there is no relation, because
it depends on the spin connection only whereas the metric depends on the
dreibein. The relation we want to establish is that, given a vector at the
beginning of some path, parallel transport with respect to the metric and then
mapping it onto a local Lorentz vector by the dreibein should give the same
result as first mapping it onto a local Lorentz vector and then transporting
with respect to the spin connection. This is the ``dreibein postulate'', which
can also be formulated locally and is then called ``torsion equation'':
\begin{exer}Show that, if the dreibein is invertible, the non-local dreibein
postulate is equivalent to the torsion equation:
\beq[Ue=eU]
     \U_\curva(a,b)_\mu\^\nu \, \e \nu b =
     \e \mu a \, \U_\curva(a,b)_a\^b ,
   \equivalent
    \D_{[\mu} \e {\nu]}a = 0.
\eeq
\end{exer}
Here one has to use that the torsion equation actually states that the total
covariant derivative of the dreibein vanishes (i.e.\ including the torsion-free
Christoffel connection, but without antisymmetrization). As the torsion
equation is an equation of motion, i.e.\ it is part of the ``dynamics'' of
dreibein gravity, it is ensured that our transport operators and geodesics
coincide with those derived from the metric in regions where it is invertible.

There are two important formulas describing how the transport operator changes
under variation of the path or the spin connection, respectively.
\begin{exer}
The functional differential of the transport operator \eref{U-int} with respect
to the connection reads
\beq[dU-do]
  \delta \U_\curva(a,b)_\ddd\^\ddd =
   \intl ab \d s \, \U_\curva(a,s)_\ddd\^\ddd \, \curva'^\mu
       \delta\o_{\mu\ddd}\^\ddd
           \, \U_\curva(s,b)_\ddd\^\ddd .
\eeq
\end{exer}
\begin{exer}
Under a smooth deformation $\curva\mapsto\curva+\delta\curva$ of the path the
variation of the transport operator is
\beq[trans-vary]
\delta \U_\curva(a,b)_\ddd\^\ddd  &= &
 \U_\curva(a,b)_\ddd\^\ddd  \, \delta\curva^\mu \o_{\mu \ddd}\^\ddd -
 \delta\curva^\mu \o_{\mu \ddd}\^\ddd \, \U_\curva(a,b)_\ddd\^\ddd \zl
  & & + \intl ab \d s \,
 \U_\curva(a,s)_\ddd\^\ddd  \, \delta\curva^\mu
      \curva'^\nu \F_{\mu\nu \ddd}\^\ddd \, \U_\curva(s,b)_\ddd\^\ddd.
\eeq
\end{exer}
Here we replaced the flat indices of the transport operator again by the dots
introduced in \eref{vec-rep} to show that these are general formulas that hold
for every representation of the transport operator, we only have to replace the
field strength and connection by the appropriate terms. The indices are
arranged such that they multiply the $\alg so(1,2)$ and $\grp SO(1,2)$ valued
fields as matrices. Remember also the convention that the points where the
fields are to be taken are indicated by the factor ordering, i.e.\ in the first
term in \eref{trans-vary} we have to take them at $\curva(b)$, at $\curva(a)$
in the second and at $\curva(s)$ inside the integral. We see that, if we fix
the end points of the path, the variation is proportional to the field
strength.

We can now extend the definition of a geodesic to singular metrics. However, it
is no longer possible to parameterize the path using its own ``proper time'' or
``proper length''.
\begin{defi}[Geodesic]
A path $\curva$ is called a geodesic if there exists a vector $c^a$ at a
(fixed) point $\curva(a)$ and a function $\tau(s)$ such that
\beq[geo-def]
  \U_\curva(a,s)^a\_b \, \e \mu b \, \curva'^\mu = \tau'(s)\,  c^a.
\eeq
\end{defi}
Hence, the tangent vector of a geodesic, projected into the local Lorentz
frames, is covariantly constant up to rescaling. If the geodesic passes through
a region where the dreibein is singular, $\tau'$ might vanish, showing that the
``proper length'' does not increase: the path remains at the same spacetime
event though it passes from one observer to another. In these regions with
singular dreibeins the new geodesics behave quite strange but as we will see in
the example below the definition still makes sense. If the dreibein is
invertible, we obviously have $\tau'\ne0$ whenever $\curva'^\mu\ne0$, and we
can reparameterize such that $\tau(s)=s$. With \eref{Ue=eU} we infer that our
definition is then equivalent to the metric definition of a geodesic given
above.

The vector $c^a$ appearing in \eref{geo-def} is something like the finite
version of $\d\X^a$ introduced above to measure the ``distance'' of two
observers in spacetime. To see what $c^a$ is, assume that $\curva$ is a
geodesic from $\curva(0)=\X$ to $\curva(1)=\Y$. We can parameterize it such
that $\tau(0)=0$ and $\tau(1)=1$. Then it is obvious that $c^a$ tells us how to
get from $\X$ to $\Y$, namely by going in the direction indicated by $c^a$ (it
is the tangent vector to the geodesic at $\X$), and the distance is given by
the length of $c^a$, just as for the infinitesimal quantity $\d\X^a$ above.
Hence, we may call $c^a$ the ``distance'' from $\X$ to $\Y$, measured in the
local Lorentz frame at $\X$.

Given the geodesic, we can calculate $c^a$ by integrating \eref{geo-def}
\beq[c-def]
  c^a = \intl01 \d s \, \U_\curva(0,s)\^a\_b \, \e \mu b \, \curva'^\mu.
\eeq
This expression is invariant under reparameterization of $s$, we just have to
know the geodesic itself to compute the distance $c^a$. In addition, the value
of this integral also exists if $\curva$ is not a geodesic. It defines another
quantity associated with a path. Let us call it the ``distance of the path
$\curva$''
\beq[V-int]
  \V_\curva(a,b)^a = \intl ab
    \d s \, \U_\curva(a,s)^a\_b \, \e \mu b \, \curva'^\mu,
\eeq
which is a vector in the local Lorentz frame at the beginning of the path.
An alternative definition is, similar to \eref{o-trans}, the differential
equation with initial condition
\beq[V-diff]
  \V_\curva(a,a)^a = 0 , \qquad
  \deldel/s/ \, \V_\curva(a,s)^a = \U_\curva(a,s)^a\_b \, \e \mu b \,
            \curva'^\mu .
\eeq
As for $\U$ we use the abbreviation $\V_\curva\^a=\V_\curva(0,1)^a$. We have
the following relations for the link or inverse of paths:
\beq[V-link]
  \V_{\curva\kette\curvb}\^a = \V_\curva\^a + \U_\curva\^a\_b \V_\curvb\^b,
  \qquad
  \V_{\curva^{-1}}^{\ a} = - \V_\curva\^b \U_{\curva,b}\^a  .
\eeq
Note that in the first equation $\V_\curvb\^b$ is a vector at
$\curvb(0)=\curva(1)$, and $\U_\curva\^a\_b$ maps this vector onto a vector at
$\curva(0)$, which can be added to $\V_\curva\^a$. Together with \eref{U-kette}
this is exactly the group multiplication on $\grp ISO(1,2)$ as introduced in
definition~\ref{poincare}. If we arrange $\U$ and $\V$ such that they build a
$4\times4$ matrix, we can summarize their definitions as
\beq
     \deldel/s/ \, \pmatrix { \U_\curva\^a\_c & \V_\curva\^a \cr 0 & 1 } =
     \pmatrix { \U_\curva\^a\_b & \V_\curva\^a \cr 0 & 1 } \,
     \pmatrix { \o_\mu\^b\_c & \e \mu b \cr 0 & 0 } \,
       \curva'^\mu
\eeq
The last matrix is the combined $\alg iso(1,2)$ connection. However, as already
mentioned, we will not use it very much, as the physical interpretation of the
two connection fields are quite different and the components are canonically
conjugate phase space coordinates and non-commuting quantum operators. Finally,
we also have to know how $\V$ transforms under deformations of the path. The
formula is similar to that given for $\U$. It will be sufficient to know how
the transformation looks like if the field equations of gravity are satisfied.
\begin{exer}
Assume that the torsion equation holds and the spin connection is flat. Then
under a smooth variation of the path $\delta\curva$ the transport operator and
the distance change by
\beq[UV-deform]
   \delta\U_\curva(a,b)_a\^b &= &
           \U_\curva(a,b)_a\^c \, \delta\curva^\mu \, \o_{\mu c}\^b
        -  \delta\curva^\mu \, \o_{\mu a}\^b \, \U_\curva(a,b)_c\^b, \zl
  \delta\V_\curva(a,b)^a &=&
    \U_\curva(a,b)^a\_b \, \delta\curva^\mu \e \mu b
 - \delta \curva^\mu \big( \o_\mu\^a\_b \, \V_\curva(a,b)\^b + \e \mu a \big).
\eeq
As indicated by the factor ordering the fields and $\delta\curva^\mu$ are to be
taken at $\curva(b)$ in the first term and $\curva(a)$ in the second term in
both equations.
\end{exer}

\subsection{Spinors, test particles, and measurements}
\label{test-particle}
So far we considered paths and geodesics in spacetime. For a given path we
defined a transport operator and a ``distance'' vector. We know want to discuss
the physical meaning of these objects and how they can be measured. To make
such measurements, our observers can use test particles. These test particles
shall have the following properties. They may be scalar, spinor, or vector like
particles, i.e.\ in a local Lorentz frame they are represented by a scalar
(this is just a point particle with not extra structure), a spinor $\spin$ as
introduced in definition~\ref{lorentz-rep}, or a vector $v^a$.

Any observer $\x$ can ``emit'' such a particle at a time $t$, it will then move
along a geodesic to another observer $\y$, where it arrives at time $t'$. The
``state'' of the particle, i.e.\ the vector $v^a$ or the spinor $\spin$, shall
stay covariantly constant while the particle is moving.
In addition, the observers can measure geodesic distances, the result of such a
measurement being the vector $c^a$ appearing in \eref{c-def}.
For physical reasons we should restrict the paths of test particles to be
timelike or lightlike, but to simplify the treatment of measurements, let us
also allow tachyons as test particles:
\begin{exer}\label{tachyon}
In principle, for all following the experiments there is a possible physical
realization of another experiment without tachyons, giving the same result if
the equations of motion are satisfied (i.e.\ the dreibein postulate must hold
and the connection must be flat). The only reason for such an experiment to
fail is that horizons might be present, so that, e.g., a real particle can
never pass along a certain loop in space. Find out which of the spacetimes
constructed in \sref{class-sol} have horizons and how this affects the
measurements described here.
\end{exer}
Consider a vector particle emitted in a state $v^a(a)$ at $\curva(a)$. The
observer at $\curva(b)$ will see it in a state
\beq
  v^b(b) = \U_\curva(b,a)\^b\_a \, v^a(a).
\eeq
Locally this can be written as
\beq
  \D_s v^a = \del_s v^a + \curva'^\mu \o_\mu\^a\_b \, v^b = 0 .
\eeq
For a spinor particle we get exactly the same, we only have to replace the
vector representation of the Lorentz algebra by the spinor representation. We
denote the spinor representation of the connection and the covariant derivative
of a spinor by
\beq[spin-oo]
  \oo_\mu = \ft12 \o_{\mu a} \gam^a , \qquad
  \D_\mu \spin = \del_\mu \spin + \oo_\mu \spin .
\eeq
The corresponding representation of the field strength is given by the
commutator of two covariant derivatives
\beq[F-spin]
   \D_{[\mu} \D_{\nu]} \spin = \ft12 \FF_{\mu\nu} \spin  \follows
  \FF_{\mu\nu} = \del_\mu \oo_\nu - \del_\nu \oo_\mu
        + \comm{\oo_\mu}{\oo_\nu}.
\eeq
We define the spinor transport operator similar to \eref{o-trans}, with
$\o_{\mu\ddd}\^\ddd$ replaced by $\oo_\mu$:
\beq[U-spin]
   \UU_\curva(a,a)=\eins, \qquad
   \deldel/s/ \UU_\curva(a,s) = \UU_\curva(a,s) \, \oo_\mu \, \curva'^\mu.
\eeq
Hence, for a spinor particle moving along a path $\curva$ we have
\beq[spin-pass]
  \D_s \spin = 0 \equivalent
  \spin(b) = \UU_\curva(b,a) \, \spin(a).
\eeq
\begin{exer}
Show that the spinor transport matrix obeys
\beq[U-spin-kette]
  \UU_{\curva} \, \UU_{\curvb} =
  \UU_{\curva\kette\curvb} , \qquad
  \UU_{\curva}^{-1} = \UU_{\curva^{-1}},
\eeq
where again $\UU_\curva$ is an abbreviation for $\UU_\curva(0,1)$,
and that it is related to the vector transport operator by
\beq[U=UU]
  \U_\curva(a,b)_{ab} = \ft12
  \Trr{ \, \UU_\curva(b,a)\, \gam_a\, \UU_\curva(a,b)\,\gam_b\, } ,
\eeq
which is the projection $\grp SL(2)\to\grp SO(1,2)$. From this relation we can
infer that we only need spinors as test particles to extract all information
about the transport operators.
\end{exer}
We can convert all our notation into the spinor representation by using the
fact the three dimensional Minkowski space $\MM^3$ is isomorphic to the Lorentz
algebra $\alg sl(2)$. Hence, every 3-vector can be converted into an $\alg
sl(2)$ matrix by expanding it in terms of the gamma matrices. Doing this with
the dreibein, we get the ``curved'' gamma matrices
\beq[curve-gam]
  \gam_\mu = \e \mu a \gam_a , \qquad
  \e \mu a = \ft12 \Trr{ \gam_\mu \gam^a } ,\qquad
  g_{\mu\nu} = \ft12 \Trr{\gam_\mu \gam_\nu},
\eeq
and the covariant derivative of an $\alg sl(2)$ valued fields is
\beq
  \v=v^a\gam_a , \qquad
  \D_\mu \v = \D_\mu v^a \, \gam_a  \follows
  \D_\mu \v = \del_\mu \v + \comm{\oo_\mu}{\v}.
\eeq
Now the gravitational fields $\gam_\mu$ and $\oo_\mu$ both take values in $\alg
sl(2)$. We can also define a matrix valued distance
$\VV_\curva=\V_\curva\^a\gam_a$, so that the two basic non-local objects are
\beq[UU.VV]
   \UU_\curva &=& \Pexp \intl 01 \d s\, \curva'^\mu\, \oo_\mu , \zl
   \VV_\curva &=& \intl 01 \d s  \, \UU_\curva(0,s) \, \gam_\mu \, \curva'^\mu
         \, \UU_\curva(s,0).
\eeq
We can even go further and introduce an $\cgrp SL(2)$ transport operator.
However, this is not a matrix and does not correspond to any physical
measurement with test particles. It is quite simple to define it as an
equivalence class of paths in $\grp SL(2)$:
\beq[csl2-trans]
   \tU_\curva(a,b) :
   s \mapsto \UU_\curva(a,a+s(b-a)).
\eeq
\begin{exer}
Show that the identities \eref{U-spin-kette} hold for $\tU$ as well. This is
not as trivial as it seems to be: for the inverse you have to show that
\beq
  \tU_\curva\inv :  s\mapsto \UU(s,0)
        \qquad\sim\qquad
  \tU_{\curva^{-1}} : s \mapsto \UU(1,1-s),
\eeq
and for the link the following paths in $\grp SL(2)$ have to be homotopic:
\beq
   \tU_{\curva\kette\curvb}:
   s &\mapsto&  \cases{ \UU_\curva(0,2s) &for $s\le\ft12 $\cr
                     \UU_\curva \, \UU_\curvb(0,2s-1) &for $s\ge \ft12$}
   \zl &\  \wr \ &  \zl \tU_\curva \tU_\curvb :
   s &\mapsto& \UU_\curva(0,s) \, \UU_\curvb(0,s).
\eeq
\end{exer}
The pair $(\tU_\curva,\VV_\curva)$ is an element of $\cgrp ISO(1,2)$, and the
relations \eref{U-spin-kette} and \eref{V-link} can be combined into
\beq[UV-kette]
  (\tU_{\curva\kette\curvb},\VV_{\curva\kette\curvb}) &=&
  (\tU_\curva,\VV_\curva) \, (\tU_\curvb,\VV_\curvb), \zl
  (\tU_{\curva^{-1}},\VV_{\curva^{-1}}) &=& (\tU_\curva,\VV_\curva)^{-1},
\eeq
with the multiplication introduced in \eref{poin-mult}. The $\VV$ component of
this relation reads explicitly
\beq[VV-kette]
  \VV_{\curva\kette\curvb} = \VV_\curva + \UU_\curva \VV_\curvb \UU_\curva\inv,
  \qquad
  \VV_{\curva^{-1}} = - \UU_\curva\inv \VV_\curva \UU_\curva.
\eeq
Let us now briefly describe how the values of these objects can be obtained by
concrete measurements. Anticipating the results from \sref{class} let us assume
that the fields are solutions of the equations of motion, which means that the
torsion equation holds and that the field strength of the spin connection
vanishes. From \eref{UV-deform} we infer that the value of the transport
operator or the distance then depends on the homotopy class of the path only.

It is quite easy to measure $\UU_\curva$ and $\VV_\curva$ if there is a
geodesic homotopic to $\curva$.
In this case one can simply send a spinor (or rather two independent spinors)
along that geodesic to measure the transport operator $\UU_\curva$, and
$\VV_\curva$ is the geodesic distance as defined in \eref{c-def}, converted
into an $\alg sl(2)$ matrix. This also means that if there is more than one
geodesic within the same homotopy class, then they have the same distance
vector $c^a$ and, provided that the metric is invertible, they are in fact
identical up to reparameterization. This is nothing but the trivial fact that
in flat space there is only one geodesic connecting two points. However, there
might be different geodesics in different homotopy classes, if spacetime is not
globally Minkowskian. On the other hand, not every homotopy class contains a
geodesic.
\begin{exer}
To measure $\UU_\curva$ or $\VV_\curva$ for a path such that there is no
geodesic homotopic to $\curva$, one can use a piecewise geodesic path homotopic
to $\curva$ and compute the final result by \eref{UV-kette}. How can such a
measurement be realized?
\end{exer}
Hence, we are now able to measure the transport operator and the distance for
all (homotopy classes of) paths. Returning to the ADM picture, we can
distinguish between paths that lie entirely in $\N$ and those in time
direction. For the former, the transport operator depends on the spatial
components $\oo_i$ only, and we can say that a measurement (passing a test
particle from $\x$ to $\y$) takes place {\em at the time $t$\/} (this needs a
tachyon, but see exercise~\ref{tachyon}). This is what we need for quantum
physics: measurements that are carried out at a fixed time. So let us stick to
paths in $\N$ in the following. For path in $\N$ all the formulas above remain
unchanged except that we have to replace the index $\mu$ by $i$, i.e\ the
tangent vector of the path becomes $\curva'^i$ etc., and all objects become
time dependent. By these measurements we cannot get information about the time
components $\gam_t$ and $\oo_t$. So let us assume that an observer $\x\in\N$ is
also able to measure the quantities $\oo_t(\x,t)$ and $\gam_t(\x,t)$ at any
time $t$, and we can define a transport operator and distance in time direction
\beq[U.V.time]
  \UU_\x(t_1,t_2) &=& \Pexp \intl{t_1}{t_2} \d t \,  \oo_t(\x,t), \zl
  \VV_\x(t_1,t_2) &=& \intl {t_1}{t_2} \d t \, \UU_\x(t_1,t) \,
             \gam_t(\x,t) \, \UU_\x(t,t_1).
\eeq
Obviously, $\UU_\x$ tells $\x$ how to compare measurements made at different
times, as it transports a spinor along the worldline of $\x$. As already
mentioned, the time components $\oo_t$ and $\gam_t$ will turn out to be pure
gauge degrees of freedom, so actually $\x$ can not only {\em measure\/} them,
he can {\em adjust\/} them as he likes.
\begin{exer}\label{free-fall}
Show that with $\oo_t=0$ and $\gam_t=\gam_0$ the observer becomes free falling,
his local Lorentz frame is co-moving (i.e.\ it is his local restframe and does
not rotate), and the time coordinate $t$ represents his proper physical time.
It is $\UU_\x(t_1,t_2)=\eins$, and $\VV_\x(t_1,t_2)=(t_2-t_1)\gam_0$ behaves
like a clock.
\end{exer}

\subsection{Singular metrics and conical singularities}
\label{cylinder}
In this section we consider a simple example to show how the typical
singularities of three dimensional gravity are described in dreibein formalism.
We start with a flat Minkowski spacetime with polar coordinates $t,r,\phi$.
Outside the origin the dreibein and spin connection satisfying the torsion
equations are given by
\beq[mink-pol-dreib]
  \e t 0 = 1, \qquad
  \e r 1 = 1, \qquad
  \e \phi 2 = r, \qquad
  \o_{\phi}\^0  =  1,
\eeq
all other components vanishing. Alternatively, we can write it in spinor
notation as
\beq[mink-pol-spin]
  \gam_t = \gam_0 , \qquad
  \gam_r = \gam_1 , \qquad
  \gam_\phi = r\gam_2, \qquad
  \oo_\phi = \ft12\gam_0.
\eeq
In the sense of definition~\ref{def-dreibein} this is not really a dreibein, as
it is not defined for $r=0$ as a map from the tangent space of the manifold
into $\MM^3$. But we can change the topology of our space to make it well
defined. Instead of having a Minkowski space with polar coordinates, we choose
the space manifold to be a cylinder $\N=\RR\times S^1$, i.e.\ we let $r$ take
any real value and do {\em not\/} identify points with different angles $\phi$
at $r=0$. The only identification is $\phi\equiv\phi+2\pi$. Thus the ``origin''
of space becomes a circle and space extends ``behind'' this origin. One can
check that \eref{mink-pol-dreib} is a well defined dreibein and spin connection
on this manifold and that the torsion equation is satisfied everywhere. The
dreibein is singular at $r=0$ and with our interpretation from above this means
that all observers with $r=0$ but $\phi$ arbitrary sit at the same physical
point. The topology of space (and spacetime) is {\em not\/} $\N\times\RR$ but
space consists of two ``cones'' (with vanishing deficit angle) glued together
at the top.

We also immediately see that the curvature or field strength vanishes
everywhere and thus the transport operator along a loop depends on its homotopy
class (which measures how often the path is wrapped around the $S^1$) only.
Given a path $\curva(s)=(t(s),r(s),\phi(s))$, solving the differential equation
\eref{o-trans} yields
\beq
     \U_\curva(a,b)_\ddd\^\ddd = \pmatrix{ 1 & 0 & 0 \cr
                                    0 &\cos\Delt\phi & -\sin\Delt\phi \cr
                                    0 &\sin\Delt\phi & \cos\Delt\phi \cr},
  \quad \Delt\phi=\phi(b)-\phi(a).
\eeq
In fact, here it actually does not depend on the homotopy of the path, as
adding $2\pi$ to the angle does not change $\U$. However, transporting a spinor
once around the circle changes its sign, as the spinor transport operator is
\beq
   \UU_\curva(a,b) = \exp \big(\, \ft12 \Delt\phi \,\gam_0 \,\big) =
                    \pmatrix{ \cos(\Delt\phi/2) & \sin  (\Delt\phi/2 )\cr
                               -\sin(\Delt\phi/2) & \cos  (\Delt\phi/2 )\cr}.
\eeq
How does a geodesic on this manifold look like? Of course, in the two regions
$r<0$ and $r>0$ spacetime is just ordinary Minkowski space and geodesics which
entirely lie in these regions are simply ``straight lines''. A possible
parameterization is
\beq[outer-geo]
t(s)=t_0+\bar t s, \quad
r(s)=\pm\sqrt{r_0^2 + \bar r^2 s^2}, \quad
\phi(s)=\phi_0 + \arctan \frac{\bar r s}{r_0},
\eeq
where $(t_0,r_0,\phi_0)$ is the nearest point to the origin and the ratio $\bar
r:\bar t$ gives the ``velocity''. Next consider a path that lies entirely
inside $r=0$:
\begin{exer}
Show that for a path $\curva(s)=(t(s),0,\phi(s))$ we have
\beq
 \U_\curva(a,s)^a\_b \, \e \mu b \, \curva'^\mu = t'(s) c^a,
                          \txt{where} c^a=(1,0,0),
\eeq
i.e.\ it is a timelike geodesic with $\tau(s)=t(s)$ as ``proper time''.
\end{exer}
We see that a geodesic is no longer determined by giving a start point and a
tangent vector. However, the definition still makes sense, because all these
paths on the manifold $\N\times\RR$ represent the same path in physical
spacetime. This was defined as the equivalence classes of points in $\M$ having
vanishing distance, and we saw that $(t,0,\phi)$ and $(t,0,\phi')$ have
vanishing distance and thus all the paths above represent the ``physical''
geodesic $t=t(s),r=0$.

Finally we have to consider geodesics that pass through $r=0$. For such a path
we must have $\phi'=0$ whenever $r\ne0$ (otherwise it would be of the type
\eref{outer-geo}). It therefore consists of three parts, the first and third
being radial lines
\beq
   t=t_- + (s-s_-) \bar t_- , \quad
   r=(s-s_-) \bar r_- , \quad
   \phi=\phi_- &&\txt{for} s\le s_-, \zl
   t=t_+ + (s-s_+) \bar t_+ , \quad
   r=(s-s_+) \bar r_+ , \quad
   \phi=\phi_+ &&\txt{for} s\ge s_+,
\eeq
and in between we take an arbitrary path inside the $r=0$ circle as an ansatz:
\beq
   t=t(s), \quad r=0, \quad \phi=\phi(s) \txt{for} s_- \le s \le s_+.
\eeq
Mapping the tangent vector of this path into the local Lorentz frames gives
\beq
  \e \mu a \curva'^\mu =
   \cases{ (\bar t_- , \bar r_-, 0 ) & for $s\le s_-$, \cr
           ( t'(s) , 0 , 0 ) & for $s_- \le s \le s_+$, \cr
           (\bar t_+ , \bar r_+, 0 ) & for $s\ge s_+$. }
\eeq
\begin{exer}\label{pass-geo}
Now you can show that the following conditions must hold for these three parts
to fit together and to build a single geodesic:
\begin{enumerate}
\item $t'(s)=0$ in the middle part and thus $t_+=t_-$;
\item the path can be reparameterized such that $\bar t_+=\bar t_-$;
\item \begin{tabular}[t]{@{}r@{~}l}
   either & $\phi_+-\phi_-=0$ (mod $2\pi$) and $\bar r_+=\bar r_-$, \\
       or & $\phi_+-\phi_-=\pi$ (mod $2\pi$) and $\bar r_+=-\bar r_-$.
       \end{tabular}
\end{enumerate}
\end{exer}
The first statement says that the path, as seen from outside, ``does not spend
time'' at $r=0$ even though the parameter might increase from $s_-$ to $s_+$.
{}From the second we can infer that $\tau=s-s_-$, $\tau=0$, $\tau=s-s_+$,
respectively for the three parts, is a possible proper length parameter for the
path. Finally the third statements tells us how the two ``outside parts'' fit
together: the first alternative gives a line that disappears from one region
being continued in the other region where $r$ has opposite sign. The second one
is a path that entirely lies within, e.g., $r\ge 0$ and is just a usual
straight line through the origin, where the angular coordinate changes by
$\pi$.

It is quite surprising that this comes out here, as our manifold is not
Minkowski space but a ``cylinder'' with a singular metric. On $\M$ the geodesic
through $r=0$ looks rather strange: It enters $r=0$ at some angle $\phi$, then
winding around the $S^1$ some integer or half integer times, and then it leaves
$r=0$ either into the region it came from (for ``half integer'') or into the
other asymptotic region (for ``integer''). So this example shows that our
definition of a geodesic is a reasonable extension to singular metrics.

Now all this might be amazing but, beside a strange extension of Minkowski
space beyond $r=0$, what is it good for? The answer is that with the same trick
we can extend spacetime beyond a conical singularity. All we have to do is to
change the values of the fields slightly. Instead of \eref{mink-pol-spin} we
choose
\beq[pol-dreib-con]
  \gam_t=\gam_0, \qquad
  \gam_r=\gam_1, \qquad
  \gam_\phi = \defangle r \,\gam_2, \qquad
  \oo_\phi = \ft12 \defangle \,\gam_0,
\eeq
with a real constant $\defangle>0$. The torsion equation is still satisfied and
the curvature vanishes, but the transport operator
\beq
     \U_\curva(a,b)_\ddd\^\ddd = \pmatrix{ 1 & 0 & 0 \cr
      0 &\cos (\defangle \Delt\phi) & -\sin(\defangle\Delt\phi) \cr
      0 &\sin (\defangle\Delt\phi) & \cos(\defangle\Delt\phi) \cr},
\eeq
is no longer independent of the homotopy class of the path. Changing
$\Delt\phi$ by $2\pi$ gives a different value for $\U_\curva$. The formula
\eref{outer-geo} for a geodesic outside the singularity does not change very
much. One just has to replace the $\arctan$ by $1/\defangle \arctan$. This
leads, e.g., to the fact that for $\defangle<1/2$ the geodesic can intersect
itself.

Also exercise~\ref{pass-geo} still holds, except that ``mod $2\pi$'' has to be
replaced by ``mod $2\pi/\defangle$''. This changes the behaviour of the
geodesic at $r=0$ drastically. It says that the path enters $r=0$ somewhere,
goes around by an integer or half integer multiple of $2\pi/\defangle$ then
going back or into the opposite region. However, going $2\pi/\defangle$ further
before leaving $r=0$ will result in a different direction of the radial line:
the geodesic is ``scattered'' at $r=0$ in any direction $\Delt\phi=2\pi
z/\defangle$ (mod $2\pi$), $z\in\ZZ$. There are only finitely many such
directions if $\defangle$ is a rational, otherwise there is a dense set of
possible directions. Here our geodesics are very different from those defined
by \eref{geo-metr} or by the extremal length condition. The first type is not
well defined at $r=0$. Paths with extremal proper length are also scattered at
$r=0$, but they exist for $\defangle>1$ only (otherwise it is always
``shorter'' to avoid the singularity), and instead of a discrete set of
possible directions there is a continuous range around $\Delt\phi=\pi$.

So after all we have learned that it is possible to introduce singular metrics
into three dimensional gravity such that at a conical singularity physics does
not change in the sense that we do not have to give up the interpretation of
the dreibein and connection field. Our example showed the simplest kind of such
a singularity, in the next section we will find other singularities with more
interesting structure.

\section{Classical Dynamics}
\label{class}
The dynamics of gravity is of course determined by the Einstein Hilbert action.
In dreibein formalism it can be most easily expressed in terms of the field
strength of the spin connection introduced in \eref{F=do+oo}. It reads
\beq
  \Wir = \ft12 \intd3\X \eps^{\mu\nu\rho} \, \e \mu a \F_{\nu\rho a}.
\eeq
Alternatively, we can express it in terms of the spinor representation and the
curved gamma matrices $\gam_\mu=\e\mu a \gam_a$. Remember that for the spin
connection there is an extra factor of one half in the corresponding relation
$\oo_\mu=\ft12\o_{\mu a}\gam^a$, so that
\beq[spin-act]
  \Wir = \ft12 \intd3\X \eps^{\mu\nu\rho} \Trr{\gam_\mu \FF_{\nu\rho}} .
\eeq
The Levi Civita tensor density $\eps^{\mu\nu\rho}$ is defined such that the
dreibein determinant $e=\det(\e \mu a)$ obeys the following relations:
\beq[eps]
 e \, \eps^{abc}=\eps^{\mu\nu\rho} \e\mu a\e\nu b\e\rho c, \qquad
 e \, \eps_{\mu\nu\rho}=\eps_{abc} \e\mu a\e\nu b\e\rho c.
\eeq
The components obey $\eps^{\mu\nu\rho}=-\eps_{\mu\nu\rho}$ (i.e.\ $\M$ has
negative signature). For the coordinates of our cylinder manifold we have
$\eps^{tr\phi}=+1$.
We will almost immediately go over to the ADM formulation, where we have a
Lagrangian on $\N$ instead of the action on $\M=\N\times\RR$. All we want to do
in this $\M$-framework is to derive the equations of motion and have a brief
look at the gauge symmetries of the action. Thereby we will ignore all the
problems that might arise from ill-defined integrals or boundary terms produced
by partial integration. We will deal with these problems in detail below. If we
just vary the action with respect to the fields we get the equations
\beq[dreib-eom]
   \D_{[\mu} \e {\nu]}a &=& \del_{[\mu} \e {\nu]}a
                          -\eps^{ab}\_c \o_{[\mu b} \e {\nu]} c = 0 , \zl
  \ft12 \F_{\mu\nu a} &=& \del_{[\mu} \o_{\nu] a}
             - \ft12  \eps_a\^{bc} \o_{\mu b}\o_{\nu c} = 0 ,
\eeq
or in the spinor representation
\beq[spin-eom]
  \D_{[\mu} \gam_{\nu]} &=& \del_{[\mu} \gam_{\nu]} +
                          \comm{\oo_{[\mu}}{\gam_{\nu]}} = 0 , \zl
  \ft12 \FF_{\mu\nu} &=& \del_{[\mu} \oo_{\nu]} +
                          \oo_{[\mu}\oo_{\nu]} = 0 .
\eeq
The first is obviously the torsion equation. We already saw that its
consequence is that our definitions of geodesics and transport operators
coincide with those in the metric formulation. The second equation than states
that spacetime is flat. So we do not get any local degrees of freedom like
gravitational waves. All the physical properties of solutions to these
equations have a ``global'' character, they are encoded in non-local quantities
like those constructed in the previous section. The aim of this section is to
identify these quantities and to give a classification of the set of physically
inequivalent solutions.
If you are not familiar with dreibein formalism, you should check that the
model here is in fact equivalent to the metric description of three dimensional
gravity~\cite{brown:88,abbott.giddings.kuchar:84,%
alpert.gott:84,barrow.burd.lancaster:86}:
\begin{exer}
Show that, if the dreibein is invertible, the torsion equation can be solved
for the spin connection uniquely. Inserting this solution into the field
strength yields the Riemann tensor $R_{\mu\nu\rho\sigma}=\F_{\mu\nu ab}\e \rho
a \e\sigma b$, and the action becomes the Einstein Hilbert action
$\ft12\sqrt{-g}R[g]$.
\end{exer}
The action also has gauge degrees of freedom, and in dreibein gravity these are
the local Lorentz rotations and diffeomorphisms. In our language introduced in
\sref{kinematics}, they directly correspond to Einstein's relativity
principles: For every observer the special relativity principle holds, i.e.\
physics looks the same in any local frame, as long as they are related by a
Lorentz rotation, and general relativity means that physics looks the same for
all observers and therefore interchanging observers (that's what
diffeomorphisms do) does not change physics.

The local Lorentz symmetry is realized by transforming the fields as
\beq[lorentz]
 \delta \oo_\mu = \L[\lp]\, \oo_\mu =  \D_\mu \lp , \qquad
 \delta \gam_\mu = \L[\lp]\, \gam_\mu = \comm{\gam_\mu}\lp ,
\eeq
where $\lp$ is an $\alg sl(2)$ valued field on $\M$. To be more precise, at
this point we only found a symmetry. Let us postpone the exact definition of
what is a gauge symmetry. It will become clear in the Hamilton Jacobi formalism
below.

Here $\L$ is introduced as a linear operator acting on fields. Diffeomorphisms
are as usual generated by the Lie derivative of the fields. However, there is
another symmetry of the action which is more basic than the diffeomorphisms and
is also more suitable for our purpose: it treats $\gam_\mu$ as a gauge field
just as $\oo_\mu$ is treated as a gauge field of the Lorentz symmetry. We call
the symmetry transformation a translation. It also has an $\alg sl(2)$ field
$\n$ as parameter:
\beq[trans]
 \delta \oo_\mu = \H[\n]\, \oo_\mu = 0 , \qquad
 \delta \gam_\mu = \H[\n]\, \gam_\mu =  \D_\mu \n  .
\eeq
To show that this is a symmetry of the action, one has to use the Bianchi
identity $\D_{[\mu}\FF_{\nu\rho]}=0$ after a partial integration. The
diffeomorphisms generated by the Lie derivative with respect to some vector
field can now be obtained by combining these two symmetries:
\begin{exer}
Show that, given any vector field on $\M$, one can choose (field dependent)
parameters $\n$ and $\lp$ such that $\L[\lp]+\H[\n]$ acts as the Lie derivative
with respect to this vector field.
\end{exer}
If there are no more gauge symmetries, then they should form a closed Lie
algebra:
\begin{exer}
The operators $\L[\lp]$ and $\H[\n]$ provide a representation of the local
Poincar\'e algebra $\alg iso(1,2)^\M$, i.e.\ their commutators read
\beq[sym-com]
\comm{\L[\lp_1]}{\L[\lp_2]} &=& \L\big[[\lp_1,\lp_2]\big] , \zl
\comm{\L[\lp]}{\H[\n]} &=& \H\big[[\lp,\n]\big] , \qquad
\comm{\H[\n_1]}{\H[\n_2]} = 0.
\eeq
\end{exer}

\subsection{Lagrangian and boundary conditions}
\label{boundaries}
So far we considered the classical equations of motion. To go over to quantum
theory we want to reformulate the theory in Hamilton Jacobi formalism and then
apply the Dirac canonical quantization method. All the necessary structure for
this purpose has already be introduced. Spacetime is given by $\M=\N\times\RR$,
hence we have a global time coordinate which can be used as the canonical time.
We just have to split off the integration over the real number $t$ from the
action integral \eref{spin-act} to get the Lagrangian.

This should yield the same equations of motion, but now some of them become
time evolution equations for the fields, others do not contain time
derivatives. These are ``constraints'', i.e.\ they restrict the initial
conditions we may choose for the remaining evolution equations. What we want to
do in the following is to provide a well defined Lagrangian, compute momenta
and Poisson brackets in the usual way and end up with a set of constraints and
a Hamiltonian, which generates the correct time evolution by taking Poisson
brackets. Even the first step will turn out not to be easy, because we want to
allow our set of observers $\N$ to be non-compact like the cylinder in the last
section. After some redefinitions of fields we will be able to compactify $\N$,
but this manifold will have boundaries. We have to introduce boundary
conditions on the fields to get a finite Lagrangian and we will see that it is
also necessary to introduce some extra degrees of freedom to get the correct
equations of motion on these boundaries, which correspond to ``spatial
infinity''.

Let us first try to get the Lagrangian directly from the action by writing it
as a time integral of some functional $\Lag$. As this becomes an integral over
the set of observers $\N$, we should also split the one-forms $\oo_\mu$,
$\gam_\mu$ on $\M$ into the one-forms $\oo_i$, $\gam_i$ and scalars $\oo_t$,
$\gam_t$ on $\N$, as already described in \sref{kinematics}. The derivatives
with respect to the coordinate $t$ become ``velocities'', denoted by a dot.  We
also need the Levi Civita tensor on $\N$, which is given by $\eps^{ij}=\eps^{t
ij}$, or $\eps_{ij}=-\eps_{t ij}$ (giving $\N$ a positive signature in contrast
to the negative signature of $\M$). Using all this we end up with the following
Lagrangian:
\beq[Lag]
\Lag = \intN{ \eps^{ij} }{  \dot \oo_i \gam_j
    + \oo_t \D_i \gam_j
    + \ft12 \gam_t \FF_{ij}}, \qquad \Wir = \int\d t \, \Lag
\eeq
To obtain this $\Lag$ we performed a partial integration. This leads us to a
problem which has been ignored up to now, namely that of the boundary
conditions for the fields. It does, of course, not occur if space is compact
without boundary, and that's the reason why most investigations on canonical
gravity in three dimensions focussed on the torus, as this is the simplest
non-trivial example for such a space.

We divide the boundary problem into two sub-problems. The first is to get a
finite Lagrangian, the second will be to get the correct equations of motion by
adding boundary terms. To get a finite action, we somehow have to restrict the
asymptotic behaviour of the the fields. Let us assume that the manifold $\N$
has a finite number of non-compact ``ends'', which asymptotically look like
cylinders (or rather cones, but at the topological level this is the same as a
cylinders), i.e.\ $\N$ consists of some compact manifold, from which a number
of ``tubes'' extend to infinity. The simplest example for such a manifold is
the cylinder introduced in \sref{cylinder} with two ends, but you can easily
imagine a more complicated manifold consisting of a sphere with $N>2$ tubes
going off, or even a torus with some of these tubes. Note that all these
manifolds have trivial tangent bundles, as long as the genus of the ``compact
part'' of $\N$ is not greater than one.

Typical boundary conditions in field theory are given as fall-off conditions,
i.e.\ the fields have to vanish suitably fast at infinity. In gravity such a
condition does not work appropriately, because vanishing of the fields means
that the metric becomes singular. That's not what we want. We want spacetime to
be Minkowskian at infinity. So we have to look for different conditions. A
first guess is to require that the difference of the fields to some set of
background fields must vanish suitably fast. However, looking at the example
considered in \sref{cylinder}, we immediately see that even this is not
possible. The fields, especially $\gam_\phi$ at $r\to\infty$, differ by an
infinite amount for different values of the parameter $\defangle$. So if we
want to have all these solutions included, we cannot restrict the asymptotic
behaviour in this way.

On the other hand, in contrast to other field theories, there is an advantage
of gravity here: The coordinate system is arbitrary. This means that, given
such an asymptotic end, instead of choosing ``local'' coordinates $r,\phi$ with
$\phi\equiv\phi+2\pi$ and $r\in[r_0,\infty)$, we may as well take another
radial coordinate $\rho$, which is bounded, $\rho\in[\rho_0,\rho_\infty)$, and
which is related to $r$ by $r=r(\rho)$ or $\rho=\rho(r)$. A typical choice
could be $\rho(r)=\arctan r$. Using the $\rho,\phi$ coordinate system, the
integral in \eref{Lag} runs over a finite range only, making the boundary
somehow simpler. We can even compactify the manifold by taking
$\rho\in[\rho_0,\rho_\infty]$. The ``ends'' of $\N$ then look like the boundary
considered in \fref{boundpath}. From the physical point of view, what we did is
to add some ``observers at infinity'' to $\N$. They will play a special role
below: there will be extra fields defined for those observers only, and they
will somehow ``glue'' a fixed background Lorentz frame to the boundary of $\N$.
You can think of this as some kind of ``fixed stars background'', and the extra
degrees of freedom to be introduced are the relative positions and orientations
of the observers with respect to this background. It will be necessary to
introduce this extra structure to get the correct equations of motion from the
Lagrangian.

However, let us return to the problem of making the Lagrangian finite. The
fields may still diverge at $\rho\to\rho_\infty$, so we have to impose boundary
conditions on them. Let us try the simplest possible condition: we require that
all the fields can be smoothly extended to the compactified manifold, i.e.\ to
$\rho=\rho_\infty$. Then the Lagrangian is certainly finite, as it is an
integral over smooth functions on a compact manifold, which from now on will be
called $\N$. Let's see whether these are reasonable physical boundary
conditions. Transferring the spin connection from the $\rho$-system to the
$r$-system we get the following limits for $r\to\infty$ (note the extra factor
$\del\rho/\del r$ appearing when transforming the cotangent index $\rho$ into
$r$):
\beq[oo-inf]
  \oo_t(r) \to \bar\oo_t, \quad
  \oo_\phi(r) \to \bar\oo_\phi, \quad
  \oo_r(r) = \rho'(r) \oo_\rho(\rho(r)) \to 0.
\eeq
The ``bared'' quantities are those at $\rho=\rho_\infty$, which may depend on
$\phi$ (and $t$, of course), and ``$\to$'' denotes convergence at $r\to\infty$
at least as fast as $\rho'(r)$ converges to $0$:
\beq
  f(r) \to f_\infty  \equivalent
  \lim_{r\to\infty} \, (f(r)-f_\infty) / \rho'(r)  \ \mbox{exists.}
\eeq
These conditions are in fact reasonable, as they just require a gauge choice:
\begin{exer}
You can always perform a Lorentz gauge transformation generated by
\eref{lorentz} such that in the $r$-system we have $\oo_r\to0$. The remaining
conditions, stating that the limits for $\oo_\phi$ and $\oo_t$ exist, then
follow by requiring $\FF_{rt}\to0$ and $\FF_{r\phi}\to0$, which does not impose
any further restrictions on the {\em solutions} of the equations of motion.
\end{exer}
What about the dreibein field $\gam_\mu$? The boundary conditions for them look
exactly like those in \eref{oo-inf}, but this is certainly not what we want:
$\gam_r\to0$ would mean that the radial lines become lines of proper length
zero, and $\gam_\phi\to\bar\gam_\phi$ is obviously not satisfied by the example
in \sref{cylinder}. What seems to be missing is a ``power of $r$'', i.e.\ we
want $\gam_r$ to approach a constant and $\gam_\phi$ to grow linearly with $r$,
so that the end looks like a cone. The trick to achieve this works as follows.
The physical dreibein field is not $\gam_\mu$ but given by
\beq[gamhut]
\widehat\gam_\mu = \gam_\mu + \D_\mu \gam_\infty,
\eeq
where $\gam_\infty$ is some $\alg sl(2)$ valued field, which we have to adjust
such that we get the correct boundary conditions. Note that this is just a
transformation of the form \eref{trans}, so $\widehat\gam_\mu$ and $\gam_\mu$
obey the same equations of motion and \eref{spin-act} is still the correct
action, though $\gam_\mu$ appearing therein is not the {\em physical\/}
dreibein field. We choose $\gam_\infty$ such that it diverges at the boundary
and thus $\widehat\gam_\mu$ does not obey the same boundary conditions as
$\gam_\mu$. What we want is that the radial lines become spacelike geodesics
with $r$ the proper length of the line, which means
$g_{rr}=\ft12\Tr(\widehat\gam_r\widehat\gam_r)\to1$ and
$\D_r\widehat\gam_r\to0$. Using the boundary conditions on $\oo_\mu$, this
leads to the requirement $\widehat\gam_r\to\bar\gam$, where $\bar\gam$ is some
(possibly $\phi$-dependent) matrix such that $\ft12\Tr(\bar\gam\bar\gam)=1$.

To achieve this, we just have to set $\gam_\infty(r,\phi)=r\bar\gam(\phi)$ for
large $r$ and smoothly falling off to $0$ when going towards the interior of
$\N$. The boundary conditions on the physical dreibein are obtained by
replacing $\oo_\mu$ with $\gam_\mu$ in \eref{oo-inf} and then inserting
\eref{gamhut}:
\beq[gam-inf]
  \widehat\gam_r \to \bar\gam, \quad
  \widehat\gam_t - r \comm{\bar\oo_t}{\bar\gam} \to \bar\gam_t , \quad
  \widehat\gam_\phi - r \,\del_\phi \bar\gam -
                    r  \comm{\bar\oo_\phi}{\bar\gam} \to \bar\gam_\phi.
\eeq
They look rather awkward but you can see that now $\widehat\gam_r$ becomes a
constant and $\widehat\gam_\phi$ becomes linear in $r$. If expressed in terms
of the fields $\gam_\mu$ instead of $\widehat\gam_\mu$, however, they are as
simple as \eref{oo-inf} and they are even simpler in the $\rho$ system, where
all the fields just have to be smooth at the boundary of $\N$. We always assume
that the fixed field $\gam_\infty$, which provides the relation \eref{gamhut},
does not depend on $t$. This is in principle possible but could cause some
trouble with an explicitly time-dependent Hamiltonian.
\begin{exer}
Show that under the conditions \eref{gam-inf} the radial lines asymptotically
become geodesics, and that the last two conditions follow from the first one by
requiring the components $\D_{[r}\widehat\gam_{t]}$ and
$\D_{[r}\widehat\gam_{\phi]}$ of the torsion to vanish for $r\to\infty$, so
that they do not impose additional conditions on the solutions.
\end{exer}
So after all we got a finite action in terms of the fields $\oo_\mu,\gam_\mu$
on a compactified manifold $\N$, which are related to the physical fields
existing on the non-compact interior of $\N$ only by \eref{gamhut}. We can now
forget how this relation was constructed using local coordinates. The relation
between the ``physical'' fields $\widehat\gam_\mu$ and the ``finite'' fields
(which appear in the Lagrangian and obey the boundary condition ``well defined
on the compactified $\N$'') is entirely described by the $\alg sl(2)$ valued
field $\gam_\infty$, which must diverge to spatial infinity, i.e.\
$\Tr(\gam_\infty\gam_\infty)\to+\infty$, at the boundaries of $\N$. Our
background structure is a compact manifold $\N$ with some boundaries looking
like \fref{boundpath}, and a fixed field $\gamma_\infty$ describing how the
fields on $\N$ are related to the physical fields on the non-compact interior
of $\N$.
\begin{exer}
Given a path $\curva$ in $\N$, then the ``physical'' distance
$\widehat\VV_\curva$ is related to the ``finite'' distance $\VV_\curva$ by
\beq[VV-finite]
  \widehat\VV_\curva = \VV_\curva + \UU_\curva \, \gam_\infty(\curva(1)) \,
        \UU_\curva\inv - \gam_\infty(\curva(0)),
\eeq
where the finite and physical distance is obtained by \eref{UU.VV} with
$\gam_\mu$ and $\widehat\gam_\mu$ inserted, respectively. Hence, the finite
distance of a path can be obtained by measuring the physical distance and
``subtracting'' the $\gam_\infty$ field (which has a fixed value for every
observer). As $\VV_\curva$ has a finite limit if the path approaches the
boundary, you can also measure the ``distance'' of the observers at infinity.
\end{exer}
Let us now investigate the dynamics induced by the finite Lagrangian $\Lag$.
What we want is, of course, to get the equations of motion \eref{spin-eom}
back. It is easy to read off the canonical variables and Lagrange multipliers,
because the Lagrangian is given in the first order standard form $\Lag=\dot q p
- \Ham$, where $q$ are the configuration variables $\oo_i$, $p$ the conjugate
momenta $\eps^{ij}\gam_j$, and $\Ham$ is the Hamiltonian, depending on the
phase space variables and the multipliers $\oo_t$, $\gam_t$, whose velocities
do not enter the Lagrangian. We could introduce Poisson brackets right here,
but we postpone it until we have our final Lagrangian. Let us first check
whether $\Lag$ really provides the correct equations of motion. Variation of
the Lagrangian with respect to the multipliers yields the constraint equations
\beq[con]
  \eps^{ij} \D_i \gam_j \weak 0 , \qquad
  \ft12 \eps^{ij} \FF_{ij} \weak 0 ,
\eeq
corresponding to the spatial components of the equations of motion
\eref{spin-eom}. They define a subset of the phase space, which is called
``physical phase space'' or ``constraint surface''. We introduce the usual
notation ``$\weak$'' for a ``weak'' equality, i.e.\ a relation between phase
space functions which becomes an equality on the physical phase space. Hence,
\eref{con} is just a definition of the symbol $\weak$.

To simplify the notation, we derive the remaining Euler Lagrange equations as
follows: if we vary a field whose time derivative enters the Lagrangian, we
write
\beq[time-parts]
  \delta\Lag[\phi,\dot\phi] = \deltadelta\Lag/\phi/ \, \delta \phi
      + \deltadelta\Lag / \dot\phi / \, \delta\dot\phi =
      \Big(  \deltadelta\Lag/\phi/ -
           \dd /t/ \deltadelta\Lag / \dot\phi / \Big) \, \delta\phi,
\eeq
which means that in all these variations we integrate the time derivative by
parts if it acts on the variation of the field, so that we can directly read
off the Euler Lagrange equation. With spatial derivatives and partial
integrations, however, we have to be more careful, because the space $\N$ has
boundaries, and the boundary conditions do not require the fields to be fixed
there.

If we now vary $\gam_i$, we find
\beq[L-var]
 \delta\Lag &=& \intN {\eps^{ij}}{ \dot \oo_i \delta\gam_j
      + \oo_t \D_i \delta\gam_j }  \zl
  &=& \intN {\eps^{ij}}{ \FF_{ti} \, \delta\gam_j }
    + \send\intdN { \oo_t \delta\gam_\phi } .
\eeq
Here we picked up a boundary term after a partial integration, using Stokes'
theorem
\beq[stokes]
   \intd2\x \eps^{ij} \del_i X_j = \send \oint \d\phi\, X_\phi ,
\eeq
where $\phi$ is the periodic coordinate on the boundary. The sum runs over all
boundaries of $\N$. They all have to be parameterized in the same direction,
and the Levi Civita tensor has to be normalized such that $\eps^{\rho\phi}=+1$,
where $\rho$ is the local coordinate pointing towards the boundary. Note that
the compactification of $\N$ now makes it very simple to deal with such
integrals, as all the fields are well defined on the boundary, where the right
hand side of \eref{stokes} is evaluated. The symbol $\oint$ denotes integration
over any range $\phi_0$ to $\phi_0+2\pi$. The fields $\oo_t$, $\gam_\phi$,
etc., are considered as periodic functions of $\phi$.

Unfortunately, we do not get the correct equations of motion. Requiring
\eref{L-var} to vanish for all $\delta\gam_j$ gives the correct equation
$\FF_{ti}=0$, but in addition we get $\oo_t=0$ on the boundary. A similar
problem arises for the variation of the spin connection, where we have
\beq[L-var2]
  \delta\Lag &=&  \intN {\eps^{ij}}{  -\delta\oo_i \dot \gam_j
      + \oo_t \comm{\delta\oo_i}{\gam_j}
      + \gam_t \D_i \delta\oo_j }  \zl
   &=& \intN {\eps^{ij}}{ ( \D_t \gam_i - \D_i \gam_t ) \, \delta\oo_j }
    + \send\intdN { \gam_t \delta\oo_\phi } .
\eeq
Here we integrated the time derivative by parts as described in
\eref{time-parts}. Again, we get the correct time components of the torsion
equation $\D_{[t}\gam_{i]}=0$, but we also get $\gam_t=0$ on the boundary. What
would it mean if the values of the multipliers are fixed at the boundary? The
freedom to choose the values of the multipliers corresponds to gauge
symmetries. A consequence would be that we loose some gauge degrees of freedom.
This would not be too bad, but to maintain the correct number of physically
equivalent states, we had to ``fix the gauge'' somehow on the boundary.
\begin{exer}
An alternative way to deal with boundary conditions is the following: find a
suitable gauge fixing for the fields $\gam_\phi$ and $\oo_\phi$ on the
boundaries such that together with $\oo_t=0$ and $\gam_t=0$ you get the same
final result for the reduced phase space. You have to make the $\gam_\infty$
field time dependent to get a non-singular metric at infinity. However, this is
rather cumbersome and at least as complicated as introducing the auxiliary
fields below.
\end{exer}
We can try to avoid these problems be adding the extra term
\beq[guess]
  \Lagb = - \send \intdN{ \gam_t \oo_\phi + \oo_t \gam_\phi }
\eeq
to the Lagrangian. But then the equations of motion for the multipliers $\oo_t$
and $\gam_t$ are no longer given by \eref{con}; we get extra constraints
$\oo_\phi=0$ and $\gam_\phi=0$ at the boundary. This is just the kind of
boundary condition we rejected above, because the simple example from
\sref{cylinder} did not satisfy them. The only way out of this dilemma is to
introduce extra auxiliary fields on the boundary. They look rather unmotivated
first, but as already mentioned we will find a nice physical interpretation for
them in the end.

For each boundary of $\N$ we introduce a pair of fields $\u(\phi)\in\grp SL(2)$
and $\v(\phi)\in\alg sl(2)$. They are not periodic in $\phi$, so consider them
as fields on a real line instead of a circle, or as fields on the boundary of
$\cov\N$ as defined by the family of paths in \fref{boundpath}, instead of
fields on the boundary of $\N$. However, they are not completely arbitrary. We
require them to be quasiperiodic (see definition~\ref{quasi-def}), i.e\ the
pair $(\u,\v)$, considered as an $\grp ISO(1,2)$ valued field, obeys the
relation
\beq
   (\u,\v)(\phi+2\pi) = (\u_\endn,\v_\endn) \, (\u,\v)(\phi),
\eeq
with the multiplication as given by \eref{poin-mult}.
Written out in components, the relation becomes
\beq[u-v-quasiper]
    \u(\phi+2\pi) = \u_\endn \u(\phi) , \qquad
   \v(\phi+2\pi) = \v_\endn + \u_\endn \, \v(\phi) \, \u_\endn\inv .
\eeq
Quasiperiodic fields of this type will play a very important role below. They
are used to classify the complete solution of the constraint equations. By
taking derivatives, there is an alternative definition for quasiperiodic
fields:
\begin{exer}
Show that \eref{u-v-quasiper} is equivalent to the requirement that the
following $\alg sl(2)$ valued derivatives of $\u$ and $\v$ are periodic:
\beq[u-v-periodic]
  \oophi =  \u^{-1}\del_\phi\u  , \qquad
   \gamphi =   \u^{-1}\del_\phi\v\,\u  .
\eeq
\end{exer}
{}From these relations we infer that $\u$ (as well as $\v$) has one degree of
freedom more than a periodic field, which is somehow encoded in the integration
constant for the differential equation $\u^{-1}\del_\phi\u=\oophi$ (or the
corresponding relation for $\v$). You can also consider the constant $\u_\endn$
(or $\v_\endn$) as this extra degree of freedom.

To get the correct equations of motion for the dreibein and the spin
connection, we now add a term to $\Lag$ which looks very much like
\eref{guess}:
\beq[extra-L]
  \Lagb =  \send \intdN{ \gam_t (\oophi- \oo_\phi )
                          + \oo_t (\gamphi - \gam_\phi ) }
\eeq
If we now vary $\oo_i$ or $\gam_i$, the boundary terms in \eref{L-var} and
\eref{L-var2} cancel against the variation of $\Lagb$. What we get is the
correct time evolution for the dreibein and the spin connection
\beq[evol-o-g]
  \dot \oo_i = \D_i \oo_t, \qquad
  \dot \gam_i = \D_i \gam_t - \comm{\oo_t}{\gam_i} .
\eeq
Variation of the multipliers no longer yields $\oo_\phi=\gam_\phi=0$ on the
boundary, as we found for \eref{guess}. Instead, we get new constraint
equations relating the boundary fields $\u,\v$ to the connection and dreibein
field tangent to the boundary:
\beq[u-v-con]
    \u^{-1} \del_\phi \u = \oophi \weak \oo_\phi , \qquad
    \u^{-1} \del_\phi \v\,\u = \gamphi \weak \gam_\phi.
\eeq
This does not impose any new constraints on $\oo_i$ or $\gam_i$, because the
differential equations are globally integrable for any arbitrary pair of
periodic fields $\oophi$ and $\gamphi$. However, it seems that we only shifted
the original problem somewhere else. As we have introduced new fields, we also
get additional equations of motion for these fields. A variation of $\v$, e.g.,
would immediately give $\oo_t=0$, which is exactly what we had before. But we
may still add terms depending on the new fields only.

What we need is a kinetic term for the new fields, which makes the field $\u$ a
configuration variable and $\v$ the conjugate momentum. It is
\beq[Lag-u-v]
  \Lagkin = \send \Big(
        \Trr{ \u_\endn\inv \dot \u_\endn \, \v(\phi_0) } -
      \intdNl{\phi_0} {\dot\u \, \u^{-1} \del_\phi \v } \Big).
\eeq
The notation $\oint_{\phi_0}$ indicates that the range of the integral is
$\phi_0$ to $\phi_0+2\pi$.
\begin{exer}
Show that the value of $\Lagkin$ does not depend on $\phi_0$. The integrand
itself is not periodic, but the first term compensates this non-periodicity.
\end{exer}
At the moment there is absolutely no motivation for this awkward looking
boundary term. The only reason to choose this $\Lagkin$ is that now we get the
correct equations of motion. It is straightforward but rather complicated to
compute these equations, as we are dealing with group valued quasiperiodic
fields. Only those variation are allowed which preserve \eref{u-v-quasiper}.
The actual derivation of the equations is given in the appendix. Requiring the
action to be stationary implies
\beq[u-v-evol]
     \dot \u =  \u \oo_t , \qquad
     \dot \v =  \u \gam_t \u^{-1} ,
    \dot \u_\endn = \dot \v_\endn = 0.
\eeq
Hence, \eref{evol-o-g} and \eref{u-v-evol} provide the full set of evolution
equations for the fields. Finally, we can check that these are consistent with
the constraints in the following sense: given any solution to the constraints
as initial condition for the time evolution equations, then the constraints are
automatically satisfied at any later time. In Dirac's formalism such
constraints are called first class, and we will see below that this is
equivalent to the fact that the Poisson algebra of the constraints is closed.
\begin{exer}
Show that under the time evolution the relations \eref{u-v-quasiper} and all
the constraint equations \eref{con} and \eref{u-v-con} are conserved for
arbitrary values of the multipliers $\oo_t,\gam_t$. As a consequence, those
cannot carry any physical information, as their time dependence is completely
undetermined. All physical information is contained in the phase space
variables $\{\oo_i,\gam_i,\u,\v\}$.
\end{exer}

\subsection{Poisson brackets, gauge symmetries and constraint algebra}
We now have a well defined Lagrangian depending on the multipliers $\oo_t$,
$\gam_t$, the configurations variables $\oo_i$, $\u$, and their conjugate
momentum variables $\gam_i$ and $\v$. The Lagrangian is first order in time
derivatives, so it should be straightforward to read off the Poisson brackets.
For the interior of $\N$ it is in fact easy, as the canonically conjugate
momentum of $\oo_i$ is $\eps^{ij}\gam_j$. To avoid confusion with the matrix
indices, we will not use brackets where both entries are matrices. Instead, we
use ``dummy'' $\alg sl(2)$ matrices and take traces to write down the brackets.
The basic bracket between the spin connection and the dreibein reads
\beq[o-g-pois]
    \pois{\Trr{\a\gam_i(\y)}}{ \oo_j(\x)} = \eps_{ij}\, \a \,
        \delta(\x,\y),
\eeq
or vice versa
\beq[g-o-pois]
    \pois{\Trr{\a\oo_i(\y)}}{ \gam_j(\x)} = \eps_{ij}\, \a \,
        \delta(\x,\y),
\eeq
and brackets between components of the connection or the dreibein alone always
vanish. Instead of tracing the momenta, we can also switch back to the vector
representation for $\alg sl(2)$ matrices (by taking $\a=\ft12\gam^a$), where
the same relation reads
\beq[o-e-pois]
    \pois{\e i a(\y)}{ \oo_j(\x)} = \ft12 \eps_{ij}\, \gam^a \,
        \delta(\x,\y),
\eeq
So far this was not too sophisticated. However, if we now consider the boundary
term, things get more involved. Not only that the fields are group valued, the
quasiperiodicity makes it hard to find the brackets. It is not easy to derive
the brackets of $\u$ with $\v$ in the usual way, and those of $\v$ with itself
are even worse. It turns out that there is only a subset of brackets that is
actually needed, namely those where at least one entry is one of the periodic
fields $\oophi$ and $\gamphi$ introduced in \eref{u-v-periodic}. The rather
technical derivation of the complete brackets is given in the appendix. All
$\u$'s commute, and thus also the bracket of $\u$ with $\oophi$ vanishes. We
have
\beq[o-v-pois]
 \pois{\u(\phi)}{\Trr{\a \oophi(\phi') }} &=& 0 , \zl
 \pois{\v(\phi)}{\Trr{\a \oophi(\phi') }} &=&
               - \u(\phi)\, \a \, \u^{-1}(\phi) \, \pdelta(\phi-\phi'),
\eeq
where $\pdelta(\phi)$ is the periodic delta function on the circle, i.e.\ it
has a peak whenever $\phi=2\pi z$, $z\in\ZZ$. The brackets with $\gamphi$ are
\beq[u-g-pois]
  \pois{\v(\phi)}{\Trr{\a \gamphi(\phi')} } &= & 0, \zl
  \pois{\u(\phi)}{\Trr{\a \gamphi(\phi')} } &= &
                   - \u(\phi) \, \a \, \pdelta (\phi-\phi').
\eeq
Note that they are periodic in $\phi'$ but quasiperiodic in $\phi$. In
particular, the quasiperiodicity parameters $\u_\endn$ and $\v_\endn$ commute
with both $\oophi$ and $\gamphi$. Differentiating them with respect to $\phi$
and using the definition \eref{u-v-periodic} we obtain the following brackets
between $\oophi$ and $\gamphi$:
\beq[bnd-pois]
  \pois{\Trr{\a \oophi(\phi)}}{\Trr{\b \gamphi(\phi')} } &= &
     \Trr{ \comm \a\b \oophi } \, \pdelta(\phi-\phi')
    - \Trr{\a\b} \, \pdelta'(\phi-\phi') , \zl
   \pois{\Trr{\a \gamphi(\phi)}}{\Trr{\b \gamphi(\phi')} } &= &
     \Trr{ \comm \a\b \gamphi } \, \pdelta(\phi-\phi') .
\eeq
We can now use these brackets to reproduce the equations of motion. We first
define the ``smeared'' constraints. Given any two $\alg sl(2)$ valued fields
$\lp$ and $\n$ on $\N$, they are
\beq[const-eq]
  \L[\lp]&= & \intN{\eps^{ij}}{\lp \D_i \gam_j}
     + \send \intdN{ \lp ( \gamphi - \gam_\phi )} \weak 0 , \zl
 \H[\n]&=& \intN{\ft12 \eps^{ij}}{\n \FF_{ij}}
     + \send \intdN{ \n ( \oophi - \oo_\phi )} \weak 0 .
\eeq
The constraint equations tell us that these quantities have to vanish for any
pair of fields $\lp$, $\n$. Remember that the integration runs over the
compactified manifold $\N$, thus $\lp$ and $\n$ have to be smooth at the
boundary and the integrals are always well defined.
Using this notation, we can write down the Hamiltonian as
\beq[Ham]
  \Ham = - \L[\oo_t] - \H[\gam_t] \weak 0 .
\eeq
Thus $\Ham$ is a linear combination of constraints, the coefficients given by
the Lagrange multipliers. It vanishes for any solution of the constraints,
i.e.\ on the whole physical phase space. This reflects the fact that our
``time'' $t$ is unphysical. As a consequence, there is no ``energy'' of a
physical state which is usually represented by the Hamiltonian defined with
respect to the physical time. However, the Hamiltonian still generates the time
evolution, but with respect to the parameter time $t$, by taking Poisson
brackets. To compute
these brackets, it is more suitable to perform a partial integration in
\eref{const-eq}, such that there are no more derivatives acting on $\oo_i$ or
$\gam_i$. Using Stokes' theorem we find
\beq[const-gen]
  \L[\lp]&=& - \intN{\eps^{ij}}{\D_i \lp \, \gam_j}
     + \send \intdN{ \lp  \gamphi }, \zl
 \H[\n]&=& - \intN{\eps^{ij}}{\del_i \n \, \oo_j - \n \oo_i \oo_j }
     + \send \intdN{ \n \oophi}.
\eeq
With the brackets above we obtain
\beq[lorentz-trafo]
  \pois{\L[\lp]}{\oo_i}  &=& \D_i \lp ,  \qquad
  \pois{\L[\lp]}{\u} =  \u \lp , \zl
  \pois{\L[\lp]}{\gam_i}  &=&  \comm{\gam_i}{\lp} ,  \qquad
  \pois{\L[\lp]}{\v} =  0 .
\eeq
and
\beq[trans-trafo]
  \pois{\H[\n]}{\oo_i}  &=& 0,   \qquad
  \pois{\H[\n]}{\u} =  0, \zl
  \pois{\H[\n]}{\gam_i}  &=& \D_i \n  ,  \qquad
  \pois{\H[\n]}{\v} = \u \n \u^{-1} .
\eeq
There are two important facts summarized in these equations. First, observe
that the right hand sides are exactly the Lorentz transformations
\eref{lorentz} and translations \eref{trans}, but now they act only on the
canonical fields on $\N$ instead of the fields on the spacetime manifold $\M$.
In addition, we get some new transformations of the extra boundary fields, and
one can check that these are indeed necessary for the complete action including
the boundary terms to be invariant under gauge transformations. Hence, we found
an explicit representation of the operators introduced in \eref{lorentz} and
\eref{trans} as Poisson brackets with the constraints. We therefore call $\L$
the Lorentz constraint, and $\H$ is called Hamilton constraint: in ADM gravity
using the metric as field variable, it splits into the generator of space and
time diffeomorphism, and the name Hamilton constraint was introduced because it
somehow replaces the Hamiltonian as the generator of time evolution.

Secondly, we see that the equations of motions are reproduced by the Poisson
brackets as the Hamiltonian is a linear combination of $\L$ and $\H$, and we
only have to replace $\lp$ and $\n$ with $\oo_t$ and $\gam_t$ to get
\eref{evol-o-g} and \eref{u-v-evol} back. For any phase space function we have
\beq
  \dot F = \pois F\Ham = \pois {\L[\oo_t]}F + \pois{\H[\gam_t]}F.
\eeq
The arbitrariness of the multipliers reflects the fact that time evolution is
unique only up to gauge transformations. In addition, as our ``time'' $t$
itself is completely arbitrary, time evolution is in fact nothing but a gauge
transformation. The action is invariant under any reparameterization of time,
which is part of the gauge symmetry, so time evolution is formally equivalent
to a gauge transformation.

We can now also check that the constraints are first class in the sense of
Dirac, i.e.\ that the Poisson algebra of $\L$ and $\H$ closes. We already did
two similar calculations: we have seen that the ``linear operators'' $\H$ and
$\L$ defined in \eref{lorentz} and \eref{trans} do form a closes algebra, and
we saw that the constraints are preserved by the time evolution.
\begin{exer}
Compute the Poisson algebra of the constraints and show that
\beq[con-alg]
\pois{\L[\lp_1]}{\L[\lp_2]} &=& \L\big[[\lp_1,\lp_2]\big] , \zl
\pois{\L[\lp]}{\H[\n]} &=& \H\big[[\lp,\n]\big] , \qquad
\pois{\H[\n_1]}{\H[\n_2]} = 0.
\eeq
You have to be careful with partial integrations and boundary terms to get the
correct results. As a consequence, given any two (smooth) phase space functions
vanishing on the physical phase space, $F\weak0$, $G\weak0$, then we also have
$\{F,G\}\weak0$.
\end{exer}
We can now give an exact definition of what we mean by a gauge transformation,
which in Dirac's~\cite{dirac:65} canonical framework is a transformation
generated by taking Poisson brackets with first class constraints.
\begin{defi}[Gauge Transformation]\label{gauge}
Two physical field configurations $\bphi_0$ and $\bphi_1$ (denoting the set
$\{\oo_i,\gam_i,\u,\v\}$) are called ``gauge equivalent'' or ``related by a
gauge transformations'' if there is a smooth deformation $\bphi_s$,
$s\in[0,1]$, such that
\beq
  \dd /s/ \bphi_s \weak \pois{ \L[\lp_s] + \H[\n_s] }{  \bphi_s},
\eeq
with some parameter fields $\lp_s,\n_s$, which may of course vary with $s$.
\end{defi}
Field configurations related by a gauge transformation are assumed to represent
the same physical state. All other symmetries of the action are called global
symmetries and they map different physical states onto each other. We will see
this quite explicitly for an example below, so it is really important to
distinguish between these symmetries. In context of gravity those global
symmetries are sometimes called ``large diffeomorphisms'' or ``large Lorentz
rotations'', because they cannot be cut into (infinitesimally) small pieces.

To identify the gauge invariant objects we will need later on, we have to know
how the finite gauge transformations described in definition~\ref{gauge} look
like, i.e.\ we have to go over from the gauge algebra to the gauge group. We
formally get it by exponentiating the ``infinitesimal generators'' $\L$ and
$\H$. This is quiet easy for the translations as the exponential series is of
first order only. A finite translation is given by
\beq[trans-fin]
  \oo_i &\mapsto& \oo_i ,
         \phantom{+ \D_i \n} \qquad    \u \mapsto \u , \zl
  \gam_i &\mapsto& \gam_i + \D_i \n , \qquad \v \mapsto \v + \u \n \u^{-1}.
\eeq
For the Lorentz symmetry exponentiation is not as simple, but of course we know
how to do it. Instead of an $\alg sl(2)$ valued field $\lp$ the parameter
becomes a group valued $\grp SL(2)$ field $\h$ and the Lorentz rotation reads
\beq[lorentz-fin]
  \oo_\mu &\mapsto& \h^{-1} (\del_\mu + \oo_\mu) \h , \qquad
   \u \mapsto \u \h,  \zl
 \gam_\mu &\mapsto& \h^{-1} \gam_\mu \h,
    \phantom{+ \oo_\mu) \h}  \qquad  \v \mapsto \v .
\eeq
However, here we have to be a little bit more careful again. In
\sref{lorentz-group} we saw that the ``Lorentz group'' is actually $\cgrp
SL(2)$ and not $\grp SL(2)$. So shouldn't the parameter of the Lorentz rotation
better be an $\cgrp SL(2)$ field $\th$? Or, the other way around, couldn't we
take an $\grp SO(1,2)$ field $h^a\_b$ as parameter, and what's the difference?
We can rewrite \eref{lorentz-fin} in the vector representation and get
\beq[lorentz-fin-vec]
  \o_{\mu ab} \mapsto h^c\_a ( \eta_{cd} \del_\mu +\o_{\mu cd}) h^d\_b ,
 \qquad
  \e \mu a \mapsto h_b\^a \e \mu b.
\eeq
However, here we already get into trouble with the transformation of $\u$.
Ignoring this problem for a moment, we could equally well replace the boundary
field $\u\in\grp SL(2)$ with $u_a\^b\in\grp SO(1,2)$), so that we don't have to
assume that $h^a\_b$ is the projection \eref{homo} of some field $\h$ onto
$\grp SO(1,2)$. The symmetry transformation, at least for the gauge fields
$\oo_i$ and $\gam_i$, is well defined for any $\grp SO(1,2)$ field $h^a\_b$.
The difference between \eref{lorentz-fin-vec} and \eref{lorentz-fin} is that
there are $\grp SO(1,2)$ valued fields which are not projections of $\grp
SL(2)$ valued fields.
\begin{exer}
Show that there are such fields on a manifold $\N$ if and only if the loop
group of $\N$ is non-trivial, i.e.\ $\LG\ne\{\loopn\}$.
\end{exer}
The question is: shall we consider those as gauge transformations or not? They
are certainly symmetries of the action. However, the definition above tells us
that they are not gauge symmetries in the sense of Dirac.
\begin{exer}\label{fake-gauge}
Show that \eref{trans-fin} and \eref{lorentz-fin} are gauge transformations if
and only if $\h$ is a projection of some $\cgrp SL(2)$ valued field $\th$ on
$\N$ (see exercise~\ref{smooth-deform}). As an example, take the cylinder
manifold with the fields given by \eref{pol-dreib-con}. With our boundary
conditions, these are in fact the $\widehat\gam_\mu$ fields. Choose
$\gam_\infty=r\gam_1$ to transform to the finite fields. Then perform a Lorentz
transformation with parameter $\h(r,\phi)=\exp(\phi\gam_0)$ (which is quite
simple). Transform back to the physical fields $\widehat\gam_\mu$. You will end
up with a different value of $\defangle$, showing that physics has changed by
choosing a field $\h$ as parameter for the Lorentz transformation which is not
a projection of any $\th$.
\end{exer}
\begin{exer}
Under the finite gauge transformations, the transport operator and the distance
of a path changes by
\beq
  \tU_\curva \mapsto \th^{-1}(\curva(0)) \, \tU_\curva \, \th(\curva(1)),
  \qquad
  \VV_\curva \mapsto \h^{-1}(\curva(0)) \, \VV_\curva \, \h(\curva(0))
\eeq
for a Lorentz rotation with paramter $\th:\N\to\cgrp SL(2)$, and
\beq[U-V-gauge]
  \tU_\curva \mapsto  \tU_\curva ,
  \qquad
  \VV_\curva \mapsto  \VV_\curva
       + \UU_\curva\,\n(\curva(1))\,\UU_\curva\inv
        - \n(\curva(0)),
\eeq
for a translation with parameter $\n:\N\to\alg sl(2)$, respectively.
\end{exer}

\subsection{Solving the constraints}
\label{solving}
To find classical solutions of the equations of motion, all we have to do is to
solve the constraints. Time evolution of the so found states is then given by
choosing suitable values for the multipliers and computing the brackets with
the Hamiltonian. A ``suitable'' choice would be $\oo_t=0$ and $\gam_t=\gam_0$,
corresponding to free falling local observers (see exercise~\ref{free-fall}). A
second step will then be to classify the equivalence classes of the solutions
modulo gauge transformations, which gives an overview over all physically
inequivalent classical states.

Let us first consider the interior equations \eref{con}. The Hamilton
constraint depends on $\oo_i$ only and states that its field strength vanishes.
The solution for such an equation is well known. It can be locally
parameterized by an $\grp SL(2)$ valued field $\g$ and reads
\beq[o=gdg]
  \oo_i = \g^{-1} \del_i \g.
\eeq
However, such a parameterization exists locally only and getting a
parameterization for the global solution is not trivial. Before showing how to
solve this problem, let us solve the Lorentz constraint locally, too. Inserting
\eref{o=gdg} we can rewrite it as
\beq
 \eps^{ij} \, \g^{-1} \del_i \big( \g \gam_j \g^{-1} \big)\, \g =0.
\eeq
It now states that the curl of the covector inside the parenthesis vanishes and
therefore there is (locally) a ``potential'' $\f\in\alg sl(2)$ such that
\beq[e=gdfg]
  \g \gam_i \g^{-1} = \del_i \f \equivalent
  \gam_i = \g^{-1} \del_i \f \g.
\eeq
How can we find the global solution? The answer is that we cannot define the
``potentials'' $\g$ and $\f$ globally on $\N$, but we can do it on the covering
manifold $\cov\N$. In addition, we already learned that it might be better to
take an $\cgrp SL(2)$ field $\tg$ instead of $\g\in\grp SL(2)$ right from the
beginning. This will avoid confusion later on, as the gauge invariants are in
fact $\cgrp SL(2)$ valued.

Remember that the elements of $\cov\N$ are homotopy classes of paths in $\N$
starting at a special point $\xo$. Hence, let us consider this point as a
special observer and what follows will describe the universe as seen by this
observer. Given a solution to the constraint equations \eref{con}, we will use
the non-local objects from \sref{test-particle} to define the following fields
on $\cov\N$:
\beq[pot-def]
  \tg(\curva)=\tU_\curva, \qquad \g(\curva) = \UU_\curva, \qquad
  \f(\curva)= \VV_\curva.
\eeq
Of course, $\g$ is just the projection of $\tg$ onto $\grp SL(2)$, so it is not
really an independent field, we just need it to derive some explicit formula.
Remember that the transport operator and the distance of a path are invariant
under smooth deformations of the path, provided that the dreibein postulate
holds and the curvature of the spin connection vanishes. Hence, these fields
are well defined on $\cov\N$ if the constraints are satisfied. If we now vary
$\curva\in\cov\N$ without keeping $\curva(1)$ fixed, we can use the spinor
representation of \eref{UV-deform} to get
\beq
   \delta \UU_\curva = \UU_\curva \, \oo_i\, \delta\curva^i
  \qquad
   \delta\VV_\curva = \UU_\curva \, \gam_i \,\delta\curva^i\,\UU_\curva\inv,
\eeq
where $\delta\curva^i$ is the variation of the end point of $\curva$. Using the
same coordinates on $\N$ and $\cov\N$, $\delta\curva^i$ is also the variation
of the {\em point\/} $\curva\in\cov\N$, so these two equations tell us how the
derivatives of $\g$ and $\f$ look like. Explicitly, we get
\beq[pot-field]
  \g^{-1}\del_i\g =  \oo_i , \qquad
  \g^{-1}\del_i\f\,\g  =  \gam_i .
\eeq
Note, however, that these equations ``live'' on $\cov\N$, and $\oo_i$ and
$\gam_i$ are the pullbacks of the gauge fields onto $\cov\N$. As $\oo_i$ and
$\gam_i$ are periodic, we are using the same symbol for a field on $\N$ and the
corresponding periodic field on $\cov\N$.

Obviously, we can recover $\gam_i$ and $\oo_i$ from $\tg$ and $\f$, and we get
exactly \eref{o=gdg} and \eref{e=gdfg}. Thus the potentials $\tg$ and $\f$
parameterize the solutions \eref{con} completely. But not every arbitrary set
of potentials on $\cov\N$ corresponds to such a solution on $\N$: the left hand
sides of \eref{pot-field} have to be periodic on $\cov\N$.
We already dealt with these kinds of conditions for the $\u$ and $\v$ fields
above, so we expect the potentials to be quasiperiodic. Indeed, this time the
condition is even stronger:
\begin{exer}
Show that the field $(\tg,\f):\cov\N\to\cgrp ISO(1,2)$ defined by
\eref{pot-def} is {\em normalized quasiperiodic,\/} i.e.\ we have
\beq[gf-quasiper]
  (\tg,\f)\circ\shift_\loopa=(\tg_\loopa,\f_\loopa)\, (\tg,\f)
\eeq
for all $\loopa\in\LG$, with $\tg_\loopa=\tg(\loopa)$ and
$\f_\loopa=\f(\loopa)$. In components:
\beq[g-f-quasiper]
  \tg(\loopa\kette\curva)= \tg_\loopa \,\tg(\curva) , \quad
  \f(\loopa\kette\curva)= \f_\loopa + \g_\loopa\, \f(\curva) \,\g_\loopa\inv  .
\eeq
\end{exer}
Every solution to the constraints is mapped onto a pair of normalized
quasiperiodic potentials. We already saw that this map is injective, as we were
able to recover the fields from the potentials. On the other hand, if $(\g,\f)$
is a set of normalized quasiperiodic potentials, we find that the gauge fields
derived via \eref{pot-field} are periodic on $\cov\N$, and of course they solve
the constraints. Hence, there is an inverse map from the set of normalized
quasiperiodic potentials to the solutions of the constraints.
This inverse map is injective, too: if there are two potentials yielding the
same gauge fields via \eref{pot-field}, then they can differ by a constant
only:
$\tg'=\tg_0\tg$, $\f'=\f_0+\g_0\f\g_0\inv$, or
$(\tg',\f')=(\tg_0,\f_0)(\tg,\f)$. But normalized quasiperiodic fields always
obey $\tg(\loopn)=\eins$, $\f(\loopn)=0$. This gives $\tg_0=\eins$, $\f_0=0$.
So the map is in fact one-to-one.

Now what about the boundary fields $\u$ and $\v$? We have to solve the boundary
constraints \eref{u-v-con}, which relates the gauge fields on the boundary to
the quasiperiodic fields $\u$ and $\v$. To insert the parameterization of the
gauge fields in terms of the potentials on $\cov\N$, we use the boundary paths
introduced in \fref{boundpath}. We get the gauge fields $\oo_\phi$ and
$\gam_\phi$ tangent to the boundary by fixing any family of paths
$\curva_{\endn,\phi}$ and using the formulas \eref{o=gdg} and \eref{e=gdfg}.
Hence, the boundary constraints are
\beq[bound-rel]
  \g^{-1} \del_\phi \g(\curva_{\endn,\phi}) &=&
  \u^{-1} \del_\phi \u(\phi) ,  \zl
 \g^{-1} \del_\phi \f \, \g(\curva_{\endn,\phi}) &=&
  \u^{-1} \del_\phi\v\,\u(\phi).
\eeq
The solution to these equations is that $(\u,\v)$ differs from $(\g,\f)$ only
by multiplication with a constant $(\b_\endn,\c_\endn)^{-1}$ from the left, or
explicitly
\beq[u-v-soln]
      \u(\phi) = \b_\endn\inv  \g(\curva_{\endn,\phi}) , \qquad
     \v(\phi) = \b_\endn\inv
           \big(  \f(\curva_{\endn,\phi}) -  \c_\endn  \big)
              \, \b_\endn ,
\eeq
where $\b_\endn\in\grp SL(2)$ and $\c_\endn\in\alg sl(2)$. These two variables
(for each end of $\N$) represent the additional physical degrees of freedom
introduced with the boundary fields $\u$ and $\v$. They depend on the special
family of paths chosen in \eref{u-v-soln}, so we always have to fix this family
before introducing $\b_\endn$ and $\c_\endn$. Nevertheless we now have a
complete parametrization of the solutions to the contraint equations, and we
can summerize the result as follows.
\begin{theo}\label{parameterization}
There is a one-to-one map from the set of physical field configurations on $\N$
onto the set of normalized quasiperiodic functions $\cov\N\to\cgrp ISO(1,2)$:
\beq[map]
  \big\{ \oo_i(\x),\gam_i(\x),\u(\phi),\v(\phi) \big\}
       \quad \longleftrightarrow \quad
  \big\{ \tg(\curva) , \f(\curva) , \b_\endn , \c_\endn \big\} .
\eeq
It is explicitly given by
\beq[map-hin]
  \tg(\curva) &=& \tU_\curva , \qquad
  \b_\endn = \UU_{\curva_{\endn,\phi}} \u^{-1}(\phi) , \zl
  \f(\curva) &=& \VV_\curva , \qquad
  \c_\endn = \VV_{\curva_{\endn,\phi}} -
   \UU_{\curva_{\endn,\phi}}
       \u^{-1}(\phi)\,\v(\phi)\,\u(\phi)\,\UU_{\curva_{\endn,\phi}}^{-1} ,
\eeq
where $\curva_{\endn,\phi}$ is a fixed family of boundary paths as defined in
\fref{boundpath}. The right hand sides are independent of $\phi$. The inverse
map reads
\beq[map-her]
  \oo_i(\x) &=& \g^{-1}\del_i \g (\alpha_\x)  ,\phantom{\f\,} \qquad
   \u(\phi) = \b_\endn\inv  \g(\curva_{\endn,\phi}) , \zl
  \gam_i(\x) &=& \g^{-1} \del_i \f \, \g (\alpha_\x) , \qquad
   \v(\phi) =  \b_\endn\inv
           \big(  \f(\curva_{\endn,\phi}) -  \c_\endn  \big)
              \, \b_\endn ,
\eeq
where $\curva_\x\in\cov\N$ is any path with $\proj(\alpha_\x)=\x\in\N$.
\end{theo}
This theorem tells us what the solutions to the constraints are, or how the
physical phase space looks like. It does not tell us whether two solutions are
related by a gauge transformation and represent the same physical state. What
we are looking for is the so called reduced phase space.
\begin{defi}[Reduced Phase Space]\label{reduced}
The set of physically inequivalent states, i.e.\ the set of equivalence classes
of solutions of the constraint equations modulo gauge transformations, is
called the reduced phase space. A physical state is uniquely given as a point
in the this space.
\end{defi}
Hence, the reduced phase space gives a complete classification of the possible
physical states of a system, and what we have to do is to find the equivalence
classes of the sets $\{\tg,\f,\b_\endn,\c_\endn\}$ under Lorentz rotations und
translations, which were given in \eref{trans-fin} and \eref{lorentz-fin}.
Remember that though $\h$ appears in the finite Lorentz rotation, the parameter
is actually an $\cgrp SL(2)$ field $\th$ on $\N$. We can use \eref{U-V-gauge}
and what we get is
\beq[lorentz-pot]
   \tg &\mapsto& \th_0\inv\tg\th, \phantom{{}_0} \qquad
   \b_\endn \mapsto \h_0\inv \b_\endn , \zl
   \f &\mapsto& \h_0\inv \f \h_0, \qquad
   \c_\endn \mapsto \h_0\inv \c_\endn \h_0 ,
\eeq
for the Lorentz rotation and
\beq[trans-pot]
   \f \mapsto \f - \n_0 + \g \n \g^{-1} , \qquad
   \c_\endn \mapsto \c_\endn - \n_0 ,
\eeq
for the translation, where $\th_0=\th(\xo)$ and $\n_0=\n(\xo)$. To simplify the
notation, we can combine both gauge transformations into a single $\cgrp
ISO(1,2)$ transformation
\beq[iso-pot]
  (\tg,\f) &\mapsto& (\th_0,\n_0)^{-1} (\tg,\f) \, (\th,\n) , \zl
  (\b_\endn,\c_\endn) &\mapsto& (\th_0,\n_0)^{-1} (\b_\endn,\c_\endn).
\eeq
Again, we are using the convention that the same symbol denotes a field on $\N$
and the corresponding periodic field on $\cov\N$. Hence, instead of using
fields on $\N$ we can use {\em periodic\/} fields $\th$ and $\n$ on $\cov\N$ as
parameters of finite gauge transformations. We can now proof a second theorem
which gives us part of the classification of gauge invariant parameters of
solutions:
\begin{theo}\label{gauge-theo}
Two configurations $\{\tg,\f,\b_\endn,\c_\endn\}$ and
$\{\tg',\f',\b'_\endn,\c'_\endn\}$ are related by a gauge transformation if
\beq
  \tg_\loopa = \tg'_\loopa , \qquad
  \f_\loopa = \f'_\loopa , \qquad
  \b_\endn = \b'_\endn, \qquad
  \c_\endn = \c'_\endn,
\eeq
for all $\loopa\in\LG$ and all ends $\endn$, i.e.\ if the potentials coincide
on the loop group and the boundary parameters are equal.
\end{theo}
The proof is quite simple. Define $(\th,\n)=(\tg,\f)^{-1}(\tg',\f')$. This is
periodic, because
\beq
  (\th,\n) \circ \shift_\loopa =
   (\tg,\f)^{-1}(\tg_\loopa,\f_\loopa)^{-1}(\tg'\_\loopa,\f'\_\loopa)
  \,(\tg'\f')
   = (\tg,\f)^{-1}(\tg'\f') = (\th,\n),
\eeq
and we also have $(\th_0,\n_0)=(\teins,0)$.
Thus $(\th,\n)$ can be used as a parameter for a gauge transformation
\eref{iso-pot}, which transforms $(\tg,\f)$ into $(\tg',\f')$.

As a result, we find that an equivalence class of solutions is entirely
determined by the discrete set of variables
$\{\tg_\loopa,\f_\loopa,\b_\endn,\c_\endn\}$. We call them the ``holonomies''.
They are still infinitely many, but they are not independent. From the
quasiperiodicity conditions \eref{gf-quasiper} we infer that they have to
provide a representation of the loop group:
\beq[LG-rel]
     (\tg_{\loopa\kette\loopb},\f_{\loopa\kette\loopb})
      =  (\tg_\loopa,\f_\loopa) (\tg_\loopb,\f_\loopb).
\eeq
It is obviously sufficient to specify the values of $\tg_\loopa$ and
$\f_\loopa$ for the {\em primitive\/} loops introduced in
definition~\ref{primloop}, all others are then given by linking them and using
\eref{LG-rel}.
So there are only finitely many parameters left to classify the physical
states, showing that the reduced phase space of three dimensional gravity on a
reasonably well-behaving manifold is finite dimensional. But we still don't
have a full classification of physical states. First of all, even the finite
set of ``primitive holonomies'' might not be independent, as there may be
relations between the primitive loops which have to be respected by the
holonomies.
A standard example is the torus, where the relation
$\loopa \kette \loopb \kette \loopa^{-1} \kette \loopb^{-1} = \loopn$ leads to
the restriction
\beq[torus-rel]
  (\tg_\loopa,\f_\loopa) (\tg_\loopb,\f_\loopb)
  (\tg_\loopa,\f_\loopa)^{-1} (\tg_\loopb,\f_\loopb) ^{-1}
   = (\teins,0)
\eeq
for the primitive holonomies, which becomes rather complicated when written out
in components. More complicated relations hold for higher genus manifolds, and
it is getting even difficult to find the range of the primitive holonomies.

In addition, different holonomies not necessarily correspond to different
physical states, as there are still gauge transformation acting on them. They
are those gauge transformation which did not occur in the proof of
theorem~\ref{gauge-theo}, i.e.\ for which the parameters $\h_0$ and $\n_0$ at
$\xo$ do not vanish. Evaluating \eref{iso-pot} for the holonomies gives
\beq[iso-gauge]
  (\tg_\loopa,\f_\loopa) &\mapsto&
  (\h_0,\n_0 )^{-1} (\tg_\loopa,\f_\loopa) (\h_0,\n_0), \zl
  (\b_\endn,\c_\endn) &\mapsto& (\h_0,\n_0)^{-1} (\b_\endn,\c_\endn).
\eeq
Hence, for the holonomies corresponding to the loops the gauge transformations
are inner automorphisms of $\cgrp ISO(1,2)$. Form definition~\ref{auto} we know
that we can use an $\grp ISO(1,2)$ paramter $(\h_0,\n_0)$ instead of the $\cgrp
ISO(1,2)$ parameter $(\th_0,\n_0)$ appearing in \eref{iso-pot}, as inner
automorphisms of the covering group are completely parameterized by elements of
$\grp ISO(1,2)$.

The boundary holonomies transform by multiplication from the left only, and
therefore it is quite easy to find the equivalence classes under gauge
transformations, provided that we have at least one end. Calling it
$\endn=\endnull$, then the quantities
\beq[mod-auto]
   (\b_{\endnull,\endn},\c_{\endnull,\endn}) &=&
  (\b_\endnull,\c_\endnull)^{-1} (\b_\endn,\c_\endn)
        \txt{for} \endn \ne \endnull , \zl
 (\tg_{\endnull,\endn},\f_{\endnull,\endn}) &=&
  (\b_\endnull,\c_\endnull)^{-1} (\tg_\loopa,\f_\loopa) \,
    (\b_\endnull,\c_\endnull)
        \txt{for} \loopa \in \LG_0 ,
\eeq
are gauge invariant. Note that again we are multiplying $\grp SL(2)$ elements
with $\cgrp SL(2)$ elements in the second line, but $\b_\endnull$ acts as an
inner automorphism so that the operation is well defined. These variables
provide a full classification of the classical solution: by
exercise~\ref{loop-free} we know that the loop group of a manifold with
boundary is free, so the reduced phase space is just a direct sum of ``number
of primitive loops'' times $\cgrp ISO(1,2)$ plus ``number of ends minus one''
times $\grp ISO(1,2)$.
If there are no ``ends'' of the space manifold $\N$, we cannot use the boundary
holonomies to obtain gauge invariant quantities. In this case, one really has
to find the equivalence classes of sets of $\cgrp ISO(2)$ elements (obeying
extra relations) modulo inner automorphism.

Finally, let us consider the time evolution of the holonomies. As we know how
they transform under gauge symmetries, we only have to replace the parameters
by the time components of the gauge fields $\oo_t$ and $\gam_t$ to get the time
evolution. We find
\beq[mod-time-evol]
  \dot \g_\loopa &=& \comm{\g_\loopa}{\oo_t(\xo)} ,
   \phantom{ {}- \gam_t(\xo)
                  + \g_\loopa \, \gam_t (\xo) \, \g_\loopa\inv} \qquad
 \dot \b_\endn = - \oo_t(\xo) \, \b_\endn , \zl
  \dot \f_\loopa &=& \comm{\f_\loopa}{\oo_t(\xo)} - \gam_t(\xo)
                  + \g_\loopa \, \gam_t (\xo) \, \g_\loopa\inv, \qquad
  \dot \c_\endn = \comm{\c_\endn}{\oo_t (\xo)} - \gam_t(\xo).
\eeq
We see that they only depend on the value of the multipliers at $\xo$, because
we eliminated all gauge degrees of freedom except those at this special point.
In the first equation we gave the time evolution for $\g_\loopa$ instead of
$\tg_\loopa$, which cannot be written as a simple commutator, as it is not a
matrix. However, the time evolution acts as a Lorentz rotation only, so it is
obvious how the time derivative of $\tg_\loopa$ looks like.
We can also integrate the time evolution by using the ``evolution operators''
\eref{U.V.time}. We then get formally the same as \eref{iso-gauge} for the
gauge transformations, and here we can insert $\tg_\loopa$ again:
\beq[time-finite]
  (\tg_\loopa,\f_\loopa)(t_2) &=&
  (\UU_\xo,\VV_\xo)(t_2,t_1) \  (\tg_\loopa,\f_\loopa)(t_1)  \
     (\UU_\xo,\VV_\xo)(t_1,t_2), \zl
  (\b_\endn,\c_\endn)(t_2) &=&(\UU_\xo,\VV_\xo)(t_2,t_1)\
  (\b_\endn,\c_\endn)(t_1).
\eeq
For free falling observers the time derivatives become very simple, as
$\oo_t=0$ and $\gam_t=\gam_0$:
\beq[free-evol]
   \dot \g_\loopa = \dot \b_\endn=0 , \qquad
  \dot \f_\loopa = \g_\loopa \gam_0 \,\g_\loopa\inv - \gam_0 , \qquad
  \dot \c_\endn = - \gam_0 .
\eeq

\subsection{Cylinder spacetimes}
\label{class-sol}
Before going over to quantum physics, in this last ``classical'' section we
will use the formalism derived above to classify the solutions of Einstein's
equations on a given manifold. We will not need the results of this section
later on. All we want to do here is to construct some interesting spacetimes,
showing that there is a huge variety of inequivalent solutions to Einstein's
equations even on the simplest possible manifold. Explicit solutions to
Einstein's equations in three dimensions were classified in
\cite{deser.jackiw.hooft:84}, and of course the solutions presented here are
closely related those given there. For a similar approach using dreibein
formalism see also~\cite{newbury.unruh:93,unruh:94} for torus and higher genus
manifolds, and recently the Klein bottle has been studied~\cite{louko:95}, too.
Our picture of spacetime as a set of observers evolving in time is a very
useful tool to visualize its structure and the various singularities. We take
the set of local observers to be the cylinder $\N=\RR\times S^1$. For
simplicity, we drop all the boundary terms here. The boundary terms were
constructed such that their existence does not affect on the gauge fields
$\oo_i$ and $\gam_i$. They are easily restored by choosing arbitrary values for
$\c_\endn$ and $\b_\endn$ and using \eref{map-her}. We will discuss their
physical interpretation in detail in the quantum section.

The coordinates on $\N$ are $r,\phi$ with $\phi\equiv\phi+2\pi$, and we have
the same coordinates without periodicity relation on $\cov\N\simeq\RR^2$. For
simplicity, we choose $\xo=(0,0)$, so that $(r,\phi)\in\cov\N$ is represented
by the path $s\mapsto(sr,s\phi)\in\N$. All the fields have trivial limits for
$r\to\pm\infty$, so there is no need to introduce the $\rho$ coordinate from
\sref{boundaries}. Nevertheless we need the  $\gam_\infty$ field which relates
the finite fields $\gam_\mu$ to the physical fields $\widehat\gam_\mu$ (see
\eref{gamhut}), in order to get the correct behaviour of the physical fields at
infinity. We simply choose $\gam_\infty=r\gam_1$, which obviously diverges to
spacelike infinity $\Tr(\gam_\infty\gam_\infty)\to+\infty$ for $r\to\pm\infty$.
In addition, we ``gauge fix'' all our observers to be free falling, i.e.\ we
always have $\oo_t=0$ and $\widehat\gam_t=\gam_t=\gam_0$.

With the results of the previous section, we know that the solutions are
parameterized by a single pair $(\tg_\loopa,\f_\loopa)$, where
$\loopa=(0,2\pi)\in\cov\N$ is the primitive loop of the cylinder. Two such
pairs represent the same physical state, if they are related by an inner
automorphism, which explicitly reads
\beq[equiv-par]
  \tg_\loopa \mapsto \h_0\inv \tg_\loopa \, \h_0, \qquad
  \f_\loopa \mapsto \h_0\inv \big( \f_\loopa + \g_\loopa \n_0 \g_\loopa\inv
           - \n_0 \big) \, \h_0.
\eeq
Using the classification of $\cgrp SL(2)$ and $\alg sl(2)$ elements, we can
split the states into three classes, depending on the value of the parameters:
if $\tg_\loopa$ is non-null, then the state is called timelike, lightlike or
spacelike depending on the value of $\tg_\loopa$, otherwise depending on the
value of $\f_\loopa$. Note that if $\tg_\loopa$ is null, then
$\g_\loopa=\pm\eins$, and $\f_\loopa$ does not transform under translations, so
that the squared length of $\f_\loopa$ is gauge invariant. The special cases
$\tg_\loopa=\pm\teins_z,\f_\loopa=0$ are included in the timelike states. An
important property of translations can be used to simplify the complete
classification of inequivalent states:
\begin{exer}\label{ortho}
Assume that $\tg_\loopa$ is non-null, i.e.\ $\g_\loopa=\g(v^a,\pm)$ for some
vector $v^a\ne0$ in \fref{sl2}. Then for any vector (or matrix) $w^a\gam_a=\w$
orthogonal to $v^a$ ($v^aw_a=0$ or $\Tr(\g_\loopa\w)=0$), and only for those,
we can find a translation that adds $\w$ to $\f_\loopa$.
\end{exer}
Note that the vector added to $\f_\loopa$ by a translation is the difference
between $\n_0$ and the Lorentz transformed ``vector''
$\g_\loopa\n_0\g_\loopa\inv$.
What the exercise says is that every vector that is orthogonal to the ``axis''
of this Lorentz rotation can be written as such a difference. This is quite
obvious in Euclidean three dimensional space, but holds in Minkowski space as
well.

Let us start with the timelike states: then either $\tg_\loopa$ is non-null and
lies in the timelike regions of \fref{sl2}, or $\f_\loopa$ is timelike. It is
obvious, that there is always a Lorentz rotation that maps a timelike vector
onto the vertical axis, so in the first case we can assume that
$\tg_\loopa\in\cgrp SO(2)$ (which is the vertical axis in \fref{sl2}), in the
second case we can assume that $\f_\loopa$ lies on the $\gam_0$ axis. If
$\tg_\loopa$ is non-null, then we still have the freedom of making a
translation acting on $\f_\loopa$. From exercise~\ref{ortho} we know that we
can add any matrix lying in the $\gam_{1,2}$ plane to $\f_\loopa$, so that
finally both parameters lie on the $\gam_0$ axis:
\beq
   \tg_\loopa = \exp ( \defangle \pi \,\tgam_0 ) , \qquad
     \f_\loopa = 2 \pi \scher \, \gam_0 .
\eeq
Remember that the exponential of the $\tgam$ ``matrices'' was introduced as a
formal notation for the exponential map $\alg sl(2)\to\cgrp SL(2)$ in
\sref{lorentz-group}. The timelike states are therefore completely
parameterized by two real numbers $\defangle,\scher$. There is no time
evolution of the holonomies, because the right hand sides of \eref{free-evol}
vanish. Hence, timelike states are static.
To find a field configuration $\oo_i,\gam_i$ leading to these holonomies, we
first have to find a pair of quasiperiodic fields $(\tg,\f)$ on $\cov\N$. This
is not difficult, we take
\beq[tg-timelike]
  \tg(r,\phi) = \exp ( \ft12 \defangle \phi \, \tgam_0 ) , \qquad
  \f(r,\phi) = \scher \phi \, \gam_0
\eeq
Using \eref{map-her} we compute the gauge fields
\beq
  \oo_\phi = \ft12 \defangle \, \gam_0 , \quad \oo_r=0, \qquad
  \gam_\phi = \scher \, \gam_0 , \quad \gam_r=0 .
\eeq
Transforming to the physical dreibein field by \eref{gamhut} gives
\beq
  \widehat\gam_t = \gam_0 , \qquad
  \widehat\gam_r = \gam_1 , \qquad
  \widehat\gam_\phi = \defangle r \, \gam_2 + \scher \, \gam_0
\eeq
Obviously, for $\scher=0$ we get the ``cones'' considered in \sref{cylinder}.
There we found that the ``central cylinder'' $(t,r=0,\phi)$ actually represents
a {\em line} in spacetime only (the worldline of the peak of the two cones
glued together at $r=0$), as points with different $\phi$ but the same $t$ can
be connected by a path of zero distance $\d\X^a=0$ everywhere. For $\scher\ne0$
this singularity is different. Writing the distance as an $\alg sl(2)$ matrix
we get
\beq
  \d\X^a \, \gam_a  =  \big( \d t + \scher \, \d \phi \big) \, \gam_0
   + \d r \, \gam_1  + \defangle r \, \d \phi \, \gam_2 .
\eeq
This vanishes if $r=0$, $\d r=0$ and $\d t=-\scher\d\phi$. Hence, there is
still a singularity at $r=0$, but this time the events to be identified lie on
a helix winding around the central cylinder, rather than being the circles with
constant $t$. The path specified by $t+\scher\phi=t_0,r=0$ represent one
spacetime event only. But $t_0$ and $t_0+2\pi/\scher$ yield the same path, so
the set of these lines, the ``singularity'', forms a circle. However, it is
still timelike, and we can measure its geodesic distance, which is
$2\pi\scher\gam_0$. Hence, the singularity is a timelike circle of ``radius''
$|\scher|$, and the sign of $\scher$ tells us in which direction time runs
around the circle.

As the singularity is a timelike closed path, one can expect that there are
such paths outside the singularity as well. In fact, the circles at constant
$r$ and $t$ become timelike for $r<\scher/\defangle$. For large $r$, the
$\defangle r$ term in $\d\X^a$ dominates the $\d\phi$ part, so far away from
the singularity the universe looks like a cone. The metric itself is obtained
by taking the square of the ``distance'' above:
\beq
  \d s^2 =  - ( \d t + \scher \, \d \phi )^2   + \d r^2
          + (\defangle r)^2 \, \d \phi^2 .
\eeq
Outside $r=0$, this describes the gravitational field of a spinning particle at
the origin. Approaching the particle, the lightcones become more and more
tilted into the angular direction in which the particles rotates. There is also
a corresponding four dimensional spacetime (and this also holds for all the
examples below), which can be obtained by adding $\d z^2$ to the metric, where
$z$ is an additional global coordinate. The timelike states then represent
spinning cosmic strings.

The spacelike case is somehow dual to the timelike, and can be obtained by
exchanging the $r$ and $t$ direction, and $\gam_1$ with $\gam_0$. However,
there are additional spacelike states leading to very peculiar spacetimes,
which have no timelike counterpart. This is due to the fact that timelike
elements of $\cgrp SL(2)$ can always be written as exponentials whereas
spacelike cannot.
However, by the same arguments as above we can always perform a gauge
transformation such that $\tg_\loopa$ and $\f_\loopa$ lie on the $\gam_1$ axis.
 Use the coordinates \eref{alg-csl2}, the parameters become two real numbers
and an additional integer $z\in\ZZ$:
\beq[tg-spacelike]
  \tg_\loopa = \teins_z \, \exp ( \defangle \pi \, \tgam_1 ) , \qquad
  \f_\loopa = 2 \pi \scher \, \gam_1 .
\eeq
These holonomies are not time-independent: $\tg_\loopa$ is still constant by
\eref{free-evol}, but $\f_\loopa$ has a constant velocity, so that
\beq[f-spacelike]
  \f_\loopa(t) = 2 \pi \scher \, \gam_1 + t \,
   ( \g_\loopa \gam_0 \g_\loopa\inv - \gam_0 )  .
\eeq
We see that the state is no longer static, as it was in the timelike case. We
can calculate the time dependence of $\f_\loopa$ explicitly using the formula
\beq[sincos]
  \exp ( - \ft12 a \,\gam_1 ) \, \gam_0 \, \exp ( \ft12 a \,\gam_1 )
    = \cosh a \,\gam_0 + \sinh a \, \gam_2.
\eeq
This gives
\beq
  \f_\loopa(t) = 2 \pi \scher \, \gam_1 +
          t \,  ( \cosh (2 \pi \defangle) - 1 ) \, \gam_0
       +  t \sinh (2 \pi \defangle) \, \gam_2.
\eeq
Remember that $\f_\loopa$ measures the geodesic distance of the circumference
of the cylinder at the base point $\xo$. We can see that for $\defangle\ne0$ it
grows linearly with $t$ and calculating its length squared, i.e
$\ft12\Tr(\f_\loopa\f_\loopa)$, shows that it is always spacelike (except for
$\defangle=\scher=0$). So this time we expect to get a non-static solution
consisting of a growing spacelike cylinder.
To find quasiperiodic potentials with the given holonomies is not so easy: for
$z\ne0$, $\tg_\loopa$ is not an exponential of any element of the algebra, so
there is no analogy of \eref{tg-timelike}. Let us consider the case $z=0$
separately first. Then we can define
\beq[gf-spacelike]
  \tg(r,\phi) &=& \exp ( \pi \defangle \, \tgam_1 ) , \zl
  \f(r,\phi)&= &\scher \phi \, \gam_1  +
   t \, ( \g(r,\phi) \, \gam_0 \, \g^{-1}(r,\phi) - \gam_0 ).
\eeq
The field $\f(\phi,r)$ has become time-dependent as well. The computation is
now straightforward and similar to the timelike case above. The result is that
the physical fields are
\beq
 \oo_\phi= \ft12 \defangle \, \gam_1 , \qquad
 \widehat\gam_t = \gam_0 , \qquad
 \widehat \gam_r = \gam_1 , \qquad
  \widehat \gam_\phi = \scher \, \gam_1 + \defangle t \, \gam_2 .
\eeq
Here, the singularity is obviously at $t=0$, so we have some kind of big bang
universe. After that, there is an expanding cylinder with growing radius
$\defangle t$ and ``offset'' $2\pi\scher$, i.e.\ if you go once around the
cylinder orthogonal to the $r$-direction, you end up $2\pi\scher$ away from
where you started. If $\scher=0$, the cylinder degenerates to a line at the big
bang, otherwise we find that the singularity is a spacelike circle of radius
$|\scher|$ similar to that for the timelike case above. For $\scher\ne0$ it is
actually the length of the cylinder that shrinks to zero at $t=0$ and not the
radius, so the singularity is in fact the circumference.
\begin{exer}
Verify this by finding a coordinate transformation such that the new $r$ and
$\phi$ directions become orthogonal. Then you have $\gam_r\to0$ for
$\scher\ne0$, but $\gam_\phi\to0$ for $\scher=0$ at the big bang.
\end{exer}
As there is no mixing of space and time, i.e.\ the space $\N$ is always
orthogonal to the time direction $t$, it is quite easy to ``see'' how the
spacelike states look like, using the ADM picture of a space evolving in time.
This also works for $z\ne0$. The simplest and most illustrative way to
construct these spacetimes is to cut the cylinder into two strips at $\phi=0$
and $\phi=\pi$. As a consequence, the two parts behave quite differently, and
spacetime is no longer symmetric under rotation of the $\phi$ coordinate. But
anyway we should not expect that spacetime has such a symmetry in general.
As $\tg_\loopa$ is not an exponential for $z\ne0$, the idea is to write it as a
product of two exponential $\teins_z=\exp(\pi z\tgam_0)$ and
$\exp(\pi\defangle\tgam_1)$, such that the quasiperiodic field $\tg$ increases
from $\teins$ to $\teins_z$ on the first strip and to $\tg_\loopa$ on the
second. The resulting field looks a bit complicated, as we have to define it on
each strip of $\cov\N$ (labeled by an integer $n$) separately:
\beq
  \tg(r,\phi) =
  \cases{ \exp\big( n \pi \defangle \, \tgam_1 \big)
          \exp\big( (\phi-n \pi) z \, \tgam_0 \big)
              & for $2\pi n \le\phi\le2\pi n + \pi $, \cr
          \exp\big( (\phi- n \pi) \defangle \, \tgam_1 \big)
          \exp\big( n \pi z \, \tgam_0 \big)
              & for $2\pi n -\pi \le\phi\le2\pi n  $. }
\eeq
For $\f_\loopa$ we could in principle insert the same expression as in
\eref{gf-spacelike}, but the metric becomes simpler if we take $\f$ to be
constant on the first strip and increasing only on the second one, but of
course  twice as fast:
\beq
  \f(r,\phi) =  t \, ( \g \, \gam_0 \, \g^{-1} - \gam_0 )
 + \cases{2n \pi \scher \, \gam_1
              & for $2\pi n \le\phi\le2\pi n + \pi $, \cr
          2(\phi-n \pi ) \scher \, \gam_1
              & for $2\pi n -\pi \le\phi\le2\pi n  $. }
\eeq
\begin{exer}
Show that $\tg$ and $\f$ are continuous, and that they are normalized
quasiperiodic with parameters $\tg_\loopa$ and $\f_\loopa$ given by
\eref{tg-spacelike} and \eref{f-spacelike}.
\end{exer}
The $\phi$ components of the gauge fields become
\beq
  \oo_\phi = \cases{ z \, \gam_0, &  \cr
                     \defangle \, \gam_1, & }
  \quad
  \gam_\phi =\cases{0 & for $0< \phi< \pi $, \cr
          2\scher\, \gam_1 - 2 \defangle t \, \gam_2
                       & for $\pi <\phi< 2\pi$. }
\eeq
They are of course periodic, but not continuous because the potentials are not
differentiable at $\phi=n\pi$. However, in principle $\tg$ and $\f$ can be
deformed slightly in the neighbourhood without destroying the quasiperiodicity,
so that the gauge fields become continuous. The components of the fields
tangent to the ``border'' $\phi=n\pi$, i.e.\ the $r$ components, are
continuous, so all the non-local objects like transport operators and lengths
of path are well defined anyway. The physical fields are
\beq
  \widehat\gam_t = \gam_0 , \qquad
 \widehat\gam_r = \gam_1 , \qquad
  \widehat\gam_\phi =\cases{2zr \, \gam_2 & for $0<\phi<\pi $, \cr
          2\scher\, \gam_1 - 2 \defangle t \, \gam_2
                       & for $\pi< \phi< 2\pi$. }
\eeq
To read off the geometry, it is again more convenient to write down the
infinitesimal distance. This contains the same information as the metric, which
is just the sum of the squares of the factors in from of the gamma matrices
(with a minus sign for $\gam_0$).
\beq
  \d\X^a \, \gam_a =
  \cases{ \d t \, \gam_0 + \d r \, \gam_1 + 2 z r \, \d \phi \, \gam_2
   & for $0<\phi<\pi $, \cr
         \d t \, \gam_0 + (\d r + 2 \scher \, \d \phi) \gam_1
               - 2 \defangle t \, \d \phi \, \gam_2
  & for $\pi<\phi<2\pi $ .}
\eeq
Now it is not very difficult to see how this space looks like. For $z=\pm1$,
the first part consists of two copies of flat Minkowski space in polar
coordinates glued together in the origin as described in \sref{cylinder}, but
there is a cut along a radial line, and $\phi$ runs from $0$ to $\pi$ only
(therefore the factor $2$ in $\gam_\phi$). For larger $z$, instead of a simple
Minkowski space, we get a screw-shaped space winding around the origin $z$
times when $\phi$ increases from $0$ to $\pi$. The second part consists of a
parallelogram shaped strip of width $2\pi\defangle t$ and ``offset''
$2\pi\scher$, which is glued to the first part along the cuts. The width of the
strip grows linearly with $t$ and vanishes for $t=0$.
\begin{exer}
Find the equivalence classes of observers connected by paths with vanishing
distance, which by definition represent a single physical point in space only.
Show that for $\scher\ne0$ there is no essential singularity, i.e.\ all
nontrivial such equivalence classes of points in $\N$ are contractible objects.
This was not so for the spacetimes considered before, where the paths with
vanishing distance  were circles or extended to infinity. Here you can also
pass from one ``end'' to the other without going through the singularity, by
passing over the strip at $r<\pi\scher$. So actually we have a wormhole
spacetime rather that a conical singularity.
If $\scher=0$, but $\defangle\ne0$, then there is only one essential singular
event at $t=0,r=0,\phi$ arbitrary, which forms a circle of observers and cannot
be contracted.
\end{exer}
So the spacelike states are already rather strange. You can try to find other
coordinates that make them ``smoother'', avoiding the rather artificial
piecewise construction. Another possibility for the quasiperiodic field $\tg$
is to use the global coordinates $(w,x,y)$ on $\cgrp SL(2)$ introduced in
exercise~\ref{smooth-deform}. This leads to a smooth but very awkward looking
metric including many trigonometric and hyperbolic functions, and it is not
easy to visualize the resulting spacetime.

What remains are the lightlike states. They are again completely different. As
in the spacelike case there is an additional integer $z$ labeling the part of
$\cgrp SL(2)$ in which the holonomy $\tg_\loopa$ lies. We only consider the
case $z=0$ here. For non-vanishing $z$ you can again cut the cylinder into
pieces exactly as we did in the spacelike case, with an even more peculiar
result.
Let us also assume that $\tg_\loopa$ lies on the upper lightcone, otherwise we
get the same type of solutions with $\phi\mapsto-\phi$. By a suitable Lorentz
rotation, we can map any point on the lightcone onto any other, so we may take
\beq
  \tg_\loopa = \exp \big( \pi ( \tgam_0 + \tgam_1)\big).
\eeq
The corresponding quasiperiodic field is
\beq
  \tg(r,\phi) = \exp \big( \ft12 \phi (\tgam_0 + \tgam_1 )\big)
  \follows
  \g(r,\phi) = \eins + \ft12 \phi \,(\gam_0 +\gam_1 ),
\eeq
where the last equation follows because the matrix $\gam_0+\gam_1$ is
nilpotent. For $\f_\loopa$ we make the general ansatz
\beq
  \f_\loopa = 2 \pi \scher\, (\gam_0 - \gam_1 )
             + 2 \pi a\, (\gam_0 + \gam_1 ) + 2 \pi b \, \gam_2 ,
\eeq
with real numbers $\scher,a,b$. By exercise~\ref{ortho} we know that we can add
any matrix to $\f_\loopa$ which is orthogonal to $\gam_0+\gam_1$. This time,
however, we can not arrange $\f_\loopa$ to be proportional to this matrix,
because it is orthogonal to itself. Instead, we can add anything to $a$ and
$b$, which also become time dependent, but we cannot change $\scher$, which is
the only gauge invariant parameter for the lightlike states. The time evolution
for $\f_\loopa$ is
\beq
   \dot\f_\loopa = \g_\loopa \gam_0 \g_\loopa\inv - \gam_0 =
         2 \pi^2 \, (\gam_0+\gam_1) - 2 \pi \, \gam_2.
\eeq
Hence, we have $\dot a = \pi$ and $\dot b= -1$. We will adjust $a$ and $b$ such
that the resulting metric becomes as simple as possible.
Next, we have to find a normalized quasiperiodic field $\f$, and we make an
ansatz similar to $\f_\loopa$:
\beq
  \f(r,\phi) = u(\phi) \, (\gam_0 + \gam_1 )
             + v(\phi)\, (\gam_0 - \gam_1 ) + w(\phi) \, \gam_2 .
\eeq
The equation $\f(r,\phi+2\pi)=\g_\loopa \f(r,\phi) \g_\loopa\inv + \f_\loopa$
now becomes a functional equation for $u$, $v$ and $w$, depending on $a$ and
$b$, and the best thing one can do is to play around with these parameters
until you get a simple final result (which is rather cumbersome, unless you use
an algebraic computer program). It turns out that a good choice is
\beq
  a = (\ft43 \pi^2 -1)\, \scher + \pi \, t , \qquad
  b = - 2 \pi \scher - t .
\eeq
Solving the functional equations leads to
\beq
  u(\phi) = \ft13 \scher \phi^3 + \ft12 t \phi^2 - \scher \phi, \quad
  v(\phi) = \scher \phi, \quad
  w(\phi) = - \scher \phi^2 - t \phi .
\eeq
and a straightforward calculation yields the following gauge fields:
\beq
  \oo_\phi = \ft12 (\gam_0 + \gam_1), \qquad
  \gam_\phi = - 2 \scher \, \gam_1 - t \, \gam_2 ,
\eeq
Finally, the physical fields are
\beq
  \widehat \gam_t = \gam_0 , \qquad
  \widehat \gam_r = \gam_1 , \qquad
  \widehat \gam_\phi = - 2 \scher \gam_1 + (r-t) \, \gam_2 .
\eeq
Again, the time direction is orthogonal to the space $\N$, and the
infinitesimal distance is
\beq
 \d\X^a \gam_a = \d t \, \gam_0 +
       (\d r - 2 \scher \d\phi) \, \gam_1 + (r-t) \d \phi \, \gam_2.
\eeq
The best way to describe this spacetime is to call it a cosmic hoover.
For $\scher=0$ and fixed $t$ we simply have the two copies of flat space known
from \sref{cylinder}, with a singular point at the origin, which is now at
$r=t$. At this point the two copies are glued together, i.e.\ the circle $r=t$,
$\phi$ arbitrary, represents one spacetime event only. But now this structure
is not static, as the singularity moves with the velocity of light in the
direction of $r$. It is not simply a singularity moving around somewhere in the
space $\N$. The situation is much worse: the positive $r$ part is totally
absorbed into the singularity with the velocity of light, and the observers
disappearing into the singularity (where they meet all the observers with the
same $r$ coordinate) reappear on the other side.

For $\scher\ne0$ there is no essential singularity any more. Spacetime becomes
totally regular as in the spacelike case, and observers never meet at a single
event. This can be seen as follows. If there is any path connecting different
observers with vanishing distance, we must have $\d t=0$, $\d r= 2 \scher
\d\phi$, and $(r-t)\d \phi=0$ everywhere of this path, which is impossible.
Hence, we are only using bad coordinates at $r=t$. Instead of the singularity
we get a wormhole connecting the two regions for positive and negative $r$, but
still the vacuum cleaner is active, as the wormhole absorbs everything with the
velocity of light.

\section{Canonical Quantization}
\label{quantization}
The aim of this section is to carry out Dirac's canonical quantization method
\cite{dirac:65} as proposed by Witten~\cite{witten:88}, starting from the
classical phase space spanned by the spatial components of the spin connection
and the dreibein, as well as the boundary fields. A quantum state is given by a
``wave functional'', which depends on one half of the phase space coordinates,
spanning the ``configuration space''. These configuration variables are
represented by multiplication operators, i.e.\ we have to choose them such that
they commute, and the conjugate momentum operators become functional derivative
operators.

The physical state space is then defined to be the kernel of the constraint
operators: acting with the constraints on physical states gives zero.
For this step it is essential that the commutator algebra of the quantized
constraints closes, i.e.\ the quantum constraints have to be first class. This
does not necessarily follow from the fact that the Poisson brackets of the
classical constraints are proportional to the constraints again. There are
mainly to problems arising with Dirac quantization at this point. The first is
that straightforward translation of the constraints may lead to ill-defined
operator products: there might appear functional differential operators acting
at the same space point. This would require a regularization of the quantum
operators. However, the great success of dreibein formulation of three
dimensional gravity is mainly based on the fact the this doesn't happen, i.e.\
the constraints will become perfectly well defined operators without any
regularization. From a physical point of view, the reason therefor is that the
theory is not really a field theory with local degrees of freedom, whose
quantization normally requires a regularization. In principle, we are dealing
with ``quantum mechanics'', our system has finitely many degrees of freedom and
we could equally well use our classical results and quantize the finite set of
holonomies. This is the so called reduced phase space method
\cite{carlip.nelson:94b}. However, constructing the physical phase space with
Dirac's method will turn out to be even simpler than classifying the classical
degrees of freedom explicitly.

After the constraints have been translated into their quantized versions, a
second problem may arise, namely the operator ordering. The classical
constraint algebra tells us that the commutator of the constraints is
proportional to themselves again, but it does not tell us whether the operators
in the commutator are in the same order as before. In general, there may be
different choices for the operator ordering leading to different
representations for the constraints. There is no reason why one of these
orderings should lead to an algebra such that taking commutators will preserve
this ordering. In addition, the structure constants appearing in the constraint
algebra may itself be phase space functions. Then the quantum algebra must be
such that these ``structure functions'' appear to the left of the constraints.
Otherwise the definition of the physical states would not be consistent, as
acting with a commutator of constraints on a physical state gives zero. But the
constraints are the {\em only\/} operators that by definition annihilate
physical states. If it is not possible to arrange the operator ordering in this
way, Dirac's method does not work and one has to choose other quantization
schemes. Again, three dimensional gravity in dreibein formulation does not
cause any problems here. We already saw that the structure constants of the
classical algebra of the constraints~\eref{con-alg} are not field-dependent. We
will also find that there is no operator ordering ambiguity, as all products of
phase space variables appearing in the constraints are such that the factors
commute.

Hence, all the typical problems occuring in field theory are absent, though we
are still dealing with an infinite dimensional phase space at the beginning.
The reduction to finitely many degrees of freedom will automatically come out
when we solve the constraint equations. Thereby, we will obtain the physical
states as wave {\em functions} depending on a finite set of variables like the
wave functions on ordinary quantum mechanics. The final step to get the Hilbert
space is then the construction of the scalar product. It will be determined by
requiring the operators corresponding to real classical quantities to become
hermitian.

\subsection{Quantum operators and constraints}
The first step in Dirac's programme is to define an operator representation for
the phase space variables, i.e.\  the spatial components of the gauge fields
$\oo_i,\gam_i$ and the boundary fields $\u,\v$. The most reasonable choice is
the ``connection representation'', where the wave functional depends on $\oo_i$
and $\u$. This is because we cannot choose $\v$ to be a multiplication
operator, as its components do not commute, and with $\oo_i$ as a differential
operator the Hamilton constraint would not be well defined, as it contains a
quadratic term in $\oo_i$. Hence, the wave functional is given by
$\Psi[\oo_i,\u]$.

As the functional depends on Lie group and algebra valued variables, we
introduce some notation to deal with derivatives with respect to such
variables.  Consider a function depending on an algebra valued variable
$\v=v^a\gam_a$. We can then define the derivative of this function in the
direction of $\gam^a$, which is simply given by
\beq
  \Trrr{ \gam^a \deldel / \v / }=  \deldel / v^a / .
\eeq
Of course, we can insert any other matrix $\a\in\alg sl(2)$ instead of the
gamma matrices. If the function is given as a power series in $\v$, then we can
calculate the derivative most easily using
\beq[alg-abl]
  \Trrr{ \a \deldel /\v/ } \, \v = \a .
\eeq
This is of course quite simple because the algebra is a vector space and we can
take derivatives with respect to the coordinates. We get the same result if we
define a matrix valued derivative operator
\beq[mat-abl]
    \Big( \deldel /\v/ \Big)_{pq} =
         \deldel /\v_{qp} / ,
\eeq
which is a derivative with respect to the matrix components. Taking the trace
of this operator with some algebra element $\a$ reproduces~\eref{alg-abl}.
However, the derivative with respect to matrix components is only defined if
the support of the function it acts on is (an open subset of) $\RR^{2\times2}$.
If we only use expressions like~\eref{alg-abl}, where $\a$ is a matrix which is
``tangent'' to the algebra (as a subset of $\RR^{2\times2}$), then we can
extend the function to be differentiated arbitrarily to $\RR^{2\times2}$ before
using the derivative~\eref{mat-abl}. The result will be independent of this
extension, as we only take the components tangent to $\alg
sl(2)\subset\RR^{2\times2}$.

Now consider a function depending on $\u\in\grp SL(2)$. We can extend it to
some open subset of $\RR^{2\times2}$ containing $\grp SL(2)$ and apply the
matrix valued differential operator~\eref{mat-abl}, with $\v$ replaced by $\u$.
We are then only allowed to take those components of the derivative which are
tangent to the group, i.e.\ we have to ``trace'' the result with $\u\a$ or
$\a\u$ for some $\a\in\alg sl(2)$. These are the matrices tangent to $\grp
SL(2)$ at $\u$. Hence, there are two complimentary sets of derivative
operators
\beq[grp-abl]
  \Trrr{ \u \a \deldel / \u / }, \qquad
  \Trrr{ \a \u \deldel / \u / } , \qquad
      \a \in \alg sl(2).
\eeq
\begin{exer}\label{tot-diff}
Show that~\eref{grp-abl} are the left and right invariant vector fields on the
Lie group, respectively, and that they can be defined on $\cgrp SL(2)$ as well,
obeying
\beq[cgrp-abl]
  \Trrr{ \tg \a \deldel / \tg / } \, \g = \g \a , \qquad
  \Trrr{ \a \tg \deldel / \tg / } \, \g = \a \g , \qquad
      \a \in \alg sl(2).
\eeq
Note that here $\g$ is a treated as a matrix valued function of $\tg$, and the
trace is a formal notation only. Moreover, the total differential of a function
$F(\tg)$ on $\cgrp SL(2)$ is given by
\beq[tot-dif]
   \d F =
  \ft12\Trr{ \g^{-1} \d \g \, \gam_a} \Trrr{ \tg \gam^a \deldel F/\tg/ }  =
  \ft12\Trr{ \d \g \, \g^{-1} \gam_a} \Trrr{ \gam^a \tg \deldel F/\tg/ } .
\eeq
Here the first traces are ``real'' traces of matrices, with $\d\g$ the
differential of the special function $\tg\mapsto\g$, whereas the right hand
factors are the formal expressions~\eref{cgrp-abl}.
\end{exer}
The derivative operators on the group do not commute, but they provide two
representation of the Lie algebra, commuting with each other:
\beq
  \Big[ \Trrr{ \gam^a \tg \deldel / \tg /  } ,
        \Trrr{ \gam^b \tg \deldel / \tg /  } \Big]  &=&
            2 \eps^{ab}\_c \Trrr{  \gam^c \tg \deldel / \tg / } \zl
  \Big[ \Trrr{ \tg \gam^a \deldel / \tg /  } ,
        \Trrr{ \tg \gam^b \deldel / \tg /  } \Big]  &= &
           - 2 \eps^{ab}\_c \Trrr{ \tg \gam^c \deldel / \tg / } \zl
  \Big[ \Trrr{ \gam^a \tg \deldel / \tg /  } ,
        \Trrr{ \tg \gam^b \deldel / \tg /  } \Big]  &= &0.
\eeq
The translation from classical Poisson brackets to operators is defined such
that the latter is $-\i\hbar$ times the classical bracket. From~\eref{o-g-pois}
we get the operator representation for the dreibein
\beq[g-op]
   \Trr{\a\gam_i(\x)} = -\i\hbar\eps_{ij} \Tr(\a \deltadelta / \oo_j(\x) / )
\eeq
which gives the commutator
\beq
      \comm{\Trr{\a\gam_i(\x)}}{\oo_j(\y)} =
        -\i\hbar \eps_{ij}\a  \, \delta(\x,\y).
\eeq
The operator representation for $\v$ is more complicated, but from the
bracket~\eref{pois-u-v} we can read off that a suitable representation is
\beq[v-true-op]
   \Trr{\a\v(\phi_0)} = \i \hbar
     \int_{-\infty}^\infty \d\phi\,  \Trrr{ \stepf(\a,\phi-\phi_0) \, \u(\phi)
           \deltadelta / \u(\phi) / },
\eeq
where $\stepf(\a,\phi)$ is the $\alg sl(2)$ valued step function introduced in
the appendix. We actually only need the following property of this operator:
\beq[comm-u-v=0]
  \comm{ \Trr{\a\v(\phi_0)}}{\u(\phi)} = 0 \txt{for} 0<\phi-\phi_0<2\pi,
\eeq
which follows from~\eref{stepf=0}, and we need the operator representation for
$\gamphi$, which is by~\eref{u-g-pois}:
\beq[v-op]
    \Trr{\a\gamphi(\phi)}  = - \i \hbar
     \sum_{z\in\ZZ}  \Trr{ \u(\phi-2\pi z) \, \a
            \deltadelta / \u(\phi-2\pi z) / }.
\eeq
We can now write down the quantized constraints. We define them as
$\i\hbar^{-1}$ times the classical quantities. As a consequence, no extra
factors of $\hbar$ appear in their algebra and they generate exactly the same
transformations on the wave functionals as the classical constraints generated
on phase space functions. The Hamilton constraint is a pure multiplication
operator and looks like~\eref{const-eq}:
\beq[H-q]
  \H[\n] = \i\hbar^{-1} \intN {\eps^{ij}}{ \n \FF_{ij} }
       + \i\hbar^{-1} \send \intdN {\n ( \u^{-1} \del_\phi \u - \oo_\phi ) }.
\eeq
The Lorentz constraint becomes a first order homogeneous differential operator.
Here it is more suitable to use the from~\eref{const-gen}, to avoid spatial
derivatives acting on functional derivatives:
\beq[L-q]
  \L[\lp] &=&  - \i\hbar^{-1} \intN {\eps^{ij}}{ \D_i \lp \gam_j}
             + \i\hbar^{-1} \send \intdN {\lp \gamphi }  \zl
          &=&  \intN {}{\D_i \lp \deltadelta / \oo_i / }
   + \send \intl{-\infty}{\infty}\!\!\!\d\phi\,
           \Trr{ \u \lp \deltadelta /\u /}.
\eeq
Note that the sum in~\eref{v-op} has been used to change the range of the
boundary integral, which runs over $\RR$ instead of once around the circle
only.
\begin{exer}
It is now easy to see that all the classical properties of the constraints are
preserved:
\begin{enumerate}
\item there is no ordering ambiguity, i.e.\ the constraints do not contain
products of non-commuting operators.
\item the commutators of the constraints with any field operator exactly
coincides with the classical brackets~\eref{lorentz-trafo}
and~\eref{trans-trafo}.
\item the algebra of the constraint operators is closed and given
by~\eref{con-alg}.
\end{enumerate}
\end{exer}

\subsection{Physical states}
\label{physical}
After defining the operators for the phase space coordinates, we now have to
construct the physical state space, which corresponds to the classical reduced
phase space. It is the set of all possible gauge-inequivalent states of our
``universe''.
\begin{defi}[Physical State Space]
The physical state space is the kernel of the constraint operators. A physical
state is represented by a wave function satisfying
\beq[con-quant]
   \H[\n] \, \Psi[\oo_i,\u] = 0 ,\qquad
   \L[\lp] \, \Psi[\oo_i,\u] = 0 .
\eeq
\end{defi}
Solving these equations is even simpler than finding the general solutions of
their classical versions, as only half of the fields are involved.
The first equation states that $\Psi$ is non-zero only for those configurations
$\oo,\u$ with $\FF_{ij}=0$ and $\u^{-1}\del_\phi\u=\oo_\phi$ on the boundaries.
The solution is some kind of ``functional delta function'', which, however,
makes the treatment of the second equation a bit complicated. Hence, let us
first eliminate all functional derivatives. We can do this by writing the
Lorentz constraint equation as a functional equation instead of a functional
differential equation.

{}From~\eref{L-q} we infer that $\L[\lp]$ acts on the fields as
\beq
  \L[\lp] \, \oo_i = \D_i \lp , \qquad
  \L[\lp] \, \u = \u \lp,
\eeq
and since it is a first order homogeneous differential operator we have the
Leibnitz rule
\beq
  \L[\lp] \, (FG) = (\L[\lp] F) \, G +  F \, (\L[\lp] G),
\eeq
where $F$ and $G$ are functions of $\oo_i$ and $\u$. Thus, $\L[\lp]$ generates
Lorentz transformations on the wave functional, and the constraint equation
tells us that $\Psi$ has to be invariant under local Lorentz rotations. As we
know how {\em finite} Lorentz rotations act on the fields, we can rewrite the
condition as
\beq[L-q-fin]
   \Psi[\oo_i,\u] = \Psi[\h^{-1}(\del_i+\oo_i)\,\h,\u\h]
\eeq
for any field $\h\in\grp SL(2)$ which is the projection of some $\cgrp SL(2)$
valued field $\th$ on $\N$. Together with the Hamilton constraint
\beq[H-q-fin]
  \Psi[\oo_i,\u] = 0 \txt{if}  \FF_{ij} \ne 0  \ \mbox{or}\
                               \oo_\phi \ne \u^{-1}\del_\phi \u ,
\eeq
we get a set of constraints without functional derivatives, so we do not have
to bother about differentiability of $\Psi$ any more. The two conditions are
consistent in the sense that the inequalities in the second condition are
preserved under Lorentz transformations, because the constraints are first
class and form a closed algebra.

To solve the Hamilton condition, remember that we know the complete solution
for the equations $\FF_{ij}=0$ and $\oo_\phi=\u^{-1}\del_\phi\u$. They are
uniquely parameterized by a normalized quasiperiodic field $\tg:\cov\N\to\cgrp
SL(2)$ and an extra variable $\b_\endn\in\grp SL(2)$ for each boundary of $\N$:
\beq[q-soln]
  \oo_i(\x) = \g^{-1} \del_i \g(\curva_\x) , \qquad
  \u(\phi) = \b_\endn\inv \g(\curva_{\endn,\phi}) .
\eeq
Due to the quasiperiodicity of $\tg$ we can insert any $\curva_\x\in\cov\N$
with $\proj(\curva_\x)=\x$ into the first equation. To get the second equations
we needed the family of boundary paths $\curva_{\endn,\phi}$ from
\fref{boundpath}. Therefore, the complete solution to the Hamilton condition
can be written as
\beq[H-soln]
  \Psi[\oo_i,\u] = \cases {  \widehat\FS[\,\tg,\{\b_\endn\}\,] &
          if~\eref{q-soln} holds, \cr 0 & otherwise. }
\eeq
{}From~\eref{lorentz-pot} we know how Lorentz rotations act of the potential
$\tg$ and the boundary variables $\b_\endn$, so we can rewrite the Lorentz
condition~\eref{L-q-fin} as
\beq
  \widehat\FS[ \,\tg,\{\b_\endn\}\,] =
  \widehat\FS[\, \th_0\inv \tg \th , \{ \h_0\inv\b_\endn \} \,] ,
\eeq
which must hold for any {\em periodic\/} field $\th:\cov\N\to\cgrp SL(2)$,
where $\th_0=\th(\xo)$ is its value at the base point of $\cov\N$.  Remember
that this extra contribution to the finite Lorentz transformation of the
potential $\tg$ was necessary because $\tg$ is {\em normalized} quasiperiodic.

We found that two potentials $\tg$ and $\tg'$ are always related by a Lorentz
rotation if there values coincide on the loop group $\LG\subset\cov\N$.
Therefore, $\widehat\FS$ is in fact a function of the values $\tg_\loopa$,
$\loopa\in\LG$ only. In addition, we know that the values $\tg_\loopa$ are not
all independent, as they have to provide a representation of the loop group
\beq
  \g_\loopa \g_\loopb = \g_{\loopa\kette\loopb}.
\eeq
So we conclude that $\widehat\FS$ can be written as a function of $\g_\loopa$,
$\loopa\in\LG_0$, which is the set of primitive loops introduced in
definition~\ref{primloop}:
\beq
    \widehat\FS[\, \tg , \{\b_\endn\}\, ] =
  \FS( \{ \tg_\loopa|\loopa\in\LG_0 \} , \{\b_\endn\}) .
\eeq
Here we have to be careful if the primitive loops are subject to relations as
described in definition~\ref{primloop}. In this case, the support of $\FS$ is
given by those sets of elements $\{\tg_\loopa\}$ which fulfill the
corresponding relation in $\cgrp SL(2)$. In case of $\N$ being a torus, this
would mean that the support of $\FS$ is given by the set of all pairs of
commuting $\cgrp SL(2)$ matrices, as from the relation
$\loopa\kette\loopb=\loopb\kette\loopa$ for the two primitive loops of a torus
we get $\tg_\loopa\tg_\loopb=\tg_\loopb\tg_\loopa$.

Yet we have not solved the Lorentz condition completely, because the variables
still transform under Lorentz rotations at the base point $\xo$, and $\FS$ has
to fulfill the relation
\beq[last-con]
  \FS( \{\tg_\loopa\} , \{\b_\endn\} )
   = \FS( \{\th_0\inv \tg_\loopa \th_0 \} , \{ \h_0\inv \b_\endn \} )
\eeq
for any $\h_0\in \cgrp SL(2)$. With definition~\ref{auto} for inner
automorphisms of $\cgrp SL(2)$, we can drop the tilde from $\th_0$ here, as the
transformation $\tg\mapsto\th_0\inv\tg\th_0$ can equally well be written as
$\tg\mapsto\h_0\inv\tg\h_0$, where $\h_0$ is the projection of $\th_0$ onto
$\grp SL(2)$. Similar to~\eref{mod-auto} for the classical holonomies, it is
not difficult to do the last step and get a full classification of all solution
to the constraint equations, provided that $\N$ has at least one boundary.
Giving this boundary the special index $\endn=\endnull$, we found that
$\tg_{\endnull,\loopa}=\b_\endnull\inv \tg_\loopa \b_\endnull$ and
$\b_{\endnull,\endn}=\b_\endnull\inv\b_\endn$ form a complete set of
independent  gauge invariant quantities. So the wave function depends on these
variables only. The final result can be summarized as follows:
\begin{theo}\label{state-space}
If the space manifold $\N$ has at least one boundary $\endn=\endnull$, the
physical state space is given by the set of wave functions
\beq[PSS-end]
 \Psi[\oo_i,\u] = \cases {
        \FS(\{\tg_{\endnull,\loopa} | \loopa\in\LG_0 \},
             \{\b_{\endnull,\endn} | \endn\ne\endnull \})
              & if~\eref{q-soln} holds,
        \cr 0 & otherwise. }
\eeq
If $\N$ has no boundary, the general solutions to the constraint equations is
\beq[PSS-noend]
 \Psi[\oo_i] = \cases {
    \FS(\{\tg_\loopa | \loopa\in\LG_0 \}) &
          if~\eref{q-soln} holds, \cr 0 & otherwise, }
\eeq
where $\FS$ is subject to the relation
\beq[FF-con]
  \FS(\{\tg_\loopa\}) =
   \FS(\{\h_0\inv\tg_\loopa\h_0\}) \txt{for any}
       \h_0 \in \grp SL(2),
\eeq
and the support of $\FS$ is restricted by relations present between the
primitive loops.
\end{theo}
For physical states we will use Dirac's bra--ket notation, the states given
above are denoted by $\kt \FS$. If there is no boundary, it is more difficult
to find the complete physical state space explicitly, especially because of the
 relations between the primitive loops, and identifying equivalence classes of
special sets of $\cgrp SL(2)$ elements can be rather awkward. Formally, a basis
of the physical state space can be given in terms of the so called moduli
space, which is defined as follows:
\begin{defi}[Moduli Space]
The set of equivalence classes of group homomorphisms $\LG\to\cgrp
SL(2):\loopa\mapsto\tg_\loopa$ modulo inner automorphisms
$\tg_\loopa\mapsto\h_0\inv\tg_\loopa\h_0$ is called the moduli space of a
manifold $\N$, of which $\LG$ is the loop group.
\end{defi}
If $\N$ has no boundary, every point in the moduli space exactly corresponds to
a basis vector of the physical state space, represented by a wave function
$\FS$ that has support on this equivalence class only. Hence, physical wave
function can be written as a function on the moduli space. Alternatively, the
moduli space can also be defined to be the set of equivalence classes of maps
$\LG_0\to\cgrp SL(2)$, obeying the relations present between the primitive
loops. In this case it is obvious that the moduli space is finite dimensional.

\subsection{Observables}
\label{obs}
Compared to ordinary quantum mechanics of point particles, we are now at the
very beginning, where the state space is introduced as the set of wave
functions $\Psi[\{x_i\}]$.
We now have to define operators acting on this physical state space, which have
to be related to physical measurement. If an operator acts on the physical
states space, it must map solutions of the constraint equations onto solutions
again. It is exactly this what defines an observable:
\begin{defi}[Observables]\label{def-obs}
A phase space function or the corresponding quantum operator $\Obs$ is called
an observables, if it commutes weakly with the constraints
\beq
  \comm{\L[\lp]}{\Obs} \weak 0 , \qquad
  \comm{\H[\n]}{\Obs} \weak 0 .
\eeq
Two such operators are considered as the same observable, if they are weakly
equal.
\end{defi}
\begin{exer}
The Poisson bracket or commutator of two observables, or any function of
observables, is again an observable.
\end{exer}
At the classical level, (weak) vanishing of the Poisson brackets with the
constraints means that $\Obs$ is (at least on the physical phase space) gauge
invariant. At the quantum level, an operator vanishes weakly if it maps all
physical states to zero, and observables are exactly those operators that map
physical states onto physical states. It is reasonable to consider weakly equal
operators as the same observable, because they take the same values on the
physical phase space, or their action on physical states are identical,
respectively, so they correspond to the same physical quantity.

In \sref{solving} we already looked for the classical gauge invariant phase
space functions. As a result we found theorem~\ref{gauge-theo}, which tells us
that all observables can be written as functions of the holonomies $\tg_\loopa$
and $\f_\loopa$ for the loops and $\b_\endn$, $\c_\endn$ for the boundary
paths. As we want the observables to be (real) numerical quantities, let us for
a moment ignore that $\tg_\loopa$ is an element of $\cgrp SL(2)$ and take
$\g_\loopa\in\grp SL(2)$ instead as the basic quantity. Using~\eref{map-hin},
we can express the holonomies directly as phase space functions:
\beq[mod-obs]
  \g_\loopa &=& \UU_\loopa , \qquad
  \b_\endn = \UU_{\curva_{\endn,\phi}} \u^{-1} , \zl
  \f_\loopa &=& \VV_\loopa , \qquad
  \c_\endn = \VV_{\curva_{\endn,\phi}} -
   \UU_{\curva_{\endn,\phi}}
       \u^{-1}\v\,\u\,\UU_{\curva_{\endn,\phi}\inv} .
\eeq
Remember that $\curva_{\endn,\phi}$ denotes the family of boundary paths from
\fref{boundpath}, and, as indicated by the ordering, that the fields $\u$ and
$\v$ are to be taken at the endpoint of the curve $\curva_{\endn,\phi}$, i.e.\
at the angular coordinate $\phi$ on the boundary. The values of $\b_\endn$ and
$\c_\endn$ as defined in~\eref{map-hin} were (weakly) independent of the
special value chosen for $\phi$, so for later convenience we keep it arbitrary
here. Similarly, the holonomies also depend on the special path representing
the homotopy class. However, they always lead to the same observables in the
sense of definition~\ref{def-obs}, because under smooth deformation the
variations of $\UU$ and $\VV$ are proportional to the constraints.

The holonomies themselves are not yet the observables, as they still transform
under gauge transformations acting at $\xo$, which were given
in~\eref{iso-gauge}. To obtain observables, we have to find quantities which
are invariant under these transformations.
In case of a manifold without boundary, we only have the loop holonomies
$\g_\loopa$ and $\f_\loopa$. To get an observable, we can take the trace of
$\g_\loopa$, which is of course invariant under inner automorphisms. Another
gauge invariant quantity is the trace of $\g_\loopa\f_\loopa$, which is a
matrix tangent to $\g_\loopa$. We denote these observables by
\beq[trace-obs]
    \UUU_\loopa = \ft12\Trr{\g_\loopa} , \qquad
    \VVV_\loopa = \Trr{\g_\loopa\f_\loopa} .
\eeq
They provide gauge invariant information contained in the holonomies
$\g_\loopa$ and $\f_\loopa$, telling us whether $\g_\loopa$ is timelike or
spacelike etc., so essentially they measures the type of the ``sigularity''
associated with the loop $\loopa$. However, there are still some
``exceptional'' gauge invariant properties of $\g_\loopa$ which are not encoded
in the trace. If, e.g., $\g_\loopa$ is timelike or lightlike, then the sign of
$\Tr(\g\gam_0)$ is invariant. We will need those observables explictly to
construct the state space of the torus.
\begin{exer}
Of course, the trace observables exist for manifolds with boundaries, too.
Find out how $\UUU_\loopa$ and $\VVV_\loopa$ are related to the parameters
$\defangle,\scher,z$ used in \sref{class-sol} to classify the solutions of the
constraints on the cylinder manifold, and which additional observables are
needed to determine $\defangle$, $\scher$ and $z$ completely.
\end{exer}
For the rest of this section we will consider a manifold with at least one
boundary. With the results of \sref{solving}, it is not difficult to find out
how observables in general look like. We have to take some loop $\loopa$
(including the trivial loop $\loopa=\loopn$), and multiply the pair
$(\g_\loopa,\f_\loopa)$ with some boundary holonomies (but not necassary the
same) from both sides to make it invariant under inner automorphisms. Hence, a
pair of observables is given by
\beq[bound-obs]
   ( \b_\endn,\c_\endn)^{-1} (\g_\loopa,\f_\loopa)\,
          (\b_\endm,\c_\endm).
\eeq
Obviously, we can label such an observable by a path $\curvb=
\curva_\endn\inv\kette\loopa\kette\curva_\endm$, whose end points both lie on a
boundary of $\N$ (possibly the same). The special observables~\eref{mod-auto},
which form a complete set of independent observables, are included
in~\eref{bound-obs} as those for paths starting at the special end
$\endn=\endnull$ and going directly to another end, or around some primitive
loop and back to the boundary $\endnull$. We can express the observables
directly in terms of the phase space variables using the transport operator and
distance of the path $\curvb$:
\beq[obs-iso]
   (\UUUU_\curvb,\VVVV_\curvb) =
    (\u,\v)\, (\UU_\curvb,\VV_\curvb)\, (\u,\v)^{-1} .
\eeq
Again, our convention applies and that the boundary fields $(\u,\v)$ are to be
taken at the beginning or end of $\curvb$, respectively. In components, the
observables read
\beq[comp-obs]
     \UUUU_\curvb &=&  \u \, \UU_\curvb \,\u^{-1} , \zl
   \VVVV_\curvb& = &
   \v + \u \, \VV_\curvb \, \u^{-1}
      - \u \, \UU_\curvb \, \u^{-1} \v\, \u\, \UU_{\curvb^{-1}} \u^{-1}.
\eeq
The ordering indicates where the fields are to be taken: $\VVVV_\curvb$ has to
be an $\alg sl(2)$ matrix at the beginning of $\curvb$, so the first $\v$ and
the ``outer'' $\u$'s are sitting there, whereas the ``inner'' boundary fields
are sitting at the end of $\curvb$.
An important feature of all these observables is that they do not refer to the
special point $\xo$ any more. They are either the trace of a holonomy along a
loop, which can be deformed arbitrarily such that it no longer passes through
$\xo$, and the same holds for a path $\curvb$ connecting two boundaries. In
addition, their end points can be shifted along the boundary, as the definition
of the boundary holonomies in~\eref{map-hin} is independent of $\phi$. In
principle we could require the paths $\curvb$ labelling the observables to
start and end at $\phi=0$, but let us allow to shift the end points to small
positive or negative angles. We will not shift them more than once around the
boundary, as this might course some trouble with the non-periodicity of the
$(\u,\v)$ fields. Some of these paths are shown for the cylinder manifold in
\fref{observables}.

The $\UUUU_\curvb$ and $\VVVV_\curvb$ are in some sense the smallest possible
complete and ``symmetric'' set of observables. More precise, they form the
smallest set of observables such that every observable is a function thereof,
but without marking any special set of ``primitive loops'' or choosing a
special boundary like $\endn=\endnull$. As we already introduced these special
objects to define the physical state space, we should better not use them again
here, which might break the symmetries of our universe. For example, we will
require  all basic observables to become hermitian, not only the already
complete set~\eref{mod-auto}. Otherwise it could happen that only for special
observables, which a priori cannot be distinguished from others, there are well
defined eigenstates with real eigenvalues etc., whereas for others are not
because they are not hermitian.

The relations present between the basic observables are the same as those for
transport operators and distances of arbitrary paths:
\beq[obs-rel]
   (\UUUU_\curva,\VVVV_\curva) \,
   (\UUUU_\curvb,\VVVV_\curvb)  =
   (\UUUU_{\curva\kette\curvb},\VVVV_{\curva\kette\curvb}),
\eeq
which, of course, only holds if $\curvb$ starts at the boundary where $\curva$
ends (but the linked path can be deformed away from this boundary).
To discuss the problems arising from quantization of these observables, it is
more convenient not to use the $\grp ISO(1,2)$ notation and switch back to the
vector representation of the $\alg sl(2)$ valued variables. Hence, the
observables are
\beq[vec-obs]
     \UUUU_\curvb &=&  \u \, \UU_\curvb  \, \u^{-1} \in\grp SL(2), \zl
   \VVV_\curvb\^a & =& \ft12 \Trr{\gam^a \VVVV_\curva} =
   v^a + u^a\_b \, \V_\curvb\^b
        - u^a\_b \, \U_\curvb\^b\_c  \, u_d\^c \, v^d \in \MM^3,
\eeq
where $u^a\_b=\ft12\Tr(\u^{-1}\gam^a\u\gam_b)$ and $v^a=\ft12\Tr(\v\gam^a)$ are
the vector representations of the boundary fields, obeying
$(u^{-1})^c\_d=u_d\^c$. They directly correspond to basic physical quantities,
which will be discussed in more detail in the next section: $\UUUU_\curvb$
discribes the transport of a spinor from one background frame into another (or
back into the same along a non-trivial loop), and $\VVV_\curvb\^a$ measures the
relative position of two background frames.
In components, the relation~\eref{obs-rel} becomes
\beq[obs-vec-rel]
   \UUUU_{\curva\kette\curvb} = \UUUU_\curva \UUUU_\curvb, \qquad
   \VVV_{\curva\kette\curvb}\^a = \VVV_\curva\^a +
       \UUU_\curva\^a\_b \VVV_\curvb\^b.
\eeq
The task is now to find the quantum representation of these observables such
that (if possible) these relations as well as their classical Poisson brackets
are preserved, and such that we can define a scalar product on the physical
state space which makes them all hermitian. Let us start with the $\UUUU$
observables, as this is quite simple.

Of course, they all commute, because they depend on $\oo_i$ and $\u$ only, and
they are represented by multiplication operators acting on the wave function
$\FS(\{\tg_{\endnull,\loopa}\},\{\b_{\endnull,\endn}\})$. In particular, the
observables associated with the variables on which the wave function depends
act as
\beq[spec-obs]
  \UUUU_{\curva_\endnull\inv\kette\loopa\kette\curva_\endnull} \ket \FS
    = \ket{ \g_{\endnull,\loopa} \FS }, \qquad
  \UUUU_{\curva_\endnull\inv\kette\curva_\endn} \ket \FS
    = \ket{ \b_{\endnull,\endn} \FS }.
\eeq
All other $\UUUU$ observables can be obtained as functions thereof, and because
they all commute it will be sufficient to render the
observables~\eref{spec-obs} hermitian, then a real function thereof becomes
hermitian as well.

As a consequence, the eigenspaces of these operators with different eigenvalues
have to be orthogonal. However, these eigenspaces are still degenerate, as the
$\UUUU$ observables fix the values of $\g_{\endnull,\loopa}$ but not that of
$\tg_{\endnull,\loopa}$. There should be an observable associated with the
winding number of $\tg_\loopa$. Introducing such additional observables, the
eigenspaces would become non-degenerate, provided that we want to make them
hermitian operators and their eigenstates normalizable. However, this is
reasonable only if they really corresponds to physical measurements, i.e.\ if
an observer inside the universe can measure them. We will see that this is not
necessarily so, and, e.g.\ for the torus, the winding numbers do not correspond
to any physical measurement. This will also have some effects on the scalar
product, whose definition becomes rather a physical than a mathematical
problem. So let us define the scalar product in the next section, after
clarifying the role of the winding number as a physical observable.

To find the action of the $\VVVV$ observables on the physical states, we will
exploit their commutation relations. To get the commutator of some
$\VVV_\curva\^a$ with $\UUUU_\curvb$, we proceed as follows. The non-commuting
fields are the dreibein $\gam_i$ appearing in $\VVV_\curva\^a$ through
$\V_\curva\^b$, which is conjugate to the spin connection $\oo_i$ in
$\UUUU_\curvb$, and the $v^a$ and $v^d$ boundary fields in $\VVV_\curva\^a$,
which do not commute with the $\u$'s in $\UUUU_\curvb$. However, we can deform
the paths such that all boundary fields commute. What we have to do is to shift
the endpoints of $\curvb$ to a small positive angle $\phi=\epsilon$, thereby
keeping the endpoints of $\curva$ at $\phi=0$ (if the two paths do not meet at
a boundary, this procedure is not necessary). As $\v(0)$ commutes with
$\u(\epsilon)$ for $0<\epsilon<2\pi$, the only remaining commutator to be
computed is
\beq
  \comm{\VVV_\curva\^a}{\UUUU_\curvb} =
  \comm{u^a\_b\, \V_\curva\^b  } { \u \, \UU_\curvb \, \u^{-1}}.
\eeq
Hence, what we actually have to compute is a commutator of a distance of one
path with the transport operator along another path. The distance is explicitly
\beq[VV-op]
  \V_\curva\^b =  \ft12 \intl01 \d s \,    \curva'^i(s)
      \Trr{ \gam^b \UU_\curva(0,s)\, \gam_i \, \UU_\curva(s,0)}.
\eeq
Obviously, there might be ordering ambiguities, because we can place the
dreibein equally well to the left or right, but this does not affect the
commutator to be computed here. A useful general formula is
\beq[comm-e-UU]
  \comm{\e ia(\x)}{ \UU_\curva(a,b) } =
    -\ft12\i  \hbar\intl ab \d t \, \eps_{ij} \curva'^j(t)\,
          \delta(\x,\curva(t)) \,
     \UU_\curva(a,t) \,  \gam^a \, \UU_\curva(t,b),
\eeq
which follows from the basic commutator relation~\eref{o-e-pois} and the
functional derivative of the transport operator with respect to the spin
connection, which has been given in~\eref{dU-do}. Inserting this, we find
\beq
  \comm{\V_\curva\^b} {\UU_\curvb} &=&
  - \ft12\i  \hbar \intl01 \d s \intl01 \d t \,
  \eps_{ij} \curva'^i(s) \curvb'^j(t) \, \delta(\curva(s),\curvb(t)) \times
       \zl
   && \hspace*{7em}  \UU_\curvb(0,t) \, \UU_\curva(s,0) \, \gam^b
    \, \UU_\curva(0,s)\,  \UU_\curvb(t,1).
\eeq
\begin{figure}[t]
\caption{The solid line corresponds to a distance operator with an index $b$
sitting at its start point, the dashed line to a transport operator. For each
intersection there is a contribution to the commutator, consisting of a
transport operator that ``picks up'' the gamma matrix. The ``sign'' is $\ft12\i
\hbar$ if you have to turn right at the intersection, and $-\ft12\i \hbar$ if
you turn left.}
\label{obscomm}
\begin{indented}
\item[] \quad
\hbox{\beginpicture
\setcoordinatesystem units <1.3em,1.3em>
\put {$\bigcirc$} at 0 0
\Pfeil -3.6 -3 -3 -2
\setquadratic
\plot -3 -2 -2 -1 0 0 /
\plot 0 0 2 1 3 2 /
\setlinear
\put {$\curva$} [br] at -3 -1.8
\put {$b$} [tr] at -3.7 -3.2
\put {$\bullet$} at -3.6 -3
\Pfeil -3.1 2 -3 2
\setdashes <.6ex>
\plot -4 2 -3.1 2 /
\setquadratic
\plot -3 2 -1.5 1.5 0 0 /
\plot 0 0 1 -1.2 3 -3 /
\setlinear
\put {$\curvb$} [bl] at -3 2.2
\put {$\comm{\V_\curva\^b}{\UU_\curvb}$} at 0  -4.5
\endpicture
$\qquad =\ \ft12\i \hbar\ \times$
\beginpicture
\setcoordinatesystem units <1.3em,1.3em>
\Pfeil -3.1 2.2 -3 2.2
\setdashes <.6ex>
\plot -4 2.2 -3.1 2.2 /
\setquadratic
\plot -3 2.2 -1.7 1.6 0 0.2 /
\plot 0 0.2 -2 -.8 -3 -1.8 /
\setlinear
\plot -3 -1.8 -3.6 -2.8 /
\plot -3.5 -3 -2.9 -2 /
\setquadratic
\plot -2.9 -2 -1.9 -1 0.1 0 /
\plot 0.1 0 1.1 -1.2 3.1 -3 /
\setlinear
\put {\ } at 3 -3
\put {\ } [bl] at -3 2.2
\put {$\bullet$} at -3.6 -3
\put {$\gam^b$} [br] at -3.7 -3.2
\put {$\UU_\curvb(0,s)\,\UU_\curva(s',0)\,\gam^b\,
    \UU_\curva(0,s')\,\UU_\curvb(s,1)$}  at 0  -4.5
\setsolid
\pfeil 2 2 3 2
\pfeil 2 2 2 1
\put {$a^i$} [lt] at 2 1
\put {$b^j$} [lt] at 3 2
\put {$\eps_{ij}a^ib^j>0$} [lt] at 3 1
\endpicture}
\end{indented}
\end{figure}%
Here we get a contribution to the commutator whenever the two paths intersect.
The term produced at each intersection consists of the transport operator along
$\curvb$ to the intersection point, then along $\curva^{-1}$ to the beginning
of $\curva$, where it ``picks up'' the matrix $\gam^b$, then back to the
intersection and further along $\curvb$. \Fref{obscomm} shows a picture of this
operation. The sign of each contribution depends on the direction in which the
paths intersect. The picture also shows the normalization of the $\eps$ tensor:
we defined $\eps_{\rho\phi}=+1$, where $\rho$ is the coordinate pointing
towards a boundary and $\phi$ runs clockwise on the boundary. For the curves
$\curva$ and $\curvb$ in the picture we then have
$\eps_{ij}\curva'^i\curvb'^j<0$ at the intersection, so that the factor in
front becomes $+\ft12\i \hbar$. What remains to be shown is that the delta
function contributes a factor $\pm1$ at each intersection, the sign depending
on the direction in which the paths meet. More general:
\begin{exer}\label{intersection}
\beq
   \intl01 \d s \intl01 \d t \,
   \eps_{ij} \curva'^i(s) \curvb'^j(t) \, \delta(\curva(s),\curvb(t))
\eeq
gives an integer that counts the number of directed intersections of the paths,
where an intersection is positive if $\curva$ ``comes from the left''.
\end{exer}
An illustating example for this graphical method to compute commutators will be
given in the next section for the cylinder manifold. What can be shown quite
general is that the $\UUUU$ and $\VVVV$ observables for the same path do not
commute:
\begin{exer}\label{comm-a-a}
Using \fref{obscomm}, compute the commutator of $\VVV_\curva\^a$ and
$\UUUU_\curva$ (with the same path). Remember that you have to shift the ends
of the $\UUUU$ path clockwise by a small positive angle at the boundaries,
which produces one intersection between this and the $\VVVV$ path if $\curva$
connects two different boundaries, but two intersection points if $\curva$
starts and ends at the same boundary. The result should be
\beq
   \comm{ \VVV_\curva\^a}{\UUUU_\curva} =
  \cases{ - \ft12\i  \hbar\,\big(\gam^a\,\UUUU_\curva
                   - \UUUU_\curva \gam^a \big) & for loops $\curva$, \cr
   - \ft12\i  \hbar\,\gam^a \,\UUUU_\curva & otherwise.}
\eeq
As a consequence, show that
\beq
   \comm{ \VVV_\curva\^a}{\UUU_{\curva,a}\^b}
  =  \cases{ \i\hbar\eps^{bcd}\, \UUU_{\curva,cd}
     & for loops $\curva$, \cr
   0 & otherwise.}
\eeq
\end{exer}
We will not derive the commuator algebra of the $\VVVV$ observables explicitly,
but we can see how they look like in general:
\begin{exer}
Show that the set of all observables $\Obs^{(1)}(\UUUU,\VVVV)$ which are linear
in $\VVVV$ form a closed algebra under the Poisson bracket and a graded algebra
together with the observables $\Obs^{(0)}(\UUUU)$ depending on $\UUUU$ only,
i.e.
\beq[graded]
  \pois{\Obs^{(1)}}{\Obs^{(1)}}= \Obs^{(1)}, \qquad
  \pois{\Obs^{(1)}}{\Obs^{(0)}}= \Obs^{(0)}, \qquad
  \pois{\Obs^{(0)}}{\Obs^{(0)}}= 0.
\eeq
Including higher order homogeneous polynomials of degree $n$ in $\VVVV$ the
observable algebra becomes
\beq
  \pois{\Obs^{(m)}}{\Obs^{(n)}}= \Obs^{(m+n-1)}.
\eeq
\end{exer}
To find the action of the $\VVVV$ observables on physical states, we first have
to fix the operator ordering somehow. The simplest possible choice would be to
place the differential operators to the right. This would define a unique
quantum operator for every $\Obs^{(1)}$ observable and the
algebra~\eref{graded} would be preserved. In addition, the
relations~\eref{obs-vec-rel} would hold, too. However, the last two factors
$\UUU_\curva\^a\_b$ and $\VVV_\curvb\^b$ in this relation will in general not
commute, so it cannot be a relation between hermitian operators. Hence, with
this simple ordering prescription we cannot have all $\VVV_\curva\^a$ operators
hermitian, whatever the scalar product is. Instead, we should have
\beq[obs-vec-rel-op]
   \VVV_{\curva\kette\curvb}\^a = \VVV_\curva\^a +
      \ft12  \UUU_\curva\^a\_b\, \VVV_\curvb\^b   +
      \ft12   \VVV_\curvb\^b \, \UUU_\curva\^a\_b.
\eeq
\begin{exer}
Show that this relation is still associative, i.e.\
$\VVV_{\curva\kette(\curvb\kette\curvc)}\^a=
\VVV_{(\curva\kette\curvb)\kette\curvc}\^a$.
\end{exer}
This suggests another prescription. Suppose that $\curva$ and $\curvb$ are two
paths such that there is no ordering ambiguity in $\VVV_\curva\^a$ and
$\VVV_\curvb\^b$ (assuming that such paths exist, see below), but there is one
for $\curva\kette\curvb$. Then \eref{obs-vec-rel-op} already fixes the ordering
for $\VVV_{\curva\kette\curvb}\^a$, which is given by the ``symmetrized''
operator, i.e.
\beq[prescription]
  \ft12(\hbox{differentials right}) +
  \ft12(\hbox{differentials left}).
\eeq
With this prescription, we get a unique quantum operator for any $\Obs^{(1)}$
(and of course also for any $\Obs^{(0)}$). To check whether this is a suitable
ordering, let us introduce the notation $\opr\Obs$ and $\opl\Obs$ for the
quantum operators corresponding to a classical observable, with the
differentials ordered to the right or left, respectively, including a factor
$\ft12$ to simplify the notation. The quantum operators are then defined by
$\Obs\mapsto\opr\Obs+\opl\Obs$.
What we have to check is that the relation~\eref{obs-vec-rel-op} holds for
these operators. The left hand side becomes
\beq
    \opr{\VVV_{\curva\kette\curvb}\^a } +
    \opl{\VVV_{\curva\kette\curvb}\^a } &=&
    \opr{\VVV_\curva\^a + \UUU_\curva\^a\_b\, \VVV_\curvb\^b }
   + \opl{\VVV_\curva\^a + \UUU_\curva\^a\_b\, \VVV_\curvb\^b } \zl &=&
        \opr{ \VVV_\curva\^a } + \opl { \VVV_\curva\^a }+
       \UUU_\curva\^a\_b\, \opr{\VVV_\curvb\^b} +
       \opl{\VVV_\curvb\^b} \,  \UUU_\curva\^a\_b.
\eeq
This must be equal to the right hand side
\beq
  \opr{\VVV_\curva\^a} +
  \opl{\VVV_\curva\^a} +
      \ft12  \UUU_\curva\^a\_b\,
      \big(\opr{\VVV_\curvb\^b } + \opl{\VVV_\curvb\^b } \big)  +
      \ft12 \big(\opr{\VVV_\curvb\^b } + \opl{\VVV_\curvb\^b } \big)
       \, \UUU_\curva\^a\_b.
\eeq
The difference between the two expressions is obviously one half of
\beq
  \comm{ \UUU_\curva\^a\_b}{ \opr{\VVV_\curvb\^b}} +
  \comm{ \opl{\VVV_\curvb\^b} }{ \UUU_\curva\^a\_b},
\eeq
which vanishes because the commutator is independent of the ordering chosen for
$\VVV_\curvb\^b$.
\begin{exer}
Show that with the prescription~\eref{prescription} the classical Poisson
bracket of the observables~\eref{graded} is preserved, i.e.\
\beq
  \comm{\opr{\Obs}+\opl{\Obs}}{\opr{\Obs'}+\opl{\Obs'}} =
  - \i\hbar \big(\opr{\pois{\Obs}{\Obs'}} +  \opl{\pois{\Obs}{\Obs'}} \big),
\eeq
for observables $\Obs,\Obs'=\Obs^{(0)},\Obs^{(1)}$ linear in or independent of
$\VVVV$.
\end{exer}
This is the best we can get: a unique quantum operator for each classical
observable that is linear in the momenta, and their classical Poisson algebra
coincides with the commutator relations. In addition, the relations between the
$\VVV_\curva\^a$ do not prevent them from being hermitian.
For higher order observables in $\VVVV$ there are no unique quantum operators.
This is well know from quantum mechanics, where unique hermitian operators for
higher order polynomials in $p$ and $x$ do not exist.

Finally, we have to find out how the $\VVVV$ operators act on a physical state.
Before doing this, let us check whether there are ordering ambiguities at all
(otherwise we could forget everything we just did). Fix some path $\curva$
connecting two ends or starting and ending at the same end. We use the same
trick as above to simplify the computation. We want to avoid products of
non-commuting boundary fields which are somewhat awkward because of the
stepfunction appearing in their commutator. This time we have to shift the {\em
beginning\/} of $\curva$ to a small positive angle $\phi=+\epsilon$, wheras the
{\em end\/} of $\curva$ remains at $\phi=0$. We then have
\beq
  \VVV_\curva\^a = v^a(\epsilon) + u^a\_b(\epsilon) \, \V_\curva\^b
          - u^a\_b(\epsilon) \, \U_\curva\^b\_c \, u_d\^c(0) \, v^d(0).
\eeq
The only product of non-commuting boundary fields in this expression is
$u_d\^c(0)v^d(0)$: the commuator of $v^a(0)$ with $u^a\_b(\epsilon)$ either
vanishes because the fields are sitting at different boundaries, or because for
$0<\epsilon<2\pi$ their commutator is zero by~\eref{comm-u-v=0}. Interchanging
$u_d\^c(0)$ with $v^d(0)$ gives
\beq
  \comm{u_d\^c}{v^d} &= &
   \ft14\comm{\Trr{\u^{-1}\gam_d \u \gam^c}}{\Trr{\gam^d \v}} \zl &=&
   -\ft14\i \hbar \Trr{\u^{-1}\comm{\gam_d}{\stepf(\gam^d,0)} \u \gam^c},
\eeq
where the operator representation~\eref{v-true-op} has be inserted.
Unfortunately, the value of the stepfunction is ill-defined (but not infinite)
at $\phi=0$. Nevertheless, if we regularize it in either direction, we get zero
for the commutator in both cases, as $\stepf(\gam^d,+\epsilon)=0$ and
$\stepf(\gam^d,-\epsilon)=-\gam^d$ (see~\eref{stepf-per} and~\eref{stepf=0}).
Hence, the only possible ordering ambiguity occurs inside $\V_\curvb\^b$, which
depends on the dreibein and the spin connection only. If we change the ordering
in~\eref{VV-op}, we always produce extra terms of the form
\beq[order-test]
   \ft12 \intl01 \d s \, \curva'^i(s)
      \Trr{ \gam^b \UU_\curva(0,s)\, \gam_c \,
          \comm{\e ic (\curva(s)) }{\UU_\curva(s,0)}}.
\eeq
\begin{exer}\label{ordering}
There is no ordering ambiguity in $\VVV_\curva\^a$, i.e.\
$\opr{\VVV_\curva\^a}=\opl{\VVV_\curva\^a}$, if all self-intersections of
$\curva$ can be removed by smooth deformations, after shifting $\curva$ such
that it starts at a boundary point with $\phi=\epsilon$ and ends at $\phi=0$.
\end{exer}
This leads to a rather peculiar result if $\curva$ is a loop, i.e.\ it returns
to the same boundary where it started: because we have to shift the {\em
beginning\/} to a small angle $\epsilon$ before looking for self-intersections,
$\VVV_\curva\^a$ and $\VVV_{\curva^{-1}}^{\ a}$ cannot be both free of ordering
ambiguities, as one of them will have a self-intersection right at the
beginning (where it meets its own end). This also follows from
exercise~\ref{comm-a-a}, as by~\eref{obs-vec-rel-op} we have
\beq[VVV-inv]
  \VVV_{\curva^{-1}}^{\ a} =
 {} - \ft12  \UUU_{\curva,b}\^a\, \VVV_\curva\^b   -
      \ft12   \VVV_\curva\^b \, \UUU_{\curva,b}\^a,
\eeq
but the two factors on the right do not commute if $\curva$ is a loop. To
obtain the action of $\VVV_\curva\^a = \opr{\VVV_\curva\^a} +
\opl{\VVV_\curva\^a}$ onto physical states, we introduce a special state $\kt1$
given by $\FS=1$. Obviously, we can then write every state as
\beq
  \ket \FS = \FS \ket 1,
\eeq
where $\FS$ is a function of the observables~\eref{spec-obs}, but possibly also
of $\tg_{\endnull,\loopa}$ (see below).
In addition, we have $\opr{\VVV_\curva\^a}\kt1=0$. To see this, take the
complete wave functional~\eref{PSS-end}, which can be considered as a phase
space function. When the differential operators are ordered to the right, the
action on $\Psi$ is given by the Poisson bracket
\beq
   \opr{\VVV_\curva\^a} \, \Psi[\oo_i,\u] = - \i \hbar \,
    \pois{ \VVV_\curva\^a }{ \Psi[\oo_i,\u] } .
\eeq
For the special state $\kt1$, $\Psi$ is a functional depending on
the Hamiltonian constraint $\H$ only: it is simply $1$ if $\H=0$ everywhere and
$0$ otherwise. Now the bracket of $\VVV_\curva\^a$ with $\H$ vanishes exactly
(not only weakly), and therefore the bracket of $\VVV_\curva\^a$ with any
function of $\H$ vanishes, too. In addition, we can write
\beq
  \VVV_\curva\^a = 2\opr{\VVV_\curva\^a} + \big(
         \opl{\VVV_\curva\^a} - \opr{\VVV_\curva\^a} \big).
\eeq
The term in the parenthesis is in fact an observable depending on $\oo_i$ and
$\u$ only: it is the commutator of the differential operators appearing in
$\VVV_\curva\^a$ with the multiplication operators therein. It also commutes
with the constaints, so it must be an observable depending on $\UUUU$ only.
Hence, we have
\beq[VVVV-act]
   \VVV_\curva\^a \,\ket {\FS} &= &
  2\opr{\VVV_\curva\^a} \ket \FS + \big(
         \opl{\VVV_\curva\^a} - \opr{\VVV_\curva\^a} \big) \ket \FS \zl &=&
     \comm{ 2\opr{\VVV_\curva\^a}}\FS  \, \ket 1 + \big(
         \opl{\VVV_\curva\^a} - \opr{\VVV_\curva\^a} \big) \ket \FS \zl &=&
        \ket{\comm{\VVV_\curva\^a}\FS + \big(
         \opl{\VVV_\curva\^a} - \opr{\VVV_\curva\^a} \big)  \FS} ,
\eeq
because the commutator of $\VVV_\curva$ with $\FS$ as well as the difference in
the parenthesis is a multiplication operator.
\begin{exer}
Here we actually need the commutator of an $\VVV_\curva\^a$ with a function of
$\tg_{\endnull,\loopa}$, but so far we only have the commutator with the matrix
$\g_{\endnull,\loopa} = \UUUU_{\curva_\endnull\inv \kette\loopa\kette
\curva_\endnull}$ and functions thereof.
Using exercise~\ref{tot-diff}, show that the commutator with a function
$F(\tg)$, where $\tg$ is some $\cgrp SL(2)$ valued function of $\oo_i$ and $\u$
(like,  e.g.\ $\tg_{\endnull,\loopa}$), is given by
\beq[comm-V-tg]
   \comm{\VVV_\curva\^a}{F(\tg)} =
       \ft12\Trr{ \g^{-1} \comm{\VVV_\curva\^a}{\g} \gam_a }
             \Trrr{ \tg \gam^a \deldel F / \tg /} .
\eeq
\end{exer}
We can interpret the last term appearing in~\eref{VVVV-act} as some kind of
``quantum correction'', which is a multiplication operator of order $\hbar$
acting on the wave function. It vanishes for paths without self-intersection
and has to be introduced for those with, because otherwise the
relation~\eref{obs-vec-rel-op} would not be satisfied. We do not have to
compute this correction term explicitly if we want to get the action of
$\VVV_\curva\^a$ for a path with self-intersections. It is always possible to
find a set of ``primitive'' paths (not necessarily identical with those used
in~\eref{spec-obs}), which are free of self-intersections and such that every
path connecting boundaries can be obtained by linking them. Then we can derive
all $\VVV_\curva\^a$ operators using~\eref{obs-vec-rel-op}. Moreover, we can
use this relation to proof any kind of property of the $\VVV_\curva\^a$
operators by induction. For example, it is sufficient to render the observables
for the primitive paths hermitian, for the rest it then follows automatically.

\section{Quantum Physics}
\label{quant}
So far we have a physical state space and know how the observables act on the
states. This alone, however, does not tell us very much about physics, as long
as we cannot compute expectation values and things like this.A physical
question which does not make use of the scalar product is: how does the
universe in some state $\kt\FS$ evolve in time? Unfortunately it doesn't evolve
at all, because the Hamiltonian is a linear combination of constraints, and by
definition it annihilates every physical state. This is obvious also because
the wave function depends on gauge invariant objects and therefore constants of
motion only. In some sense we are dealing with a Heisenberg picture, but the
observables acting on the physical states are time-independent as well. This is
the famous ``problem of time'' in quantum gravity
\cite{kuchar:92,isham:93,unruh.wald:89}: the state as well as the operators are
time-independent, but nevertheless they represent a non-static universe in
general. The question is how to reconstruct the time evolution. To investigate
this in more detail, and to clarify the role of our observer in quantum physics
is the aim of this section.

In principle, the problem of time is also present in classical canonical
gravity, where the state of the universe is given as a point in the reduced
phase space (see definition~\ref{reduced}). Strictly speaking, a classical
state is a point in this space, which is a gauge invariant object just like the
quantum state, and it does not evolve in time.
In the classical theory it was quite easy to recover all the physical
structures including time evolution. The reduced phase space was defined to be
a set of equivalence classes of dreibein and connection fields. Hence, we could
simple choose {\em a particular\/} representative for each class to see how the
corresponding state looks like. That's exactly what we did in \sref{class-sol}.
For the quantum wave function this is not possible: it is just a function on
the configuration space, sometimes called the ``wave function of the
universe''.
The only way to recover things like ``time'' and to make physical statements is
to formulate everything in terms of gauge invariant ``observables''. This is a
well know feature of theories with an unphysical time parameter (see,
e.g.,~\cite{matschull:doc,wald:93} for a general discussion). What we need to
get a time evolution is a special phase space variable that behaves like a
clock. The time shown on this clock is a dynamical variable rather than an
background parameter, because the physical time {\em is\/} a dynamical quantity
in general relativity. A time evolution can then be recovered be reading off
the clock whenever a measurement is made. In fact, we already have such a
clock, and using this will ``solve'' the problem of time in our toy model.

\subsection{Astronomy, clocks, and time evolution}
The main idea to solve the problem of time is to consider time as something
that exists for an observer inside the universe as a dynamical quantity, but
which has no meaning to somebody looking at the universe from outside, who only
``sees'' the quantum state, or the point in the reduced phase space. So let us
return to our observer $\xo$ sitting inside the manifold and making
astronomical observations. A nice feature of this point of view is that it is
possible to consider the same observer both as the {\em relativistic\/}
observer (making experiments with test particles, clocks etc.) and as the {\em
quantum\/} observer (interpreting the wave function as probability amplitude).
We will also see that Dirac's formalism is a very convenient way to describe
this situation. It will easily keep apart and show the relations between the
objects related to ``the universe'', which is, e.g., the quantum state and the
observables derived in the last section, and quantities which are related to a
single observer. These ``local'' quantities are the holonomies~\eref{mod-obs}.
They are associated with paths starting at the special point $\xo$ and, from
the physical point of view, they correspond to astronomical measurements. To
see that some of them may be used as clocks, we have to find out their physical
meaning.

As seen from the observer $\xo$, $\g_\loopa$ is the transport operator along
$\loopa$ back to himself, and $\f_\loopa$ is the distance of this loop, which
both can be measured as described in \sref{test-particle}. However, $\g_\loopa$
was actually a projection of an $\cgrp SL(2)$ valued holonomy $\tg_\loopa$,
which cannot be measured by passing spinors along loops. To distinguish between
different values of $\tg_\loopa$ corresponding to the same $\g_\loopa$, the
observer has to use another method. From \sref{cylinder} we know that different
values of $\tg_\loopa$ really correspond to different physics, in that case to
different deficit angles of the conical singularity. So it should be possible
to measure this from inside the universe. In fact, we need the $\gam_\infty$
field from~\eref{gamhut}, or, equivalently, we have to now how the physical
dreibein behaves at infinity. Without going into details, here is the trick how
to measure $\tg_\loopa$:
\begin{exer}\label{tg-g}
Assume that $\loopa=\loopa_\endn$ is the loop going once around an end as shown
in \fref{boundpath}. The observer $\xo$ now asks all the observers on the paths
$\curva_{\endn,\phi}$ to send him the value of $\gam_\infty$ along these paths.
As a result, he can define
\beq
   \bar\gam_\infty(\phi,s) =
     \UU_{\curva_{\endn,\phi}}(0,s) \,  \gam_\infty(\curva_{\endn,\phi}(s))
  \, \UU_{\curva_{\endn,\phi}}(s,0) ,
\eeq
which certainly depends on the special paths and even their parametrization,
but which always diverges to spacelike infinity for $s\to1$. Choosing $s$ large
enough such that $\bar\gam_\infty(\phi,s)$ is everywhere spacelike, he can
count how often it winds around the $\gam_0$ axis when $\phi$ increases by
$2\pi$ (which is not necessarily an integer). This is independent of the
parametrization of the paths and also invariant under finite redefinitions of
the $\gam_\infty$ field. Show that this winding number is twice the winding
number of $\tg_\loopa$, interpreted as a path in $\grp SL(2)$, around the
non-contractible loop $\grp SO(2)\subset\grp SL(2)$. On the other hand, if
$\loopa$ is not a loop winding around a boundary, and especially if there is no
boundary at all, then there is no way to measure $\tg_\loopa$, which uses only
the background structure given and is in agreement with general relativity.
\end{exer}
Hence, though from the mathematical point of view the winding numbers are
observables, we will not consider them as physical observables in the sense
that they can be measured by an observer inside a universe without asymptotic
ends.
This will have some important consequences for the scalar product and the state
space of the torus. However, for manifolds consiting of a sphere with some
holes, including the cylinder as our standard example, it is in fact possible
to measure the winding numbers of the holonomies.

What is the physical meaning of the boundary holonomies $\b_\endn$ and
$\c_\endn$? In~\eref{mod-obs}, the $\grp SL(2)$ valued matrix $\b_\endn$ is
obtained by ``pulling'' the field $\u$ from the boundary along the path
$\curva_\endn$ to $\xo$. Assume there is somebody sitting in the background
emitting spinors $\spin$ ``at infinity'' such that they appear in a state
$\u^{-1}(\phi)\spin$ at the boundary of $\N$, and move along the path
$\curva_{\endn,\phi}\inv$. Then $\xo$ will find them in the state
$\b_\endn\spin$. Let us call this ``somebody'' a fixed stars background, or you
can also think of a microwave background, with ``microwaves'' constisting of
spinors. Then it is the relative orientation of the local frame of $\xo$ with
respect to this background that is encoded in $\b_\endn$. The vector
representation $b_{\endn,a}\^b$, which gives a proper Lorentz rotation,
actually tells us how the local frame of $\xo$ is rotated and boosted against
the background. This can be measured via some kind of ``generalized redshift''
of particles in the microwave radiation, provided one knows in which state they
are emitted (the redshift is nothing but the Lorentz-transformed wave vector of
the photon, and if the microwaves consist of spinors of know polarization, then
measuring the ``redshift'' is exactly what has been described).

There is also an interpretation for $\c_\endn\in\alg sl(2)$ or the
corresponding vector $c_\endn\^a\gam_a=\c_\endn$. The expression for $\c_\endn$
above looks very much like the ``addition rule''~\eref{VV-kette} for distances.
It is the geodesic distance to the boundary, minus a vector that is obtained by
pulling $\u^{-1}\v\u$ from the boundary to $\xo$. Hence, if $\v$ is interpreted
as the relative position of the fixed stars background with respect to the
boundary, then $\c_\endn$ measures the relative position of $\xo$ with respect
to the background, or simply the geodesic distance to some special event in the
background, a supernova, say. Remember that $\VV_\curva$ is not really the
physical distance but $\widehat\VV_\curva$ given by~\eref{VV-finite}, which of
course diverges at the boundary, whose physical distance is infinite.
Nevertheless $\c_\endn$ can be interpreted as the physical distance to some
special event. For simplicity, we can assume that $\gam_\infty(\xo)=0$. Then
you can think of $\v$ as some renormalized ``bare'' field $\widehat\v$ such
that the last equation in~\eref{mod-obs} is a regularized version of the same
expression with $\VV$ replaced by $\widehat\VV$ and $\v$ replaced by
$\widehat\v$, which both correspond to real physical but infinite quantities:
both the supernova and $\xo$ are infinitely far away from the boundary but
their relative position is finite.

As this ``relative position'' of $\xo$ with respect to the background
represents a position in both space and time, the variables $\c_\endn$
essentially become clocks, measuring the time elapsed since the supernova. We
can infer this already from the time evolution equations~\eref{free-evol} for
free falling observers. In this case the spacelike part of $\c_\endn$ is
constant (the place where the supernova took place) and the time component
increases linearly in $t$. It is not unusual that in a way like this time
reappears automatically. It is a general feature of canonical gravity on spaces
that are either non-compact or have boundaries, that additional physical
degrees of freedom appear, which are normally not necessary to introduce if one
is just dealing with Einstein's field equations. Sometimes they arise from
extra boundary fields being introduced to obtain a well defined Lagrangian,
like in our case. In other examples, like, e.g., the quantization of
spherically symmetric gravity in four dimensions~\cite{kastrup.thiemann:93,%
kastrup.thiemann:94}, they can also arise from boundary conditions to be
imposed on the multipliers and therefore on the parameters of gauge
transformations. Then some of the symmetries normally considered as gauge
symmetries no longer fit into the precise definition~\ref{gauge}. As a result,
there are more gauge invariants than expected. In the case mentioned one finds
that in addition to the Schwarzschild mass, the only ``classical parameter'' of
spherical symmetric vacuum Einstein gravity, there is a canonically conjugate
variable representing the ``time at infinity''. It is a ``conserved current''
corresponding to time rescaling symmetry at infinity, which in classical
gravity is considered as a gauge transformation, but which cannot be generated
by taking Poisson brackets with constraints.

How can one use the clocks to recover a time evolution? Without going into
detail, let us sketch the main idea, using point particle quantum mechanics as
a simple example once again. For a free particle we can introduce the first
order action with parameter time $\Lag=\dot xp+\dot TE -\Ham$, where the
Hamiltonian $\Ham=q\const=q(p^2/2m - E)$ with Lagrange multiplier $q$ is, like
in gravity, given by a ``linear combination'' of (only one) constraint
$\const$. All the variables depend on $t$, and the equations of motion
\beq
     \dot x = \frac {qp}m, \quad
     \dot p = 0 , \quad
     \dot T = q , \quad
     \dot E = 0 , \quad
     \const = \frac{p^2}{2m} - E \weak 0
\eeq
obviously describe a particle in one dimension and a clock $T$, where the free
parameter $q$ indicates how fast the physical time runs with respect to $t$,
and the conjugate variable to $T$ is the energy $E$. The constaint $\const$
generates a gauge transformation in the sense of definition~\ref{gauge}, whose
finite version has a real parameter $ \gpar$ and reads
\beq[gpar]
    x \mapsto x^{(\gpar)} = x +   \frac { \gpar p}m , \qquad
    T \mapsto T^{( \gpar )} = T +   \gpar ,
\eeq
where the superscript $(\gpar)$ denotes the phase space function obtained by a
finite gauge transformation with parameter $\gpar$.
The reduced phase space consists of equivalence classes of quadrupels
$\{x,p,T,E\}$ modulo this transformation.
Canonical quantization is staightforward, we have a wave function $\Psi[x,T]$
and
\beq
  p = \i \hbar \deldel / x / , \qquad
  E = \i \hbar \deldel / T / .
\eeq
A physical state has to satisfy
\beq
   \const \Psi =
  \Big( \frac {p^2}{2m} - E \Big) \Psi =
  \Big( - \frac {\hbar^2}{2m} \frac{\del^2}{\del x^2}
         - \i \hbar \deldel / T / \Big) \Psi[x,T] = 0,
\eeq
which formally looks like the Schr\"o\-din\-ger equation, but $T$ is not the
canonical time coordinate $t$. There is no time evolution with respect to this
coordinate: the state is a fixed function of the configuration variables $x$
and $T$. To recover time evolution, one has to ask questions like: what is the
value of $x$ ``when'' $T$ takes the value $T_0$. At the classical level, the
answer can be given as follows. The state is an equivalence class as described
above, so choose a particular representative with the property $T=T_0$, then
the value of $x$ for this representative is what you want to know. We can
define a corresponding phase space function
\beq[gauge-integral]
   x|_{T=T_0} &=& \int \d  \gpar  \, \deldel T^{( \gpar )} /   \gpar  / \,
         \delta(T^{( \gpar )}-T_0) \, x^{( \gpar )} \zl &=&
      \int \d  \gpar  \, \delta(T +   \gpar  - T_0 ) \,
             \Big( x +   \frac { \gpar p}m \Big) =
         x + (T_0-T) \frac pm .
\eeq
Note that (for a fixed value of $T_0$) this gives a phase space function which
is in fact an observable: its bracket with the constraint $\const$ vanishes. We
obtained this quantity by intergrating over the gauge group $(\RR,+)$, picking
up only those configurations with $T=T_0$. The extra term $\del T^{( \gpar
)}/\del  \gpar $ has been introduced to get the correct factor, because the
intergral runs over the gauge parameter $  \gpar $, wheras the argument of the
delta function is a special phase space function, the  ``clock'' $T$. It is the
{\em interpretation\/} of $T$ as a clock that makes the last expression look
like a time evolution of $x$. In quantum physics we can now ask for the
expectation value of $x$ ``at the time'' $T=T_0$, and we just have to insert
the corresponding operator
\beq
  \bra \Psi \big( x|_{T=T_0}\big) \ket \Psi =
  \bra \Psi \Big( x + (T_0-T) \frac pm \Big)  \ket \Psi.
\eeq
But what is the scalar product? It has to be such that all real observables
become hermitian, and up to rescaling there is only one such product:
\beq[ex-sp]
  \braket \Phi \Psi =
  \int \d x \, \Phi^*[x,T] \Psi[x,T] .
\eeq
There is no integration over $T$, as otherwise physical states would not be
normalizible. The value of the product is independent of the special $T$
inserted, which follows from the ``Schr\"o\-din\-ger equation''
$\const\Phi=\const\Psi=0$.
\begin{exer}
The scalar product is defined on the {\em physical\/} state space only. Show
that it is not possible to define a product on the whole state space such that
the operators $x,p,T,E$ are hermitian and restricting it to the physical state
space gives \eref{ex-sp}.
\end{exer}
Now we can compute the expectation value
\beq[ex-obs]
  \bra \Psi \big( x|_{T=T_0}\big) \ket \Psi =
  \int \d x \, \Big(  x \,\Psi^* \, \Psi
      + \frac {\i\hbar}m (T_0-T) \, \Psi^* \, \deldel \Psi / x / \Big).
\eeq
Knowing that this is independent of $T$, we choose $T=T_0$ and what we get is
the usual expression for the expectation value of $x$ at a time $T_0$. Hence,
formally the wave function $\Psi[x,T]$ behaves exactly like a solution of the
time-dependent Schr\"o\-din\-ger equation, but $T$ is not the canonical time.
Neither the state nor the observables depend on the parameter time $t$, which
totally disappears from the canonical framework. Nevertheless there is a time
evolution, which is encoded in $\Psi[x,T]$ as a {\em correlation\/} bewteen $T$
and $x$. The state does not make any predictions about a measurement of, say,
$T$ alone, the result is just random. If you read off a clock for the first
time in your life there is no ``expectation value'' for the result. The same
holds for $x$, which might be considered as the position of the sun in the sky,
which is completely random if you see it for the first time, too. However,
there is a correlation between the two measurements, and it makes sense to ask
for the position of the sun seen by somebody who just read off a specific time
$T_0$ from a clock. It is exactly this what the {\em observable\/}
\eref{ex-obs} represents.

What does this mean for our gravity model? Here we have some more gauge degrees
of freedom than just the time rescaling above. It is not enough to ask for the
value of, e.g., a holonomy $\f_\loopa$ at a specific time, we also have to
specify where the observer is when he makes this measurement. It turns out that
the following is a reasonable question: what is the value of some function $X$
of the holonomies \eref{mod-obs} for an observer being in a relative position
$\c_\endnull=\c$ and orientation $\b_\endnull=\b$ with respect to the supernova
in the background $\endn=\endnull$. Similar to \eref{gauge-integral} we can
define a corresponding observable
\beq
  X \big|_{{\b_\endnull=\b}\atop{\c_\endnull=\c}} =
  \int \d \h \, \d \n \,
     \Big| \deldel (\c_\endnull^{(\h,\n)}, \b_\endnull^{(\h,\n)}) /
                    (\h,\n) /  \Big| \,
    \delta ( \c_\endnull^{(\h,\n)} - \c) \,
    \delta (\b_\endnull^{(\h,\n)} \b^{-1} ) \,
         X^{(\h,\n)}.
\eeq
We are integrating over the gauge group $(\h,\n)\in\grp ISO(1,2)$ acting on the
holonomies. According to \eref{iso-gauge} the transformed holonomies are
\beq
   \b_\endnull^{(\h,\n)} = \h^{-1} \b_\endnull , \qquad
   \c_\endnull^{(\h,\n)} = \h^{-1} \big( \c_\endnull - \n \big) \, \h .
\eeq
The first term in the integral is the Jacobian of this transformation, which
becomes one if we choose the measure to be a Haar measure on $\grp ISO(1,2)$,
and define the delta functions with respect to this measure. Independent
thereof we can solve the integral, as the delta functions pick up exactly one
term, namely that for
\beq
   \h =  \b_\endnull \b^{-1} , \qquad
   \n =  \c_\endnull - \b_\endnull \b^{-1} \c\, \b\, \b_\endnull\inv.
\eeq
As a special example choose $X=\f_\loopa$, so that by \eref{iso-gauge} we have
\beq
   \f_\loopa^{(\h,\n)} = \h^{-1} \big( \f_\loopa
    + \g_\loopa \n \g_\loopa\inv - \n \big ) \h .
\eeq
Inserting all this, we end up with a sightly complicated expression
\beq[f-at]
   \f_\loopa \big|_{{\b_\endnull=\b}\atop{\c_\endnull=\c}} &=&
      \b\,\b_\endnull\inv \big( \f_\loopa + \g_\loopa (
           \c_\endnull - \b_\endnull \b^{-1} \c\,\b\,\b_\endnull\inv )
      \, \g_\loopa\inv
      - \c_\endnull +\b_\endnull \b^{-1} \c\,\b\,\b_\endnull\inv \big)
     \, \b_\endnull \b^{-1} \zl &=&
     \b \f_{\endnull,\loopa} \b^{-1}
     - \b \,\g_{\endnull,\loopa} \b^{-1} \c\,\b\,
         \g_{\endnull,\loopa}\inv \b^{-1} + \c,
\eeq
where $\g_{\endnull,\loopa}$ and $\f_{\endnull,\loopa}$ are the gauge invariant
quantities introduced in \eref{mod-auto}. Hence, the result is indeed an
observable, as $\b$ and $\c$ are just numerical parameters. But where is the
time evolution? It is not so obvious as in our point particle example, but we
have to remember that an observer is actually represented by a worldline, so
its relative position with respect to the supernova, i.e.\ the $\c$ parameter
above, will change. Assume that the observer is free falling, then by
\eref{free-evol} his position and orientation with respect to the background is
given by
\beq[bar-b]
  \b(\tau)= \bar\b , \qquad \c(\tau) = \bar\c - \tau \gam_0,
\eeq
where $\tau$ is his own proper time. Once he found out which values of $\bar\b$
and $\bar\c$ apply to him, there is an observable $\f_\loopa^{(\tau)}$ obtained
by inserting \eref{bar-b} into \eref{f-at}, which is parameterized by $\xo$'s
local time and obeys
\beq
   \dd / \tau / \f_\loopa^{(\tau)} =
   \bar\b\, \g_{\endnull,\loopa} \bar\b^{-1} \gam_0 \,
      \bar\b\, \g_{\endnull,\loopa}\inv \bar\b^{-1} - \gam_0.
\eeq
This is the time evolution of the holonomy $\f_\loopa$ as seen by $\xo$. It
becomes a well defined gauge invariant phase space function or quantum
observable after specifying the values for $\bar\b$ and $\bar\c$, which can be
obtained by measuring the geodesic distance and the ``redshift'' of the
supernova in the background once at $\tau=0$.
Hence, for every {\em special\/} observer we can define observables describing
the time evolution of the universe as seen by this observer, but there is
nothing like the time evolution of the universe as a whole, expressed in terms
of gauge invariant quantities.

Now this consideration was rather classical, but is should be clear how to
transform it to quantum physics. As there is no prediction about the values of
$\bar\b$ and $\bar\c$ to be measured first, there is also nothing like an
uncertainty relations or something like this between them, though the
holonomies $\b_\endnull$ and $\c_\endnull$ do not commute. The result of these
measurement are random anyway, so $\xo$ can measure them and he will get some
values. Then $\f_\loopa^{(\tau)}$ becomes a well defined operator on the
physical state space and one can compute an expectation value of this
observable, or that of an projection operator onto some eigenspace of
$\f_\loopa^{(\tau)}$ (provided one has the scalar product, which will be
derived in the next section). The interpretation of this expectation value is
the probability for the observer on the worldline given by \eref{bar-b} to
measure a specific value of $\f_\loopa$ at a time $\tau$. This probability will
in general depend on $\tau$, so the special observer $\xo$ sees a non-static
universe.

\subsection{Hermitian operators and the Hilbert space}
After clarifying the physical meaning of the holonomies and observables, let us
come back to the state space and the last problem to be solved, namely the
construction of the scalar product. We already mentioned it in \sref{obs}, but
we saw that it depends on which {\em mathematical\/} observables we want to
become {\em physical\/} observables represented by hermitian operators having
normalizable eigenstates. In particular the possibility to measure the winding
number of $\tg_\loopa$ will play a crucial role. By exercise~\ref{tg-g} we now
know that it depends on the existence of a boundary of the space $\N$ whether
$\tg$ can be measured or not. We will give examples for both cases, the
cylinder as a manifold with boundaries and the torus as the simplest without
boundary in the next section.

\begin{figure}[t]
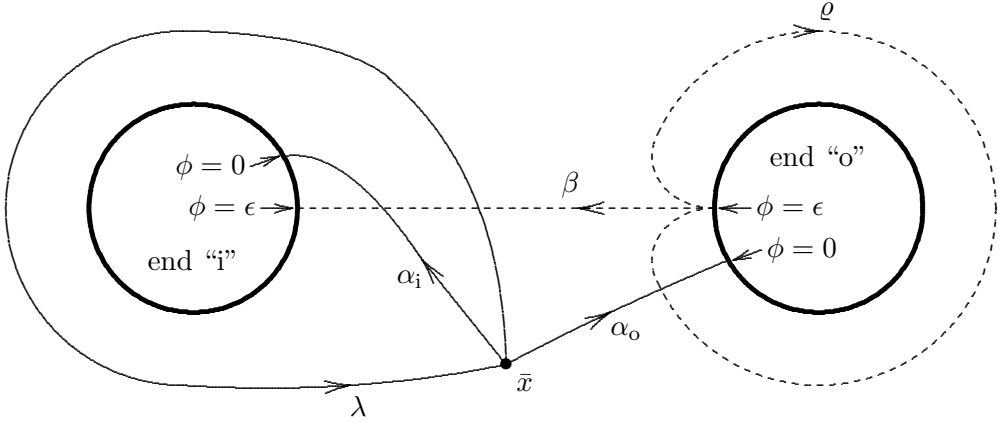

\caption{The basic observables of the cylinder manifold. The wave functional
depends on the two transport operators $\tg$ and $\b$ corresponding to the
paths $\loopc$ and $\curvb$ in the picture. The observables are constructed
from the holonomies corresponding to the paths $\curva_\endnull$,
$\curva_\endeins$ and $\loopa$, all starting at the special observer $\xo$.}
\label{observables}
\begin{indented}
\item[] \quad
\beginpicture
\setcoordinatesystem units <3.6em,3.6em>
\multiput {\ } at -1 2 -1 -2 9 2 9 -2 /
\setdashes <.5ex>
\circulararc 157 degrees from 8.7 0 center at 7 0
\circulararc -156 degrees from 8.7 0 center at 7 0
\ellipticalarc axes ratio 12:10 115 degrees from 6 0 center at 6 -.5
\ellipticalarc axes ratio 12:10 -115 degrees from 6 0 center at 6 .5
\plot 6 0 2 0 /
\setsolid
\Pfeil 6.9 1.7 7 1.7
\Pfeil 4.8 0 4.7 0
\put {$\loopc$} [bl] at 7 1.8
\put {$\curvb$} [br] at 4.7 .1
\put {$\bullet$} at 4 -1.5
\put {$\xo$} [tl] at 4.1 -1.6
\setquadratic
\plot 3.2 -.5 2.5 .3 1.87 .5 /
\plot 4 -1.5 5 -1 6.13 -.5 /
\plot 4 -1.5 3.7 .1 2.86 1.26 /
\plot 4 -1.5 2.5 -1.7 .9 -1.7 /
\setlinear
\Pfeil 2.45 -1.704 2.5 -1.7
\put {$\loopa$} [tl] at 2.5 -1.8
\Pfeil 4.9 -1.05 5 -1
\put {$\curva_\endnull$} [tl] at 5 -1.1
\Pfeil 4 -1.5 3.2 -.5
\put {$\curva_\endeins$} [tr] at 3.2 -.6
\circulararc 180 degrees from .9 1.7 center at .9 0
\ellipticalarc axes ratio 3.03:1 -70 degrees from .9 1.7  center at .9 1
\put {$\phi=\epsilon$} [l] at 6.4 0
\put {$\phi=\epsilon$} [r] at 1.6 0
\put {$\phi=0$} [l] at 6.5 -.4
\put {$\phi=0$} [r] at 1.5 .4
\put {end ``$\endnull$''} at 7 .5
\put {end ``$\endeins$''} at 1 -.5
\pfeil 6.35 0 6.05 0
\pfeil 6.45 -.4 6.18 -.5
\pfeil 1.65 0 1.95 0
\pfeil 1.55 .4 1.82 .5
\setplotsymbol ({\bf .})
\circulararc 360 degrees from 0 0 center at 1 0
\circulararc 360 degrees from 6 0 center at 7 0
\endpicture
\end{indented}
\end{figure}

\Fref{observables} shows a picture of the compactified cylinder manifold.
Instead of drawing a cylinder, it is more convenient to consider it as a sphere
with two holes, and the picture shows part of the sphere. On the two boundaries
``$\endnull$'' and ``$\endeins$'' we have the angular coordinate $\phi$ running
clockwise (note that on the ``negative $r$'' end this is not the $\phi$
coordinate used in sections~\ref{cylinder} and~\ref{class-sol}: here we need
both coordinates running clockwise, as otherwise the formulas for the brackets
of the boundary fields $\u$ and $\v$ would not be correct). We then choose a
point $\xo$ inside $\N$ as our observer, and select two paths $\curva_\endnull$
and $\curva_\endeins$ from $\xo$ to the boundaries at $\phi=0$. They provide
the family of boundary paths as introduced in \fref{boundpath}, but we will
actually only need these two. In addition, there is the primitive loop
$\loopa$.

The wave functional of a physical state of the cylinder, as shown in
theorem~\ref{state-space}, is an arbitrary function of the two quantities
$\tg=\tg_{\endnull,\loopa}$ and $\b=\b_{\endnull,\endeins}$, corresponding to
the paths $\loopc$ and $\curvb$ in the picture, which are already shifted to a
small positive angle $\epsilon$ to compute the commutators below.
To obtain the action of some observable $\UUUU_\curva$ on a state
$\kt{\FS(\tg,\b)}$, we only have to deform the path such that it becomes a
combination of $\loopc$ and $\curvb$, then the wave function is multiplied by
the corresponding combination of $\g$ and $\b$. For example, we have
$\curva_\endeins\inv \kette \loopa^z \kette
\curva_\endeins\sim\curvb^{-1}\kette \loopc^z\kette\curvb$ and therefore
\beq
     \UUUU_{\curva_\endeins\inv \kette \loopa^z \kette \curva_\endeins}
           \, \ket \FS =
     \UUUU_{\curvb^{-1}\kette\loopc^z\kette\curvb} \, \ket \FS
   = \ket { \b^{-1} \g^z \b \, \FS }.
\eeq
As we want all these observables to be hermitian, the scalar product has to
obey
\beq
  \braket{\FS_1}{\g \FS_2} = \braket{\g\FS_1}{\FS_2}, \qquad
  \braket{\FS_1}{\b \FS_2} = \braket{\b\FS_1}{\FS_2}.
\eeq
Or, equivalently, the eigenstates for different values of $\g$ or $\b$
(represented by delta functions in $\tg$ and $\b$) have to be orthogonal. So
far we saw this already for a general manifold, but now we can apply
exercise~\ref{tg-g}, i.e.\ we know that there is another observable that
measures the winding number of $\tg_\loopa$ and hence distinguishes between the
different $\tg$ corresponding to the same value of $\g$. Of course, we want
this observable to be hermitian, too. As a result, all the eigenspaces become
one dimensional, and to make them orthogonal the scalar product must be
\beq[scal-1]
  \braket{\FS_1}{\FS_2} =
   \int \d \tg \, \d \b \, \, \prodf(\tg,\b)\,
    \FS_1\^*(\tg,\b) \, \FS_2(\tg,\b),
\eeq
where $\prodf$ is a function to be adjusted such that the conjugate observables
become hermitian as well. The measure chosen for this integral is in principle
arbitrary, as any redefinition can be absorbed into the function $\prodf$, but
the most natural choice is to take a Haar measure on $\grp SL(2)$ and $\cgrp
SL(2)$, respectively.

To fix the function $\prodf$, we have to compute the action of the $\VVVV$
observables. As we know from \sref{obs}, we only need to compute the action of
two special $\VVVV$ observables corresponding to two independent
self-intersection free paths, and make them hermitian. Their action on physical
states is obtained by~\eref{VVVV-act}, where the ``quantum correction'' term
vanishes. So what we need are two paths such that they have no
self-intersections and, if possible, simply commutators with $\g$ and $\b$, so
that the commutator with $\FS(\tg,\b)$ becomes simple, too.
A good choice is
\beq
 \curvc=\curva_\endnull\inv\kette\loopa^{-1}\kette\curva_\endeins, \qquad
 \loopd=\curva_\endeins\inv\kette\loopa^{-1}\kette\curva_\endeins.
\eeq
The first is obviously self-intersection free, as it passes from one boundary
to the other. The second is a loop, but if we shift its beginning, as required
in exercise~\ref{ordering}, to $\phi=\epsilon$ (see \fref{observables}), all
intersections can be removed. Here it is essential that $\loopa^{-1}$ appears
and not $\loopa$, which whould lead to a self-intersection. The reason for
choosing these two paths is that each intersects with one of the paths $\curvb$
and $\loopc$ once only, so we get simple commutators. With the graphical rules
from \fref{obscomm} we have
\beq[prim-comm]
  \comm{ \VVV_\loopd\^a } { \UUUU_\curvb } =
      \ft12\i  \hbar\, \UUUU_\curvb \, \gam^a , \qquad
  \comm{ \VVV_\curvc\^a } { \UUUU_\loopc } =
      \ft12\i  \hbar\, \UUUU_\curvc \, \gam^a .
\eeq
This immediately leads to the representation (using~\eref{comm-V-tg})
\beq
  \VVV_\loopd\^a \, \ket \FS =
    \Ket{ \ft12\i  \hbar \Trrr{ \b \gam^a \deldel \FS / \b / } } , \quad
  \VVV_\curvc\^a \, \ket \FS =
    \Ket{ \ft12\i  \hbar \Trrr{ \tg \gam^a \deldel \FS / \tg / } }.
\eeq
Hence, the observables act by generating multiplication of $\tg$ or $\b$,
respectively, from the right. This is exactly the transformation under which
the Haar measure in~\eref{scal-1} is invariant. To render these operators
hermitian, we have to choose $\prodf=1$ (or any other constant).  Hence, the
scalar product takes the simplest possible form
\beq[scal]
  \braket{\FS_1}{\FS_2} =
   \int \d \tg \, \d \b \, \FS_1\^*(\tg,\b) \, \FS_2(\tg,\b).
\eeq
To illustrate some of the results of \sref{obs}, let us see how other
observables act and which kind of ordering ambiguities appear. Consider the
path $\curvb$ in \fref{observables}. It is obviously free of
self-intersections, so we expect $\VVV_\curvb\^a$ to be free of ordering
ambiguities. There are two ways to derive the quantum operator.
\begin{exer}
Using the graphical rules from \fref{obscomm}, show that
\beq
  \comm{\VVV_\curvb\^a}{\UUUU_\curvb} =
   -\ft12\i  \hbar \, \gam^a \, \UUUU_\curvb , \qquad
  \comm{\VVV_\curvb\^a}{\UUUU_\loopc} =
   \ft12\i  \hbar \, \UUUU_\loopc \, \gam^a .
\eeq
Hence, assuming that there are no ``quantum corrections'' in~\eref{VVVV-act}
for this operator, we have
\beq
  \VVV_\curvb\^a \, \ket \FS =
  \Ket{ \ft12\i  \hbar \Trrr{\tg\gam^a \deldel \FS /\tg/ }
       -\ft12\i  \hbar \Trrr{\gam^a\b \deldel \FS/ \b /} }.
\eeq
\end{exer}
This is obviously hermitian, too, because the Haar measure in invaraint under
multiplication from the left as well. Another way to obtain the same operator
is to use that $\curvb\kette\loopd=\curvc$, so the
relation~\eref{obs-vec-rel-op} gives
\beq
  \VVV_\curvb\^a + \ft12 \UUU_\curvb\^a\_b \, \VVV_\loopd\^b
                 + \ft12 \VVV_\loopd\^b \, \UUU_\curvb\^a\_b
     = \VVV_\curvc\^a .
\eeq
{}From~\eref{prim-comm} it is easy to infer that
\beq
  \comm{\VVV_\loopd\^b }{ \UUU_\curvb\^a\_b } =
  \ft12 \comm{\VVV_\loopd\^b }
  { \Trr{\UUUU_\curvb\inv \gam^a\UUUU_\curvb \gam_b }} = 0,
\eeq
so we can reorder the factors as we like and get
\beq
  \VVV_\curvb\^a  \, \ket \FS & = &
 \VVV_\curvc\^a \, \ket \FS -
  \UUU_\curvb\^a\_b \, \VVV_\loopd\^b \ket \FS \zl &=&
   \Ket{ \ft12\i  \hbar \Trrr{ \tg \gam^a \deldel \FS / \tg / } }
  - \Ket{ \ft12\i  \hbar \, b^a\_b
     \Trrr{ \b \gam^b \deldel \FS / \b / } } .
\eeq
With $b^a\_b \, \b \gam^b = \gam^a \b$ this becomes exactly the same as the
operator given above. As another example, we can compute the observable
$\VVV_{\loopd^{-1}}^{\ a}$ using~\eref{VVV-inv}:
\beq
  \VVV_{\loopd^{-1}}^{\ a} &=&
 {} - \ft12  \UUU_{\loopd,b}\^a\, \VVV_\loopd\^b   -
      \ft12   \VVV_\loopd\^b \, \UUU_{\loopd,b}\^a \zl & = &
 {} -   \UUU_{\loopd,b}\^a\, \VVV_\loopd\^b   -
      \ft12 \comm{  \VVV_\loopd\^b }{ \UUU_{\loopd,b}\^a} .
\eeq
\begin{exer}
Evaluate the commutator, using
$\loopd=\curvb^{-1}\kette\loopc^{-1}\kette\curvb$ and the identity
$\gam^a\g\gam_a=-\g+2\Tr(\g)\eins$ for $\grp SL(2)$ matrices. The result is
\beq
  \ft12 \comm{  \VVV_\loopd\^b }{ \UUU_{\loopd,b}\^a} =
   \ft12\i  \hbar \Trr{\UUUU_\loopc}
    \Trr{\UUUU_\curvb\inv\UUUU_\loopc\,\UUUU_\curvb\,\gam^a},
\eeq
and the resulting quantum operator becomes
\beq
  \VVV_{\loopd^{-1}}^{\ a} \, \ket \FS & =
   & \Ket { \ft14\i\hbar \Trr{\b^{-1}\g^{-1}\gam^a\g\,\b\,\gam_b}
                 \Trrr{\gam^a\b\deldel\FS/\b/ } } \zl
       & & {}+ \Ket{\ft12\i  \hbar \Trr{\g} \Trr{\b^{-1}\g\,\b\,\gam^a} \FS }
\eeq
\end{exer}
Here we get a quantum correction in form of an extra imaginary multiplication
term of order $\hbar$, and it is also obvious that there is an ordering
ambiguity in the first part of this expression. If we bring the differential
operator to the left, then it also acts on the prefactor, which depends on
$\b$. But the extra term is exactly what is needed to make the operator
hermitian under the scalar product~\eref{scal}, which can be checked by partial
integration. Hence, we see that the ``quantum corrections'' appearing
in~\eref{VVVV-act} are really necessary to obtain a full set of hermitian
$\VVVV$ operators.

There are many other interesting questions about the physics described by this
state space, and most of them are still to be explored. For example, one can
now ask for the eigenstates of the $\VVVV$ operators and their spectrum, look
for complete sets of commuting observables other than the $\UUUU$'s to provide
another basis for the state space, or find the uncertainty relations for the
observables and construct semi-classical universes with minimum uncertainty.
All typical questions arising in quantum mechanic can be translated to similar
questions for our toy model of quantum gravity, and they will certainly also be
interesting for quantum gravity in general. We will not go into more details
here, as all this is far beyond the scope of this article. However, an article
about three dimensional canonical gravity would not be complete without
dicussing {\em the\/} standard example, so the last section will be a rather
mathematical derivation of the physical Hilbert space for the torus.

\subsection{Torus quantization}
Quantization of three dimensional canonical gravity has mostly focussed on the
torus, because, form a mathematical point of view, this is the simplest
non-trivial manifold (for a really huge collection of references
see~\cite{carlip:95}). In contrast to the cylinder manifold, there are no
boundary conditions for the fields and we don't have to introduce the extra
fields $\u$ and $\v$, which makes the canonical formalism easier. On the other
hand, it is the lack of the background structures associated with the boundary
fields that makes it more difficult to extract physical information.
We considered this in detail for the cylinder manifold above, so let us focus
on a mathematically rigorous construction of the Hilbert space in this section.

On the torus we have two primitive loops $\loopa$ and $\loopb$, and they obey
the relation $\loopa\kette\loopb=\loopb\kette\loopa$. Hence, the loop group is
Abelian and the wave function is $\FS(\tg_\loopa,\tg_\loopb)$, which has
support on pairs of commuting $\cgrp SL(2)$ elements, and must be invariant
under inner automorphisms~\eref{FF-con}. Before constructing the state space
explicitly, let us remember what we know about observables measuring the
``winding number'' of the holonomies $\tg_\loopa$ and $\tg_\loopb$. As the
torus has no boundary, exercise~\ref{tg-g} tells us that there is no way for an
observer sitting on the torus to distinguish between the holonomies
$\tg_\loopa$ and $\teins_{2z}\tg_\loopa$, which are projected onto the same
matrix $\g_\loopa\in\grp SL(2)$ (see~\eref{alg-csl2}, where the elements
$\teins_{2z}\in\cgrp SL(2)$ have been introduced as the ``loops'' of $\grp
SL(2)$; the maps $\tg\mapsto\teins_{2z}\tg$ are the ``shift'' maps on $\grp
SL(2)$ as introduced in \sref{covering}). Only the matrices $\g_\loopa$ and
$\g_\loopb$ can be measured by passing a spinor once around the primitive
loops. As a consequence, there are operators
\beq
  {\cal P}_\loopa \ket{\FS(\tg_\loopa,\tg_\loopb)} &=&
   \ket{\FS(\teins_2 \tg_\loopa, \tg_\loopb) } , \zl
  {\cal P}_\loopb \ket{\FS(\tg_\loopa,\tg_\loopb)} &=&
   \ket{\FS(\tg_\loopa, \teins_2 \tg_\loopb) } ,
\eeq
commuting with all physical observables: if there was any observable which did
not commute with ${\cal P}_\loopa$, then it would be possible to measure the
winding number of $\tg_\loopa$. The operators $\cal P$ are very much like the
the permutation operators for identical particles in quantum mechanics. They
also  commute with all observables, as it is exactly this what ``identical
particle'' means. In both cases the state space splits into ``superselection
sectors''. They are the eigenspaces of the $\cal P$ operators, which can always
be diagonalized simultaneously with any other set of commuting observables.

In multi-particle quantum mechanics, the square of any primitive permutation
operator is the identity, which restricts its eigenvalues to be $\pm1$, leading
to bosonic and fermionic states. In our case the algebra of the $\cal P$
operators is free. Thus any eigenvalue is allowed except zero, because the
$\cal P$'s are invertible, and for any pair of non-vanishing complex numbers
$q_\loopa$, $q_\loopb$ we get a superselection sector. Note that we do not even
have a restriction for the eigenvalues to be imaginary or real, because a
priori there is no reason for the $\cal P$ to be (anti)hermitian.
\begin{exer}
Show that all sectors are isomorphic: for any pair
$q_\loopa,q_\loopb\in\CC-\{0\}$, one can define (quasiperiodic!) functions
$C_\loopa, C_\loopb : \cgrp SL(2)\to\CC-\{0\}$ such that
$C_\loopa(\teins_{2}\tg)=q_\loopa C_\loopa(\tg)$ and similar for $\loopb$. Then
\beq
  \ket{\FS(\tg_\loopa,\tg_\loopb)} \mapsto
  \ket{C_\loopa(\tg_\loopa) C_\loopb(\tg_\loopb)
      \FS (\tg_\loopa,\tg_\loopb)}
\eeq
is a one-to-one map from the sector with eigenvalues $(1,1)$ onto the sector
$(q_\loopa,q_\loopb)$.
\end{exer}
Hence, we only need to consider one special superselection sector, knowing that
all others are isomorphic (in contrast to quantum mechanics, where the bosonic
anf fermionic sectors are not isomorphic). The simplest choice is of course the
sector with eigenvalues $1$ for both $\cal P$ operators. In this sector the
wave function $\FS$ depends on the $\grp SL(2)$ matrices $\g_\loopa$ and
$\g_\loopb$ only.

To get a basis for the state space, we have to find the moduli space of the
torus, but with respect to $\grp SL(2)$ instead of $\cgrp SL(2)$. This is not
very difficult, the relevant properties of $\grp SL(2)$ have been studied
already in \sref{lorentz-group}. We found (exercise~\ref{commute}) that two
matrices commute if and only if they lie on the same line through the origin in
\fref{sl2}, but they are allowed to lie in different parts. This line is unique
if not both matrices are ``null'' elements of $\grp SL(2)$, i.e.\ either
$\eins$ or $-\eins$. In this case they do not transform under inner
automorphisms, and we get four special equivalence classes of pairs of
commuting matrices called
\beq
\Klasse_{z_\loopa,z_\loopb}(\eins) =
\big\{(z_\loopa\eins,z_\loopb\eins)\big\},
\eeq
each consisting of a single pair only. In all other cases we can find a
non-vanishing vector $v^a$, an angle $\winkel$ and two signs
$z_\loopa,z_\loopb=\pm$ such that the matrices are
\beq[torus-par]
   \g_\loopa = \g(v^a \cos\winkel , z_\loopa ) , \qquad
   \g_\loopb = \g(v^a \sin\winkel , z_\loopb ) .
\eeq
Here we used the special coordinates $\g(v^a,\pm)$ on $\grp SL(2)$ introduced
in~\eref{sl2-par}. The vector has to be of length squared greater or equal to
$-1$, so the range for $v^a$ and the angle $\winkel$ is restricted by
\beq[range]
  \lange \ge -1 / \cos^2\winkel, \qquad
  \lange \ge -1 / \sin^2\winkel, \qquad v^a \ne 0,
\eeq
where $\lange=v^av_a$. There is still a sign ambiguity in this parameterization
as $v^a\mapsto-v^a$, $\winkel\mapsto\winkel+\pi$ does not change the
holonomies. We can fix this partly by requiring that $v^a$ lies inside or on
the {\em upper\/} lightcone if it is timelike or lightlike, then the only
remaining ambiguity occurs if it is spacelike or lies on the boundary, where
one of the signs is arbitrary.

How do the equivalence classes look like? In \sref{lorentz-group} we found that
an inner automorphism acts on the vector $v^a$ like a proper Lorentz rotation,
and this is of course still true with extra factors of $\cos\winkel$ or
$\sin\winkel$ here. Given two vectors $v^a$ with the same length squared
$\lange$, then we can always find a proper Lorentz rotation mapping them onto
each other. This is obviously true for spacelike vectors as the hyperboloids
with $\lange>0$ are connected. For timelike (or lightlike) vectors $v^a$ this
follows because then by definition the vector always lies on the {\em upper}
hyperboloid (or lightcone). Hence, an equivalence class of pairs of commuting
non-null holonomies is determined by the length $\lange$, the angle $\winkel$,
and the two signs $z_\loopa,z_\loopb$. We denote it by
\beq
\Klasse_{z_\loopa,z_\loopb}(\lange,\winkel) =
  \big\{ (\g(v^a \cos\winkel , z_\loopa )
         ,\g(v^a \sin\winkel , z_\loopb )) \big\}.
\eeq
Here $v^a$ runs over all non-vanishing vectors with the given length squared
$\lange$, lying inside or on the upper lightcone if $\lange\le0$.
\begin{figure}[t]
\caption{The set of equivalence classes of pairs of commuting $\grp SL(2)$
elements. Each class is parameterized by a length $\lange$, an angle $\winkel$,
 and two signs $z_\loopa,z_\loopb$. The picture shows the part $\Klasse_{++}$,
which forms a cylinder with a rather peculiar looking boundary. The complete
moduli space is obtained by identifying $\winkel$ with $\winkel+\pi$ in the
spacelike region $\lange>0$ and gluing the parts for different $z_\loopa,
z_\loopb$ together along the boundaries as indicated.}
\label{moduli}
\begin{indented}
\item[] \quad
\beginpicture
\setcoordinatesystem units <3em,3.5em>
\put {$\winkel=0$} [l] at 5.1 0
\put {$\winkel=\pi$} [l] at 5.1 2
\put {$\winkel=2\pi$} [l] at 5.1 4
\put {$-2$} [t] at -2 -.15  \plot -2 0 -2 -.1 /
\put {$-1$} [t] at -1 -.15  \plot -1 0 -1 -.1 /
\put {$0$} [t] at 0 -.15  \plot 0 0 0 -.1 /
\put {$1$} [t] at 1 -.15  \plot 1 0 1 -.1 /
\put {$2$} [t] at 2 -.15  \plot 2 0 2 -.1 /
\plot 3 0 3 -.1 /
\plot 4 0 4 -.1 /
\put {$\lange$} [t] at 4.5 -.1
\setdashes <.5ex>
\plot 0 0 0 4 /
\setsolid
\put {$\bullet$} at 0 4.4
\put {$\Klasse_{++}(\eins)$} [lb] at 0.1 4.4
\put {$\bullet$} at 2 0.7
\put {$\Klasse_{++}(\lange,\winkel)$} [lt] at 2.1 0.7
\put {$\bullet$} at 2 2.7
\put {$\Klasse_{++}(\lange,\winkel+\pi)$} [lb] at 2.1 2.7
\put {identify} [l] at 3.1 1.7
\Pfeil 3 1.7 2.1 0.8
\Pfeil 3 1.7 2.1 2.6
\put {$\Klasse_{+-}$} [r] at -1.4 1
\put {$\Klasse_{-+}$} [r] at -1.4 2
\put {$\Klasse_{+-}$} [r] at -1.4 3
\put {$\Klasse_{-+}$} [r] at -1.4 4
\put {$\circ$} at 0 2
\setplotsymbol ({\bf .})
\multiput {\coskurve} at 0 0 0 1 0 2 0 3 /
\plot -1 0 5 0 /
\plot -1 4 5 4 /
\plot 0.05 2 5 2 /
\endpicture
\end{indented}
\end{figure}
Fixing the two signs $z_\loopa,z_\loopb$, we can draw a picture of the
corresponding region of the moduli space, which is shown in \fref{moduli}. It
consists of the timelike region $\lange<0$, which is bounded by~\eref{range}, a
lightlike circle at $\lange=0$ and the spacelike part $\lange>0$ forming a
cylinder that extends to infinity. In addition, there is an extra point
$\Klasse_{z_\loopa,z_\loopb}(\eins)$ in each region.

\begin{figure}[t]
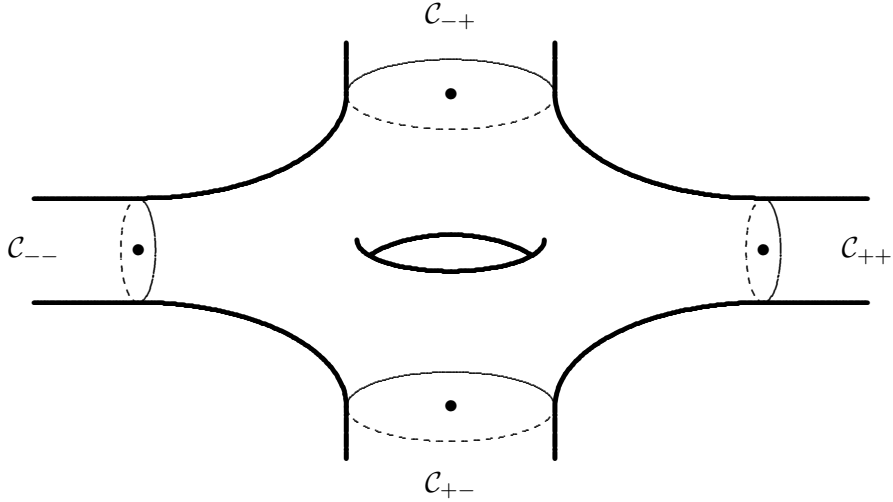

\caption{The moduli space of the torus, consisting of the equivalence classes
of commuting $\grp SL(2)$ matrices modulo inner automorphisms. It consists of
four spacelike sectors in form of cylinders extending to infinity, where
opposite points are identified, glue to a timelike sector, which is a torus
with four holes. In between there are four lightlike circles and null points.
Every point in this picture (except for the identifications in the spafcelike
sectors) corresponds to exactly one basis element of the physical state space.}
\label{torus-moduli}
\begin{indented}
\item[] \quad
\beginpicture
\setcoordinatesystem units <3.6em,1.8em>
\ellipticalarc axes ratio 3:1 180 degrees from 1 -3 center at 0 -3
\ellipticalarc axes ratio 3:1 180 degrees from 1  3 center at 0  3
\ellipticalarc axes ratio 1:3 180 degrees from -3 -1 center at -3 0
\ellipticalarc axes ratio 1:3 180 degrees from 3 -1 center at 3 0
\setdashes <.5ex>
\ellipticalarc axes ratio 3:1 -180 degrees from 1 -3 center at 0 -3
\ellipticalarc axes ratio 3:1 -180 degrees from 1  3 center at 0  3
\ellipticalarc axes ratio 1:3 -180 degrees from -3 -1 center at -3 0
\ellipticalarc axes ratio 1:3 -180 degrees from 3 -1 center at 3 0
\setsolid
\multiput {$\bullet$} at 0 3 3 0 -3 0 0 -3 /
\put {$\Klasse_{++}$} at 4 0
\put {$\Klasse_{--}$} at -4 0
\put {$\Klasse_{-+}$} at 0 4.5
\put {$\Klasse_{+-}$} at 0 -4.5
\setplotsymbol ({\bf .})
\ellipticalarc axes ratio 3:1  92 degrees from  0 -.4 center at  0 .2
\ellipticalarc axes ratio 3:1 -92 degrees from  0 -.4 center at  0 .2
\ellipticalarc axes ratio 7:4  48 degrees from  0 .3 center at  0 -.9
\ellipticalarc axes ratio 7:4 -49 degrees from  0 .3 center at  0 -.9
\ellipticalarc axes ratio 2:1 90 degrees from 3 -1 center at 3 -3
\ellipticalarc axes ratio 2:1 90 degrees from 1 3 center at 3 3
\ellipticalarc axes ratio 2:1 90 degrees from -3 1 center at -3 3
\ellipticalarc axes ratio 2:1 90 degrees from -1 -3 center at -3 -3
\plot -1 -3 -1 -4 / \plot 1 -3 1 -4 /
\plot -1  3 -1  4 / \plot 1  3 1  4 /
\plot -3 -1 -4 -1 / \plot -3 1 -4 1 /
\plot  3 -1  4 -1 / \plot  3 1  4 1 /
\endpicture
\end{indented}
\end{figure}

The coordinates are not yet unique. In the spacelike region we find that the
two classes for $\winkel$ and $\winkel+\pi$ are actually the same. The reason
is that we can rotate a spacelike vector into its negative (this is not
possible for a timelike of lightlike vector). We have
$\Klasse_{z_\loopa,z_\loopb}(\lange,\winkel) =
\Klasse_{z_\loopa,z_\loopb}(\lange,\winkel+\pi)$ for $\lange>0$. This leads to
the identification shown in the picture. Finally, the parts for different
values of $z_\loopa,z_\loopb$ have to be glued together.
\begin{exer}
By using the relation~\eref{csl2-par}, which tells us how group elements that
lie on the boundaries are identified, show that the cylinders in \fref{moduli}
are glued together as indicated in the picture, i.e.\ the timelike region of
$\Klasse_{++}$ is glued to $\Klasse_{-+}, \Klasse_{+-}$ etc.\ along the
boundary, so that all four parts together form a torus, with four cylinders
extending to infinity and four extra points.
\end{exer}
A picture of the compete moduli space of the torus is shown in
\fref{torus-moduli}.
An interesting, but for our purpose not very important feature of this moduli
space is its non-Hausdorff topology:
\begin{exer}
The moduli space has a natural topology. It is a quotient space of a closed
subset of $\grp SL(2)\times \grp SL(2)$ modulo an equivalence relation. Show
that this topology is locally like $\RR^2$ at a generic point (otherwise we
could not draw a picture), but non-Hausdorff at the lightlike circles and the
singular points $\Klasse_{z_\loopa,z_\loopb}(\eins)$. In particular, {\em
every} open neighbourhood of such a point contains the {\em whole} lightlike
circle, and two opposite points on this circle cannot be separated by open
neighbourhoods.
\end{exer}
We can now give a basis for the physical state space. There is one state for
each point in the moduli space, and the corresponding wave functional is given
by a distribution with support on the corresponding equivalence class of
commuting $\grp SL(2)$ elements only. As we will not integrate or differentiate
with respect to the holonomies, we do not have to introduce distributions etc.\
on the moduli space. If you want to consider the states as wave functions, just
take $\FS(\g_\loopa,\g_\loopb)$ to be one on the considered equivalence class
and zero elsewhere. The basis elements of the state space are
\beq
  \ket{\eins,z_\loopa,z_\loopb} , \qquad
  \ket{z_\loopa,z_\loopb,\lange,\winkel},
\eeq
where the range of the parameters and the identifications are of course the
same as for the equivalence classes. Let us introduce another set of
coordinates on the moduli space, which will be more convenient later on. The
states $\kt{\eins,z_\loopa,z_\loopb}$ will be called null states, and the
remaining will be divided into  timelike, lightlike, and spacelike states,
depending on $\lange$. Of course, this means that the holonomies are timelike,
lightlike or spacelike in the sense introduced in \sref{lorentz-group}. On each
of these sectors we will introduce separate coordinates and ``standard
representatives'' for the matrices.

Let us start with the timelike states. By a proper Lorentz rotation, we can
always achieve that the vector $v^a$ is vertical in \fref{sl2}, i.e.\ it lies
on the $\gam_0$ axis and the holonomies are elements of the $\grp SO(2)$
subgroup. Then we can find two real numbers $a,b$ such that
\beq[time-class]
  \g_\loopa = \pmatrix{ \cos a & \sin a \cr -\sin a & \cos a }, \qquad
  \g_\loopb = \pmatrix{ \cos b & \sin b \cr -\sin b & \cos b },
\eeq
The parameters $a,b$ are obviously periodic with period $2\pi$, but apart from
this they are unique, because every timelike hyperboloid intersects the $\grp
SO(2)$ subgroup exactly once. However, $a$ and $b$ cannot be both multiples of
$\pi$, as then we are back in the null sector. The states will be denoted by
\beq[time-state]
  \ket{\Stime,a,b}, \qquad
  (a,b) \ne  (m\pi,n\pi), \ m,n\in\ZZ.
\eeq
The new parameters $a,b$ are very different from the old $\lange,\winkel$
coordinates, as we now have one set of coordinates for the whole timelike
sector, which was glued together in a complicated way before. Now we can easily
see that it is a torus with four holes, but the boundaries are shrunk to
points.

For the spacelike classes the result is similar. We can always rotate a
spacelike vector such that it lies on the horizontal axis in \fref{sl2}, i.e.\
the $\gam_1$-axis. In contrast to the timelike sector, the spacelike sectors as
shown in \fref{torus-moduli} are not connected, so we have to keep the two
signs as parameters. The commuting matrices representing a spacelike class are
\beq[space-class]
  \g_\loopa = {z_\loopa}
              \pmatrix{ \cosh a & \sinh a \cr \sinh a & \cosh a }, \quad
  \g_\loopb = {z_\loopb}
              \pmatrix{ \cosh b & \sinh b \cr \sinh b & \cosh b },
\eeq
and the corresponding states are called
\beq[space-state]
  \ket{\Sspace,z_\loopa,z_\loopb,a,b} , \qquad
    (a,b) \ne  (0,0)
\eeq
Here we have to exclude $a=b=0$, which yields a null class. Of course, this
corresponds to the boundary of the spacelike region, which is also shrunk to a
point. In addition, this time the parameters are not unique, as a given
spacelike hyperboloid intersects the $\gam_1$ axis always twice and we can
rotate a spacelike vector into its negative by a continuous transformation.
This leads to the identification
\beq[space-id]
  \ket {\Sspace,z_\loopa,z_\loopb,a,b}
 = \ket {\Sspace,z_\loopa,z_\loopb,-a,-b}.
\eeq
Finally, every lightlike vector can be rotated into the special vector
$v^a=(\ft12,\ft12,0)$, because every vector on the upper lightcone (by
definition $v^a$ lies always there) can be transformed into every other. We
then keep the old coordinate $\winkel$ and get the representatives
\beq[light-class]
  \g_\loopa =  {z_\loopa}
              \pmatrix{ 1 & \cos \winkel \cr 0 &  1 }, \qquad
  \g_\loopa =  {z_\loopb}
              \pmatrix{ 1 & \sin \winkel \cr 0 &  1 },
\eeq
and the state functionals are denoted by
\beq[light-state]
  \ket{\Slight,z_\loopa,z_\loopb,\winkel} .
\eeq
To summarize, we now have a complete basis of the state space of the torus, and
a general element of this state space can be expanded as
\beq[torus-expand]
  \ket{\FS} & = &
  \sum_{z_\loopa,z_\loopb}  \FS_\eins(z_\loopa,z_\loopb)
                 \ket{\eins,z_\loopa,z_\loopb}
   +  \oint \d a\,\d b \, \FS_\Stime(a,b) \, \ket {\Stime,a,b}  \zl &&
 + \sum_{z_\loopa,z_\loopb}   \int \d a\, \d b \,
        \FS_\Sspace(z_\loopa,z_\loopb,a,b)
    \, \ket {\Sspace,z_\loopa,z_\loopb,a,b}  \zl &&
 + \sum_{z_\loopa,z_\loopb}   \oint \d \winkel \,
     \FS_\Slight(z_\loopa,z_\loopb,\winkel)
    \, \ket {\Slight,z_\loopa,z_\loopb,\winkel} ,
\eeq
where the wave function $\FS_\Sspace$ is assumed to be symmetric
$\FS_\Sspace(z_\loopa,z_\loopb,a,b) = \FS_\Sspace(z_\loopa,z_\loopb,-a,-b)$,
and $\FS_\Stime$ and $\FS_\Slight$ are periodic in $a,b$ and $\winkel$,
respectively.
\begin{exer}
The $\lange,\winkel$ coordinates above together with the $a,b$ coordinates on
the timelike and spacelike sectors provide a full atlas of the moduli space in
\fref{torus-moduli} (ignoring the singular points and the identifications in
the spacelike sector to make it a proper manifold). Find the transition
functions.
\end{exer}
Now we have to find out how the observables act on the basis of the physical
state space. As there is no boundary, we only have the trace observables and
some exceptional operators representing gauge independent properties of the
holonomies which are not contained in the trace. Because the loop group of the
torus is Abelian, every loop can be written as $\loopc=\loopa^m \kette
\loopb^n$ for some $m,n\in\ZZ$. The corresponding trace observables are
\beq[tor-obs]
  \UUU(m,n) &=& \ft12 \Trr{ \g_\loopc } =\ft12  \Trr{ \UU_\loopc(0,1) }, \zl
  \VVV(m,n) &=&  \Trr{ \g_\loopc \f_\loopc } =
       \intl01 \d s \Trr{ \UU_\loopc(0,s) \, \loopc'^i \gam_i
               \, \UU_\loopc(s,1) } .
\eeq
The $\UUU$ operators are obviously pure multiplication operators and they
commute with each other. In fact, they form an algebra, as every product of
them is again a linear combination. Explicitly, we have
\beq[UUU-alg]
  2 \, \UUU(m,n) \, \UUU(p,q) = \UUU(n+p,m+q) +  \UUU(n-p,m-q),
\eeq
and, as well as the $\VVV$'s, they are not independent. The most important
relations are
\beq[UUU-sym]
  \UUU(m,n)= \UUU(-m,-n), &\qquad &\UUU(0,0)=1, \zl
  \VVV(m,n)= \VVV(-m,-n), &\qquad &\VVV(0,0)=0.
\eeq
All these relations are derived from the following useful identities for $\grp
SL(2)$ matrices:
\beq[sl2-formulas]
  &&\g + \g^{-1} = \Trr{\g} \, \eins, \qquad
  \g - \g^{-1} = \Trr{\g\gam_a} \, \gam_a , \zl
  &&\Trr{\g} \, \Trr{\g'} = \Trr{ \g\g'} + \Trr{\g^{-1}\g'}.
\eeq
It is now straightforward to find out how the $\UUU$ operators act on the
physical states. They multiply them by the trace of the corresponding matrix
$\g_\loopa\^m\g_\loopb\^n$, and since the trace is gauge-invariant, we may
insert the ``standard representatives'' given above for the holonomies,
i.e~\eref{time-class} for the timelike states etc. The result is
\beq
  \UUU(m,n) \ket {\eins,z_\loopa,z_\loopb}
              &=& z_\loopa\^m z_\loopb\^n
               \ket {\eins,z_\loopa,z_\loopb}, \zl
 \UUU(m,n) \ket {\Slight,z_\loopa,z_\loopb,\winkel}
              &=& z_\loopa\^m z_\loopb\^n
               \ket {\Slight,z_\loopa,z_\loopb,\winkel}, \zl
 \UUU(m,n) \ket {\Sspace,z_\loopa,z_\loopb,a,b}
              &=& z_\loopa\^m z_\loopb\^n \cosh(ma+nb)
               \ket {\Sspace,z_\loopa,z_\loopb,a,b}, \zl
 \UUU(m,n) \ket {\Stime,a,b}
              &=& \phantom{z_\loopa\^m z_\loopb\^n }\cos(ma+nb)
               \ket {\Stime,a,b}.
\eeq
The basis states are obviously eigenstates of the $\UUU$ operators, but some
eigenspaces are still degenerate. Only the spacelike eigenspaces are one
dimensional, because by~\eref{space-id} $a,b$ and $-a,-b$ represent the same
state. In the timelike sector, the eigenspaces are two dimensional consisting
of the states $\kt{\Stime,a,b}$ and $\kt{\Stime,-a,-b}$, and the null and
lightlike sectors are completely degenerate.
\begin{exer}
Show that the following observables together with the $\UUU$ operators form a
complete set of commuting operators such that all eigenspaces are
non-degenerate:
\beq[Winkel]
   {\cal E} &=& \cases { 1 & for $\g_\loopa,\g_\loopb=\pm\eins$, \cr
                       0 & otherwise, }  \zl
  {\cal Z}_{\loopa} &=& \cases { \sign \Trr{\g_{\loopa} \gam_0 }
                 & if $\g_{\loopa}$ is lightlike or timelike \cr
                        0 & otherwise ,}  \zl
   &&\sin \Winkel \, \Trr{\g_\loopa \gam_a} =
           \cos \Winkel \, \Trr{\g_\loopb \gam_a }.
\eeq
The first checks whether the state is a null state, the second, together with a
similar observable for $\loopb$, tells us whether the matrices lie in the upper
or lower lightcone if they are timelike or lightlike. In the third equation the
angle $\Winkel$ is defined modulo $\pi$ only, but together with the $\cal Z$
observables we can fix its value on the lightlike states by requiring
\beq
   \Winkel \ket{\Slight,z_\loopa,z_\loopb,\winkel} =
   \winkel \ket{\Slight,z_\loopa,z_\loopb,\winkel}.
\eeq
\end{exer}
As we want all these observables to be hermitian, the scalar product on the
physical state space has to be such that the basis states are orthogonal.
Of course, we can then normalize the basis to be orthonormal and the scalar
product reads
\beq[sp]
  \braket{\eins,z_\loopa,z_\loopb}{\eins,z_\loopa,z_\loopb} &= &1, \zl
  \braket{\Slight,z_\loopa,z_\loopb,\winkel}
         {\Slight,z_\loopa,z_\loopb,\winkel'}
    & =& \delta(\winkel-\winkel') , \zl
  \braket{\Sspace,z_\loopa,z_\loopb,a,b}
         {\Sspace,z_\loopa,z_\loopb,c,d}
    & =& \delta(a-c)\, \delta(b-d)
         + \delta(a+c)\,\delta(b+d) , \zl
  \braket{\Stime,a,b}
         {\Stime,c,d}
    & =& \delta(a-c)\, \delta(b-d)  ,
\eeq
and vanishes if the signs $z_\loopa,z_\loopb$ do not coincide. Note that this
is actually not the final definition of the scalar product, it just defines the
product as a function of the basis elements. As the basis was only required to
consist of eigenstates of the $\UUU$ operators, we can still multiply all basis
elements by arbitrary complex factors. If we want the $\VVV$ observables to be
hermitian as well, we now have to choose the basis states themselves
appropriate.

The procedure to derive the product is slightly different from that used for
the cylinder. What we are going to do can be explained most easily using our
standard example again. Assume we have a particle in one dimension and let
$\UUU$ be the position operator. Then what we defined up to now is a basis
consisting of eigenvectors
$\UUU\kt x = x \kt x$, but we actually did not fixed the normalization of the
basis vectors $\kt x$. Nevertheless we can {\em define\/} a scalar product
depending on the basis by $\brkt xy=\delta(x,y)$: for each choice of basis
states we get a different scalar product. Now there is another operator $\VVV$
with $[\VVV,\UUU]=\i\hbar$.
The task is to fix the {\em basis\/} $\kt x$ such that $\VVV$ becomes
hermitian. From the commutator we learn that
\beq
  \comm\VVV\UUU \ket x = \i\hbar\, \ket x \follows
  \VVV \ket x = \i\hbar \dd /x/ \ket x - \hbar g'(x) \, \ket x,
\eeq
where $g$ is an unknown complex function. We can change the basis $\kt x
\mapsto \exp(ig(x))\,\kt x$, and in the new basis $\VVV$ simply acts as an
differential operator, and it also becomes hermitian with respect to the scalar
product derived from the new basis. In principle, it would be sufficient to
render $g$ real, but then we can still multiply the basis states by a phase
factor, which does not change the scalar product any more, but simplifies the
action of $\VVV$ on the basis. As a result, we get a basis and a scalar product
for the state space, without having to know their explicit representations in
terms of wave functions.

Hence, all we have to do is to compute the commutator algebra of the $\UUU$ and
$\VVV$ observables, then we know their action on the basis states up to some
function like $g$. Provided that it is possible to make all $\VVV$ observables
hermitian, there must be a choice for the $g$ function such that~\eref{sp} is
the correct product. Computing the commutator of $\UUU$ and $\VVV$ is
straightforward. We rewrite the $\VVV$ operators as
\beq[VVV-op]
  \VVV(m,n) = \intl01 \d s \,
         \Trr{ \U_\loopc(0,s) \gam_a \U_\loopc(s,1) } \,
                 \loopc'^i(s) \,  \e i a (\loopc(s)) ,
\eeq
where $\loopc$ is any loop homotopic to $\loopa^m\kette\loopb^n$.
Choosing another loop $\loopd\sim\loopa^p\kette\loopb^q$ such that it
intersects $\loopc$ at finitely many points only, we can use the commutator
\eref{comm-e-UU} to get
\beq
  \comm{\VVV(m,n)}{\UUU(p,q)} &=&
 \intl01 \d s \,  \Trr{ \U_\loopc(0,s) \gam_a \U_\loopc(s,1) } \,
            \loopc'^i(s) \comm{\e i a (\loopc(s)) }{ \Trr{\UU_\loopd(0,1)}}\zl
  &=& -\ft12\i  \hbar
      \intl01 \d s \, \d t \, \eps_{ij} \loopc'^i(s) \loopd'^j(t) \,
         \delta(\loopc(s),\loopd(t)) \,\times \zl &&
   \qquad\qquad
      \Trr{ \UU_\loopc(0,s) \, \gam_a \, \UU_\loopc(s,1) } \,
      \Trr{ \UU_\loopd(0,t) \, \gam^a \, \UU_\loopd(t,1) } .
\eeq
Using the second formula in \eref{sl2-formulas} the product of the two traces
becomes
\beq
  && \Trr{\UU_\loopc(0,s) \, \UU_\loopd(t,1) \,
        \UU_\loopd(0,t) \, \UU_\loopc(s,1) }   \zl
 && \qquad {}- \Trr{\UU_\loopc(0,s) \, \UU_\loopd(t,0) \,
        \UU_\loopd(1,t) \, \UU_\loopc(s,1) } .
\eeq
The delta function picks up a term only at intersection points of the two
curves. Hence, the first of the two traces is the trace of a transport operator
going along $\loopc$ to an intersection point, then around $\loopd$ to the same
point again, and finally further along $\loopc$, the second term is the same
but the direction of $\loopd$ is reversed. The values of these transport
operators are in fact independent of the special intersection point, as on the
torus they depend on the two winding numbers of the loop only. In particular,
they take the same values for every intersection point. The integral then
simply counts the intersections of the two loops, as already seen in
exercise~\ref{intersection}. Choosing the orientations of the loops such that
the intersection number of $\loopa$ with $\loopb$ is $+1$, the intersection
number of $\loopa^m\kette\loopb^n$ with $\loopa^p\kette\loopb^q$ is $mq-np$. We
finally end up with the commutator
\beq[comm-UU-VV]
  \comm{\VVV(m,n)}{\UUU(p,q)} =
   \ft12\i  \hbar (np-mq)  \big( \UUU(m+p,n+q) - \UUU(m-p,n-q) \big) .
\eeq
\begin{exer}
Show that a similar relation holds for the commutator of two $\VVV$ operators:
\beq[comm-VV-VV]
  \comm{\VVV(m,n)}{\VVV(p,q)} =
   \ft12\i \hbar (np-mq)  \big( \VVV(m+p,n+q) - \VVV(m-p,n-q) \big) ,
\eeq
either by computing it straightforward using~\eref{comm-e-UU}, or by showing
that it is linear in $\VVV$ again and using the Jacobi identities.
\end{exer}
We can now compute the action of the $\VVV$ operators on the basis states. For
the timelike states, we have
\beq
  \comm{\VVV(m,n)}{\UUU(p,q)} \ket {\Stime,a,b}  &=&
    \ft12\i \hbar (np-mq) \big( \UUU(m+p,n+q) - \UUU(m-p,n-q) \big)
             \ket{\Stime,a,b} \zl
  &=& -\i\hbar  (np-mq) \sin(ma+nb) \sin(pa+qb)  \ket{\Stime,a,b},
\eeq
where we used the formula
$\cos(x+y) - \cos(x-y) = -2 \sin x \sin y $.
We can rewrite the eigenvalue appearing in the last expression as
\beq
  \i\hbar \sin(ma+nb) \Big( n \deldel/a/ - m \deldel/b/ \Big) \cos(pa+qb),
\eeq
so that the action of the $\VVV$ operator is fixed to be
\beq[VV-time]
  \VVV(m,n) \, \ket {\Stime,a,b} = -\i \hbar
   \sin(ma+nb) \Big( n \deldel/a/ - m \deldel/b/ \Big) \ket {\Stime,a,b},
\eeq
up to an additional multiplicative term. However, one can easily check that the
commutator relations between the $\VVV$ operators are satisfied without
``quantum corrections'', and in this form they are also hermitian with respect
to the scalar product~\eref{sp}. Using our argument from above, any possible
multiplicative term can now be absorbed into a redefinition of the basis, so
that the last equation gives the correct action of the $\VVV$ operators on
timelike states. The computation for the spacelike sector is similar and gives
\beq[VV-space]
  \VVV(m,n) \ket {\Sspace,z_\loopa,z_\loopb,a,b}  =
  - \i \hbar z_\loopa\^m z_\loopb\^n
     \sinh(ma+nb) \Big( n \deldel/a/ - m \deldel/b/ \Big)
    \ket {\Sspace,z_\loopa,z_\loopb,a,b}.
\eeq
Another important fact to note here is that the vector fields represented by
the $\VVV$ operators on the timelike and spacelike sectors of the moduli space
all vanish at the excluded points $(0,0)$ for the spacelike sector and
$(m\pi,n\pi)$ for the timelike sector. Hence, acting with any observable on
spacelike and timelike states will never produce a state outside these sectors.
We get another split into superselection sectors. The lightlike sectors should
then of course also be mapped onto themselves by the $\VVV$ observables. To see
how they act there, we have to use some formulas:
\begin{exer}
Compute the commutator of the $\VVV$ observables with the matrices $\g_\loopa$
and $\g_\loopb$. It is essential that they commute (as matrices), otherwise the
result would depend on the special loop chosen in~\eref{VVV-op}:
\beq
  \comm{\VVV(m,n)}{ \g_\loopa } &=
    &   \ft12\i  \hbar n \big( \g_\loopa^{1+m} \g_\loopb^n
                     - \g_\loopa^{1-m} \g_\loopb^{-n} \big)\zl
  \comm{\VVV(m,n)}{ \g_\loopb } &=
    & -    \ft12\i  \hbar m \big( \g_\loopa^m \g_\loopb^{1+n}
                    - \g_\loopa^{-m} \g_\loopb^{1-n} \big).
\eeq
Show also that for lightlike matrices we have
\beq[light-trace]
  \Trr{\g_\loopa\^m \g_\loopb\^n \gam_a } =
   m \Trr{\g_\loopa \gam_a } + n \Trr{\g_\loopb\gam_a} .
\eeq
\end{exer}
We can now derive the commutator of $\VVV(m,n)$ with the angular variable
$\Winkel$ introduced implicitly by~\eref{Winkel}. We just compute the
commutator with the whole equation, which of course has to vanish. This gives
\beq
   \lefteqn{-\comm{\VVV(m,n)}{ \Winkel }
          \big( \cos \Winkel \Trr{\g_\loopa\gam_a}
                    + \sin \Winkel \Trr{\g_\loopb \gam_a} \big)}\zl
  &=& \sin \Winkel  \comm{ \VVV(m,n)}{ \Trr{\g_\loopa\gam_a } }
  - \cos \Winkel  \comm{ \VVV(m,n)}{ \Trr{\g_\loopb\gam_a } } \zl
  &=& \i\hbar \, \big( n \sin \Winkel + m \cos \Winkel \big) \,
      \big( m \Trr{\g_\loopa\gam_a } + n \Trr{\g_\loopb\gam_b} \big),
\eeq
where we used the formulas given in the exercise and the fact the the matrices
are lightlike. Now~\eref{Winkel} implies that there is a vector $v_a$ such that
$\Tr(\g_\loopa\gam_a)=2v_a\cos\Winkel$ and
$\Tr(\g_\loopb\gam_a)=2v_a\sin\Winkel$ (which of course is the vector
introduced in~\eref{torus-par}). Inserting this into the last equation gives
\beq[VV-Winkel]
  \comm{\VVV(m,n)}{\Winkel} =
     \i\hbar \, (n \sin\Winkel + m \cos \Winkel)^2
\eeq
{}From this we could infer that $\VVV(m,n)$ has to acts on lightlike states as
\beq
   \VVV(m,n) \, \ket {\Slight,z_\loopa,z_\loopb,\winkel} =
    -\i\hbar (n \sin\winkel + m \cos \winkel)^2  \deldel/\winkel /
  \ket {\Slight,z_\loopa,z_\loopb,\winkel} ,
\eeq
but this not hermitian. Here we have to add an imaginary multiplicative term,
or even simpler we have to ``symmetrize'' the operator:
\beq
   \VVV(m,n) \, \ket {\Slight,z_\loopa,z_\loopb,\winkel}  =
    - \i \hbar\, (n \sin\winkel + m \cos \winkel)  \deldel/\winkel /
        (n \sin\winkel + m \cos \winkel)\,
   \ket {\Slight,z_\loopa,z_\loopb,\winkel} ,
\eeq
which is obviously hermitian with respect to the product~\eref{sp}.
This completes the construction of the physical Hilbert space for the torus.
All observables now have well defined action on the basis states and we saw
that it is possible to render all these operators hermitian with respect to a
unique scalar product. The state space splits into 13 superselection sectors,
the four spacelike, lightlike and null sectors and the single timelike sector
(in addition to the split into superselection sectors with respect to the $\cal
P$ operators above, leading to sectors which cannot be distinguished by an
observer sitting on the torus). A general state functional is given by
expanding it as shown in~\eref{torus-expand}, and restricting the wave
functions of the non-compact spacelike sectors to be square integrable, i.e.
\beq
   \int \d a\, \d b \, \FS_\Sspace\^* \FS_\Sspace < \infty,
\eeq
we finally get a well defined Hilbert space. Using that the scalar product is
unique up to rescaling, we can answer some questions arising in a naive
description of the state space of the torus, where one assumes that the wave
function can be written as function of the traces of holonomies.
First of all, we immediately see that in this way we cannot cover the whole
state space, because the traces alone cannot distinguish between lightlike and
null states. So already then we loose some states. Some approaches go even
further trying to construct a Fock-like space based on the $\UUU$ operators as
creation operators. This leads to the so called loop states or loop
representation, providing a discrete basis for the state space. There have been
some discussions about the existence of such a
basis~\cite{smolin:90,marolf:93,ashtekar.loll:94,matschull:94a}, so let us
close this section with a brief look at these states.

We first have to find a suitable definition for loop states in our notation.
Usually they are defined as wave functions $\FS=\Tr(\g_\loopc)$, but as we do
not know how our basis above looks like when expressed in terms of wave
functions, we have to construct them using some basic properties. In fact, the
loop states can be generated as follows. There is a ``vacuum'' $\kt{0,0}$,
represented by a constant wave function. In our formalism, this means that
acting on it with the differential operators $\VVV(m,n)$ gives zero. Then the
loop states are
\beq[loop]
  \ket{m,n} = \UUU(m,n) \ket{0,0}.
\eeq
They obviously have a nice generalization to higher genus manifolds, because we
get a state for each homotopy class of loops on $\cov\N$. By using commutation
relations, one can derive the action of the $\UUU$ and $\VVV$ operators on
these states, and they are quite simple. However, in general the loop states
are not normalizable, and they actually provide a very inconvenient basis of a
subset of the state space. If $\VVV(m,n)\kt{0,0}=0$ for all $m,n$, then all the
functions $\FS$ in the expansion~\eref{torus-expand} must be constants, but not
necessarily equal. If we want the vacuum to be normalizable, the spacelike
parts must vanish, because otherwise we would get an infinite contribution to
the norm of the state. So loop states can only be timelike, spacelike, or null,
if we want them to be normalizable. On the other hand, in the null and
lightlike sectors we cannot build up a Fock structure, because acting with the
$\UUU$ operators on the vacuum just reproduces it. Hence, the only useful loop
states are timelike. We can compute them explicitly and get
\beq
  \ket{m,n} = \oint \d a\, \d b\,
               \cos(ma +nb) \, \ket {\Stime,a,b}.
\eeq
Note that in a similar construction for the spacelike sector a $\cosh$ would
appear, so that the norms would be even more divergent than the norm of the
vacuum. For the timelike states, however, the norm is finite, and they are in
fact orthonormal:
\beq
  \braket{m,n}{p,q} =2 \pi^2 (\delta_{m,p} \delta_{n,q} +
      \delta_{m,-p} \delta_{n,-q} ) ,
\eeq
the symmetrization appearing here because the $\UUU$ operators have the
symmetries~\eref{UUU-sym}. But still they cover only half of the timelike
sector, the remaining states with $\cos$ replaced by $\sin$ cannot be written
as loop states. Hence, we have to conclude that loop states do not provide a
suitable basis for the physical state space. They only exist when restricted to
the timelike sector, but this means that they are not represented by wave
functions of the form $\FS=\Tr(\g_\loopa\^n \g_\loopb\^m)$, which doesn't
vanish on the spacelike sector. Even the existence of the timelike loop states
is rather accidental and a special feature of the torus. It is essential that
there is a superselection sector of the state space corresponding to a finite
region of the moduli space. For the cylinder there are no such sectors and the
wave functions $\FS(\tg,\b)=\Tr(\g^n)$ are obviously not normalizable.

Coming back once again to simple point particle quantum mechanics, it is quite
obvious to see what goes wrong. If you want to find a Fock-like basis for the
state space for a particle in one dimension, you can, e.g., take the harmonic
oscillator eigenfunctions to be your basis states, and this of course works
perfectly, because you have the ``correct'' relations between the creation and
annihilation operators, which are the hermitian conjugate of each other. For
the loop states this is totally different, as the creation operators themselves
are hermitian. The quantum mechanical analogy would be to define the ``vacuum''
by $p\kt{0}=0$ and to create the basis states by $x\kt n=\kt {n+1}$. This
obviously leads to the wave functions $\Psi_n(x)=x^n$, which are not
normalizable and do not provide any reasonable basis of the state space. The
only exception occurs if the range of $x$ is finite, and this corresponds to
our timelike loop states. \appendix
\section*{Appendix}
\def\theequation{A.\arabic{equation}}
\setcounter{equation}{0}
In \sref{class} we did not give the details of how to compute the equations of
motion and the Poisson brackets for the boundary fields $\u$ and $\v$. The
technical details of these computations are rather cumbersome, though the
results can be summarized in the simple equations \eref{u-v-evol} and the
brackets~(\ref{o-v-pois}--\ref{bnd-pois}). In principle only one of the two
calculations is necessary, as once we have the Poisson brackets we could obtain
the equations of motions by computing the brackets with the Hamiltonian.
However, let us nevertheless do both independently and check whether the
results are correct.

To compute the equations of motion for $\u$ and $\v$, we split
the Lagrangian into an interior and a boundary term. We take the interior term
to be everything which does not depend on $\u$ or $\v$, i.e.~\eref{Lag}, plus
the $\oo_\phi$ and $\gam_\phi$ part of \eref{extra-L}. With Stokes' theorem we
can write this as a single integral over the interior of $\N$:
\beq[Lag-int]
\Lagint  =  \intN{ \eps^{ij} }{ \dot \oo_i \gam_j  - \D_i  \oo_t \,  \gam_j
            - \del_i \gam_t \, \oo_j +  \gam_t \oo_i \oo_j }.
\eeq
The boundary term is the rest of \eref{extra-L} plus the kinetic term
\eref{Lag-u-v}. By using \eref{u-v-periodic} we can rewrite it as
\beq[Lag-bnd]
  \Lagbnd &=& \send \Big(
        \Trr{ \u_\endn\inv \dot \u_\endn \, \v(\phi_0) } \zl
       && \hspace*{3em} {}- \intdNl{\phi_0} {\u^{-1} \dot\u \, \gamphi
               - \oo_t \gamphi - \gam_t \oophi } \Big),
\eeq
where $\gamphi$ and $\oophi$ are periodic functions of $\u$ and $\v$:
\beq[u-v-phi]
    \oophi = \u^{-1} \del_\phi \u , \qquad
    \gamphi= \u^{-1} \del_\phi \v\,\u.
\eeq
Only the multiplier fields appear in both parts and we can derive the equations
of motion for the remaining fields independently. Varying $\oo_i$ and $\gam_i$
in $\Lagint$ immediately gives \eref{evol-o-g}. Now consider a variation of
$\v$ in $\Lagbnd$. As $\v$ is completely (and uniquely) determined by
$\v(\phi_0)$ and $\gamphi$ via the differential equation \eref{u-v-phi}, we can
equally well vary $\gamphi$ and $\v(\phi_0)$ independently, which gives us the
equations of motion
\beq[eom-u]
   \u^{-1} \dot \u = \oo_t , \qquad   \u_\endn\inv \dot \u_\endn = 0 .
\eeq
So we already found the first part of \eref{u-v-evol}. The variation of $\u$ is
slightly more complicated, because we cannot vary $\u$ without varying $\v$:
the relation \eref{u-v-quasiper} has to remain invariant. However, as before we
can vary $\oophi$ and $\u$ at some special point independently, keeping
$\gamphi$ and $\v(\phi_0)$ fixed. To find the variation of $\Lagbnd$, we can
use
\beq[ddu]
  \delta( \u^{-1} \dot \u) = \u^{-1} \del_t ( \delta\u \, \u^{-1} ) \, \u ,
\eeq
and a similar formula for $\u_\endn$, which follows from the identity $\del
\u^{-1}=-\u^{-1}\del\u\,\u^{-1}$ for any matrix $\u$ and any differential
operator $\del$. Inserting this and integrating time derivatives by parts if
they act on variations, we find
\beq
  \delta \Lagbnd = \send \Big(
         - \Trr{ \u_\endn\inv \delta \u_\endn \, \dot\v(\phi_0) }
      + \intdNl{\phi_0} {\delta\u \,\u^{-1} \del_\phi \dot \v
               +\gam_t \, \delta\oophi } \Big),
\eeq
where we also dropped terms with $\dot\u_\endn$ vanishing by \eref{eom-u}.
Integrating the $\del_\phi$ derivative by parts, we can use
\beq
  \del_\phi( \delta \u \, \u^{-1} ) =
  \u \,  \delta ( \u^{-1} \del_\phi \u) \, \u^{-1} = \u^{-1} \delta\oophi \u  ,
\eeq
and we have to take a boundary term into account:
\beq
   \delta \Lagbnd &=& \send \Big(
        \Trr{     \delta\u\,\u^{-1} \dot\v \bei{\phi_0}{\phi_0+2\pi}
               -  \u_\endn\inv \delta \u_\endn \, \dot\v(\phi_0)  }   \zl
      && \hspace*{3em}
   {}+ \intdNl{\phi_0} {\delta\oophi ( \gam_t -  \u^{-1}  \dot \v \u ) }
               \Big).
\eeq
Making use of \eref{u-v-quasiper} again, we find that the first term can be
converted into
\beq
  \delta \Lagbnd = \send \Big(
        \Trr{     \delta\u \,\u^{-1}(\phi_0+2\pi) \dot\v_\endn  }
      - \intdNl{\phi_0} {\delta\oophi ( \u^{-1}  \dot \v \, \u - \gam_t )  }
               \Big).
\eeq
Thus we may consider $\delta\oophi$ and $\delta\u(\phi_0+2\pi)$ as independent
and get the equations
\beq
  \u^{-1}  \dot \v \, \u = \gam_t, \qquad
  \dot\v_\endn=0,
\eeq
which is the second part of \eref{u-v-evol}.

The Poisson brackets of the boundary fields are also slightly complicated to
obtain, mainly because they are group and algebra valued and quasiperiodic.
There are various methods to derive the brackets. For example, one could treat
the conditions \eref{u-v-quasiper}, defining $\u$ and $\v$ to be quasiperiodic,
as second class constraints, as well as the requirement that $\u\in\grp SL(2)$
and $\v\in\alg sl(2)$. Then $\u$ and $\v$ would become matrix fields on a real
axis and one could derive the Dirac brackets in the usual way \cite{dirac:65}.
For group valued fields, there are methods to avoid this detour via Dirac
brackets \cite{matschull.nicolai:93b}, but still we have to deal with the
quasiperiodic character of the fields.
However, after all the only requirement to be imposed in the brackets is that
they reproduce the correct equation of motion, which have to coincide with the
Euler Lagrange equations just derived. This leads to a nice trick to check
whether some ``suggested'' brackets are correct, and we will use this trick to
show that the brackets for the boundary fields given below are the right ones.
This is much easier than calculating them by extracting the canonically
conjugate momentum of $\u(\phi)$ and solving second class constraints.

To use this trick, we define a very special bracket, namely that of a phase
space function $F$ with the kinetic part of the Lagrangian, which is always
assumed to be of first order, so that the kinetic part simply consists of the
terms containing a time derivative. The Lagrangian then splits into
$\Lag=\Lagkin-\Ham$, where $\Ham$ is the Hamiltonian. In general, the Poisson
bracket can be written as
\beq
  \pois FG = \sum_{r,s} \deldel F/q_r/ \pois{q_r}{q_s} \deldel G / q_s /,
\eeq
where $q_r$ is any set of coordinates on the phase space. The bracket with
$\Lagkin$ is defined by replacing
\beq
  \deldel \Lagkin / q_r / \to \deldel \Lagkin / q_r /
              - \dd /t/ \deldel \Lagkin /\dot q_r /.
\eeq
It just means that whenever we compute a Poisson bracket with a velocity, we
first integrate the kinetic Lagrangian by parts. This is reasonable because
$\Lagkin$ is anyway defined up to a total time derivative only. The result of
such a bracket is again an expression which is of first order in time
derivatives.

How does this help us to find brackets? Suppose we have a suggestion for the
Poisson brackets, then is is quite obvious that
\beq
  \pois { F }{ \Lagkin } - \pois F{\Ham}
\eeq
is proportional to the equations of motion. Hence, if for every phase space
function $F$ we have
\beq[F->dF]
  \pois {F}{\Lagkin} = \dot F,
\eeq
then the brackets {\em do\/} provide the correct equations of motion. In our
case, the Lagrangian has a rather simple structure, and we can even proof that
under the condition \eref{F->dF} the brackets coincide exactly with the correct
Poisson brackets. If we summarize the degrees of freedom of $\u$ by a set of
coordinates $q_\alpha$ and those of $\v$ by $p_a$, then the kinetic Lagrangian
has the form $\Lagkin=p_\alpha M_{\alpha a}\dot q_a$, i.e.\ it is linear in
both $\dot q_a$ and $p_\alpha$, and the matrix $M_{\alpha a}$ depends on the
$q$'s only. We do not assume that $M_{\alpha a}$ is quadratic or even
invertible.

Now suppose we have the Poisson brackets such that the $q_a$ commute and
\eref{F->dF} holds. Take $F=q_b$:
\beq
   \dot q_b =
  \pois{q_b}{p_\alpha M_{\alpha a}\dot q_a} =
   \pois{q_b}{p_\alpha}  M_{\alpha a}\dot q_a .
\eeq
Thus $M_{\alpha a}$ is left invertible and $\{q_b,p_\alpha\} M_{\alpha
a}=\delta_{ab}$. Then choose $F=p_\beta$:
\beq
   \dot p_\beta &= &
  \pois{p_\beta}{p_\alpha M_{\alpha a}\dot q_a} \zl
  &= &  \pois{p_\beta}{p_\alpha} M_{\alpha a}\dot q_a
   + p_\alpha \pois{p_\beta}{q_b} \del_b M_{\alpha a} \dot q^a  \zl
  && {} - p_\alpha \del_b M_{\alpha a} \dot q_b \pois{p_\beta}{q_a}
        - \dot p_\alpha  M_{\alpha a} \pois{p_\beta}{q_a}
\eeq
where $\del_b$ denotes the derivative with respect to the variable $q_b$. The
first three terms must cancel and from the last term we infer that $M_{\alpha
a}$ is also right invertible, the bracket $\{q_a,p_\alpha\}$ being the inverse.
Some more algebra then shows that cancelation of the first three terms implies
\beq
  \pois{p_\alpha}{p_\beta} M_{\alpha a} M_{\beta b} =
   p_\beta \del_a M_{\beta b} - p_\alpha \del_b M_{\alpha a} .
\eeq
Using this and the previous results we find
\beq
 \pois{p_\alpha M_{\alpha a} }{p_\beta M_{\beta b}}=0, \qquad
 \pois{p_\alpha M_{\alpha a} }{q_b} = \delta_{ab}.
\eeq
Now $p_\alpha M_{\alpha a}$ is exactly the ``canonically conjugate'' momentum
of $q_a$, so the brackets are indeed the correct ones, which one can
equivalently obtain directly from $\Lagkin$.

Obviously, deriving the brackets from $\Lagkin$ in the usual way seems to be
much simpler. However, observe that \eref{F->dF} does not require a suitable
basis of the phase space, like $q_a,p_\alpha$ above, to check whether the
brackets are correct or not. In fact, it is not easy to find a simple basis for
quasiperiodic fields. However, there is a simple (over-complete) set of phase
space functions to test \eref{F->dF}, namely the values $\u(\phi),\v(\phi)$,
$\phi\in\RR$. A disadvantage of this method is that it does not help us to
derive the brackets. The ansatz is, however, motivated by using the general
methods for group valued canonical fields~\cite{matschull.nicolai:93b}.

Because of the derivative acting on $\v$ in the Lagrangian, we expect the
brackets to be non-local, and in fact the full brackets of $\u$ with $\v$ and
of $\v$ with itself are rather awkward. They simplify considerably (but are
still not as simple as those for $\oo_i$ and $\gam_i$) if we fix $\phi_0$
appearing in \eref{Lag-u-v}, and consider $\v(\phi_0)$ and $\gamphi(\phi)$ as
independent variables instead of $\v(\phi)$, as we already did above. The
kinetic Lagrangian   expressed in terms of these variables is
\beq[Lag-u-v']
  \Lagkin = \send \Big(
        \Trr{ \u_\endn\inv \dot \u_\endn \, \v(\phi_0) }  -
      \intdNl{\phi_0} {\u^{-1} \dot\u \, \gamphi } \Big).
\eeq
Let us start with the bracket of $\u(\phi)$ with $\v(\phi_0)$. We make the
ansatz
\beq[pois-u-v]
    \pois{\u(\phi)}{\Trr{\a \v(\phi_0)}} =
     \stepf(\a,\phi-\phi_0) \, \u(\phi) ,
\eeq
where $\stepf(\a,\phi)$ is an $\alg sl(2)$ valued function depending linearly
on $\a$, and possibly also on $\u$. Note that $\phi_0$ was arbitrary in
\eref{Lag-u-v'}, so this is the complete bracket of $\u$ and $\v$. We can now
compute the bracket of $\v(\phi_0)$ with the periodicity parameter $\u_\endn$
and get
\beq
   \pois{ \u_\endn}{\Trr{\a \v(\phi_0)}} &=&
   \pois{ \u(\phi+2\pi) \u^{-1}(\phi) } { \Trr{\a \v(\phi_0)}} \zl &=&
   \stepf(\a,\phi-\phi_0+2\pi) \,\u_\endn - \u_\endn \,\stepf(\a,\phi-\phi_0).
\eeq
Now the right hand side must be independent of $\phi$. Hence, there must be a
matrix $\stepf_0(\a)\in\alg sl(2)$ such that
\beq[stepf-per]
  \stepf(\a,\phi+2\pi)\, \u_\endn  = \u_\endn \stepf(\a,\phi)
       + \u_\endn \stepf_0(\a).
\eeq
As the last bracket is then also inpendent of $\phi_0$, we infer that the
bracket of $\u_\endn$ with $\gamphi=\u^{-1}\del_\phi\v\u$ vanishes. This leads
to the following bracket of $\u_\endn$ with $\Lagkin$, which has to reproduce
the velocily of $\u_\endn$:
\beq
      \pois{\u_\endn}{\Lagkin} =
   \pois{\u_\endn}{\Trr{\u_\endn\inv \dot \u_\endn \v(\phi_0)}} =
   \u_\endn \stepf_0(\u_\endn \dot \u_\endn ) = \dot \u_\endn.
\eeq
Thus we must have $\stepf_0(\a)=\a$. To get the correct bracket of $\u$ with
the kinetic Lagrangian, it turns out that the only additional requirement is
\beq[stepf=0]
  \stepf(\a,\phi)=0 \txt{for} 0 < \phi < 2 \pi.
\eeq
Together with the quasiperiodicity \eref{stepf-per} this defines $\stepf$
uniquely as a stepfunction in $\phi$, up to the indeterminate values at the
points $\phi=2\pi z$. One can easily check (by calculating the hights of the
steps recursively) that the derivative of the stepfunction is
\beq
  \del_\phi \stepf(\a,\phi) =
    \sum_{z\in\ZZ} (\u_\endn)^z \a \, (\u_\endn)^{-z} \,
           \delta(\phi-2\pi z) .
\eeq
Differentiating \eref{pois-u-v} with respect to $\phi$ or $\phi_0$ and using
the quasiperiodicity relation for $\u$ gives
\beq[pois-u-d-v]
  \pois{\u(\phi)}{\Trr{\a \u^{-1}\del_{\phi}\v\,\u(\phi_0)}}  &= &
    {} -  \u(\phi) \a  \, \pdelta(\phi-\phi_0) , \zl
  \pois{ \u^{-1} \del_\phi \u (\phi) }{ \Trr{\a\v(\phi_0)} } &= &
    \u^{-1}(\phi_0)\, \a \, \u(\phi_0) \, \pdelta(\phi-\phi_0),
\eeq
where $\pdelta(\phi)=\sum \delta (\phi-2\pi z)$ is the periodic delta function
on the circle. This is already the bracket given in \eref{u-g-pois}, as
$\u^{-1}\del_\phi\v\,\u=\gamphi$ and $\u^{-1}\del_\phi\u=\oophi$.
Using these formulas we can compute the bracket of $\u(\phi)$ with $\Lagkin$.
As we can choose $\phi_0$ in \eref{Lag-u-v'} arbitrarily, we take
$\phi_0<\phi<\phi_0+2\pi$, i.e.\ $\phi$ lies inside the range of the integral
and the bracket of $\u(\phi)$ with $\v(\phi_0)$ vanishes by \eref{stepf=0}:
\beq
  \pois{\u(\phi)}{\Lagkin} &= &
  \pois{\u(\phi)}{ - \oint_{\phi_0}\d\phi'
                 \,\Trr{\u^{-1}\dot\u\, \gamphi} } \zl & = &
   \oint_{\phi_0}
             \d\phi'\, \u(\phi) \, \u^{-1}(\phi') \, \dot\u(\phi') \,
         \delta(\phi-\phi')
       = \dot \u(\phi).
\eeq
Finally, we have to define the brackets of $\v$ with itself. They are fixed by
requiring that any bracket of $\v$ with itself is again linear in $\v$, and of
course the Jacobi identities must hold. Taking the bracket of \eref{pois-u-v}
with $\Tr(\b\v(\phi_0))$ (with the same angular coordinate again), and
antisymmetrizing in $\a,\b$, leads to the bracket
\beq[v-v]
  \pois{ \Trr{\a\v(\phi_0)}}{\Trr{\b\v(\phi_0)}} =
        -  \Trr{\comm\a\b \v(\phi_0)}.
\eeq
Note that this is valid only if the two $\v$'s are taken at the same point.
{}From the first line in \eref{pois-u-d-v} we infer that $\Tr(\a\gamphi)$ acts
on $\u$ by multiplying it from the right with $\a$, whereas $\Tr(\a\v)$ in
\eref{pois-u-v} acts by multiplication from the left with some other matrix.
Hence, these two actions commute and using the Jacobi identities gives
\beq
  \pois{ \Trr{\a\,\u^{-1}\del_\phi\v\,\u(\phi)}}{\Trr{ \b\v(\phi_0)}} = 0 .
\eeq
This is the first line of \eref{u-g-pois} and can be transformed into a
differential equation for the commutator of $\v(\phi)$ with $\v(\phi_0)$, which
is obviously linear in $\v$ again, because we have the initial condition
\eref{v-v}. Note that, in constrast to the bracket \eref{pois-u-v}, that
between the $\v$'s is an {\em integrated} stepfunction, and hence it is
continuous in the angular coordinate, so that its value at the special point
\eref{v-v} is well defined. We do not need the full bracket explicitly to check
whether they are correct. If we want to compute the bracket of $\v(\phi)$ with
the kinetic Lagrangian, we can take $\phi=\phi_0$, because $\phi_0$ is
arbitrary in \eref{Lag-u-v'}. Then $\v(\phi_0)$ commutes with the integral, as
it commutes with $\gamphi$ and by \eref{pois-u-v} it also commutes with
$\u(\phi)$ within the range of the integral $\phi_0<\phi<\phi_0+2\pi$ (there is
no {\em distributional\/} contribution at the boundary, just a finite one which
does not enter into the value of the integral). So we only have to consider the
bracket with the first term. A formula similar to \eref{ddu} can be used to
simplify the calculation:
\beq
  \pois X{\u_\endn\inv \dot \u_\endn} =
    \u_\endn\inv
    \del_t \big( \pois X{ \u_\endn } \u_\endn \inv \big)\, \u_\endn,
\eeq
where $\del_t$ acts on $\u_\endn$ only.
Using this we find with a partial integration of the $\del_t$:
\beq
   \lefteqn{\pois{ \Trr{\a\v(\phi_0)} } {
       \Trr{ \u_\endn\inv \dot \u_\endn \, \v(\phi_0)}}}
  \zl& =& \Trr{  \u_\endn \a\, \u_\endn\inv
         \del_t (\u_\endn \v(\phi_0) \u_\endn\inv ) -
         \comm{\a}{ \u_\endn\inv \dot \u_\endn } \v(\phi_0) }
       = \Trr{ \a \dot  \v(\phi_0) }.
\eeq
Hence, with the brackets introduced, \eref{F->dF} holds and they are the
correct Poisson brackets.

\end{document}